\documentclass[11pt]{article}
\pdfoutput=1
\usepackage{jheppub}

\newcommand\fft[2]{{\frac{#1}{#2}}}

\begin{document}

\preprint{MCTP-15-27}

\title{\boldmath High-temperature asymptotics of supersymmetric partition functions}

\author{Arash Arabi Ardehali}

\affiliation{Michigan Center for Theoretical Physics, Randall Laboratory of Physics,\\
The University of Michigan, Ann Arbor, MI 48109--1040, USA}

\emailAdd{ardehali@umich.edu}

\abstract{We study the supersymmetric partition function of 4d
supersymmetric gauge theories with a U($1$) R-symmetry on Euclidean
$S^3\times S_\beta^1$, with $S^3$ the unit-radius squashed
three-sphere, and $\beta$ the circumference of the circle. For
superconformal theories, this partition function coincides (up to a
Casimir energy factor) with the 4d superconformal index.

The partition function can be computed exactly using the
supersymmetric localization of the gauge theory path-integral. It
takes the form of an elliptic hypergeometric integral, which may be
viewed as a matrix-integral over the moduli space of the holonomies
of the gauge fields around $S_\beta^1$. At high temperatures
($\beta\to 0$, corresponding to the hyperbolic limit of the elliptic
hypergeometric integral) we obtain from the matrix-integral a
quantum effective potential for the holonomies. The effective
potential is proportional to the temperature. Therefore the
high-temperature limit \emph{further localizes} the matrix-integral
to the locus of the minima of the potential. If the effective
potential is positive semi-definite, the leading high-temperature
asymptotics of the partition function is given by the formula of
Di~Pietro and Komargodski, and the subleading asymptotics is
connected to the Coulomb branch dynamics on $R^3\times S^1$. In
theories where the effective potential is not positive
semi-definite, the Di~Pietro-Komargodski formula needs to be
modified. In particular, this modification occurs in the SU($2$)
theory of Intriligator-Seiberg-Shenker, and the SO($N$) theory of
Brodie-Cho-Intriligator, both believed to exhibit ``misleading''
anomaly matchings, and both believed to yield interacting
superconformal field theories with $c<a$.

Two new simple tests for dualities between 4d supersymmetric gauge
theories emerge as byproducts of our analysis.}

\maketitle \flushbottom

\section{Introduction}

Knowledge of the high-temperature asymptotics of the elliptic genera
of 2d superconformal field theories (SCFTs) has allowed micro-state
counting of certain supersymmetric Black~Holes
\cite{Strominger:1996,Breckenridge:1997,Kraus:2007}. In this work we
study the high-temperature asymptotics of the 4d analog of the
elliptic genus. This is the supersymmetric (SUSY) partition function
$Z^{\mathrm{SUSY}}(b,\beta)$, defined by the path-integral of the
theory on Euclidean $S_b^3\times S_\beta^1$, with $\beta$ the
circumference of the circle, and $S_b^3$ the unit three-sphere with
squashing parameter $b$; the round three-sphere corresponds to
$b=1$, and we assume $b$ to be a positive real number throughout
this paper. The superscript SUSY is added to emphasize that $i)$ the
path-integral is computed with periodic boundary conditions around
the circle, $ii)$ the Lagrangian used for path-integration is made
compatible with supersymmetry on $S_b^3\times S_\beta^1$, and $iii)$
a background U($1$)$_R$ gauge field is turned on along $S_\beta^1$
in order to make the supercharges independent of the ``time''
coordinate parameterizing the circle (see
\cite{Romelsberger:2005eg,Festuccia:2011}). In analogy with thermal
quantum physics we refer to $\beta$ as the ``inverse
temperature''---even though our fermions do not have thermal (i.e.
anti-periodic) boundary condition around $S_\beta^1$.

Because of the condition $iii$ above, for $Z^{\mathrm{SUSY}}$ to be
well-defined we need a U($1$)$_R$ symmetry in the theory, whose
existence we take for granted below; the presence of the greater
superconformal symmetry is not necessary. For superconformal
theories, however, $Z^{\mathrm{SUSY}}$ becomes more significant, and
coincides (up to a Casimir energy factor) with the 4d superconformal
index of \cite{Romelsberger:2005eg,Kinney:2005ej}, which counts the
protected operators in the theory.

Unlike the 2d elliptic genera, the 4d SUSY partition functions---or
alternatively the 4d superconformal indices---of holographic SCFTs
do not seem to encode Black~Hole physics \cite{Kinney:2005ej}, but
they may aid the microscopic counting of supersymmetric
Giant~Gravitons \cite{Bourdier:2015}.\\

The asymptotics of the elliptic genera of 2d SCFTs are well-known,
thanks to their simple modular properties (see for instance
\cite{Kraus:2007}). In four dimensions, on the other hand, analogous
general results for the asymptotics of $Z^{\mathrm{SUSY}}$ have only
begun to appear recently. Di~Pietro and Komargodski have combined
ideas from supersymmetry and hydrodynamics to argue
\cite{DiPietro:2014} that the SUSY partition functions of 4d
Lagrangian theories exhibit the following universal behavior at the
leading order:
\begin{equation}
\ln Z^{\mathrm{SUSY}}(b,\beta)\approx
-\frac{\pi^2}{3\beta}\left(\frac{b+b^{-1}}{2}\right)\mathrm{Tr}R,\quad
\quad (\text{as } \beta\to 0) \label{eq:dk}
\end{equation}
with $\mathrm{Tr}R$ the U($1$)$_R$-gravitational-gravitational
`t~Hooft anomaly of the theory. We refer to the above relation as
the Di~Pietro-Komargodski formula.

The supporting arguments for (\ref{eq:dk}) are, however, somewhat
indirect, and contain some intuitive elements that we would like to
scrutinize by a more direct analysis.

In \cite{DiPietro:2014,Ardehali:2015b} the relation (\ref{eq:dk})
was directly verified for free chiral and U($1$) vector multiplets.
In the present work we extend the analysis to interacting
supersymmetric gauge theories with a semi-simple gauge group. [Our
approach enables us to study also the non-Lagrangian $E_6$ SCFT; see
subsection \ref{sec:e6}.] The SUSY partition function of such
theories can be computed exactly using the supersymmetric
localization \cite{Assel:2014}. We write this as
\begin{equation}
Z^{\mathrm{SUSY}}(b,\beta):=\int e^{-S}\mathcal{D}\phi\mathcal{D}A\
\xrightarrow{\text{SUSY localization}}\int\mathrm{d}^{r_G} x\
f(x_1,\dots,x_{r_G};b,\beta), \label{eq:susyLoc}
\end{equation}
with $r_G$ the rank of the gauge group, $f$ some complicated
function of its arguments, and the integral on the (far) right over
the range $-1/2\le x_i\le 1/2$. The path-integral of the theory on
$S_b^3\times S^1_\beta$ is displayed schematically as $\int
e^{-S}\mathcal{D}\phi\mathcal{D}A$, with $\phi$ and $A$ representing
the matter fields and the gauge fields respectively.

The $x_i$ in (\ref{eq:susyLoc}) parameterize the unit hypercube in
the Cartan subalgebra of the gauge group; we denote this hypercube
by $\mathfrak{h}_{cl}$. The exponential function $z_i=e^{2\pi i
x_i}$ maps $\mathfrak{h}_{cl}$ to the moduli space of the
eigenvalues of the holonomy matrix $P\exp(i\oint_{S^1_\beta}A_0)$,
with $A_0$ the component of $A$ along $S^1_\beta$. The (matrix-)
integral on the (far) right of (\ref{eq:susyLoc}) is thus over the
``classical'' moduli space of the holonomies around the circle;
hence the subscript $cl$ in $\mathfrak{h}_{cl}$.

The matrix-integrals appearing in $Z^{\mathrm{SUSY}}(b,\beta)$ are
known in the mathematics literature as \emph{elliptic hypergeometric
integrals} (EHIs) \cite{Dolan:2008,Spirido:2009}. The
high-temperature limit corresponds to the \emph{hyperbolic limit} of
the EHIs. This limit can be rigorously analyzed with the machinery
that Rains has developed in \cite{Rains:2009}. Following Rains's
approach, we find that at high temperatures the integrand of the
matrix-integral in (\ref{eq:susyLoc}) simplifies as\footnote{In the
present section we assume the theories under study have non-chiral
matter content. Otherwise, some of the following expressions need to
be slightly modified. We will comment on the case of theories with
chiral matter content in section~\ref{sec:N=1Asy}.}
\begin{equation}
f(x_1,\dots,x_{r_G};b,\beta)\overset{\beta\to
0}{\longrightarrow}\exp\left[-\left(\mathcal{E}_0^{DK}(b,\beta)+V^{\mathrm{eff}}(x_1,\dots,x_{r_G};b,\beta)\right)\right],
\label{eq:betaMatIntegrand}
\end{equation}
where
\begin{equation}
\mathcal{E}_0^{DK}(b,\beta)=\frac{\pi^2}{3\beta}\left(\frac{b+b^{-1}}{2}\right)\mathrm{Tr}R,\label{eq:dkE}
\end{equation}
and $V^{\mathrm{eff}}$ is a real, continuous, piecewise linear
function of the $x_i$ (examples can be found in
Figures~\ref{fig:A1}, \ref{fig:SO5}, and \ref{fig:ISS} below). We
interpret $V^{\mathrm{eff}}$ as a quantum effective potential for
the interaction of the holonomies. This is of course not a
low-energy effective potential from the perspective of the
three-sphere; rather, it is loosely a ``high-energy effective
potential'', as it governs the high-temperature behavior of the SUSY
partition function. In section \ref{sec:Dis} we will discuss the
extent to which an alternative viewpoint (roughly speaking, from a
crossed channel) allows considering $V^{\mathrm{eff}}$ as a
conventional (low-energy) quantum effective potential.

We find that $V^{\mathrm{eff}}$ is inversely proportional to
$\beta$. Therefore \emph{the high-temperature limit further
localizes the matrix-integral} to the locus of the minima of
$V^{\mathrm{eff}}$. This locus is a subspace of $\mathfrak{h}_{cl}$
that we denote by $\mathfrak{h}_{qu}$. We can thus combine
(\ref{eq:susyLoc}) and (\ref{eq:betaMatIntegrand}) to write
\begin{equation}
Z^{\mathrm{SUSY}}(b,\beta)\overset{\beta\to
0}{\longrightarrow}\int_{\mathfrak{h}_{qu}}
e^{-(\mathcal{E}_0^{DK}(b,\beta)+V^{\mathrm{eff}}(x_1,\dots,x_{r_G};b,\beta))}\approx
e^{-(\mathcal{E}_0^{DK}(b,\beta)+V^{\mathrm{eff}}_{min}(b,\beta))},
\label{eq:betaLoc}
\end{equation}
with $V^{\mathrm{eff}}_{min}$ the minimum of $V^{\mathrm{eff}}$ over
$\mathfrak{h}_{cl}$---or alternatively the value of
$V^{\mathrm{eff}}$ on $\mathfrak{h}_{qu}$.\\

A similar ``high-temperature localization'' of path-integrals has
long been known to occur in non-supersymmetric pure gauge theories
on Euclidean $R^3\times S^1$ \cite{Svetitsky:1982b,Yaffe:1982a}. In
our case the problem is more under control for two reasons. Firstly,
since the spatial manifold that our theories live on is compact, our
path-integrals are finite and do not need IR regularization.
Secondly, thanks to the supersymmetric localization, we have the
luxury of having at our disposal the exact partition function of the
interacting gauge theory, which we can then study using standard
methods of asymptotic analysis. In the non-supersymmetric cases of
\cite{Svetitsky:1982b,Yaffe:1982a}, on the other hand, the
high-temperature limit is employed to seek approximate results.\\

It turns out that $V^{\mathrm{eff}}$ vanishes at the origin of
$\mathfrak{h}_{cl}$ (corresponding to $x_i=0$). Therefore its
minimum $V^{\mathrm{eff}}_{min}$ is guaranteed to be $\leq 0$. In a
large set of examples we find that $V^{\mathrm{eff}}_{min}=0$, and
consequently recover the Di~Pietro-Komargodski formula (\ref{eq:dk})
from (\ref{eq:betaLoc}).

For some interacting theories, however, we find that
$V^{\mathrm{eff}}_{min}<0$ (see Figures~\ref{fig:ISS} and
\ref{fig:BCI}). In such cases (\ref{eq:betaLoc}) implies that the
formula (\ref{eq:dk}) receives a modification:
\begin{equation}
\ln Z^{\mathrm{SUSY}}(b,\beta)\approx
-\frac{\pi^2}{3\beta}\left(\frac{b+b^{-1}}{2}\right)\mathrm{Tr}R\
-V^{\mathrm{eff}}_{min}(b,\beta)\quad \quad (\text{as } \beta\to 0).
\label{eq:dkCorrected}
\end{equation}

We are aware of only two examples where this modification occurs.
One is the SU($2$) Intriligator-Seiberg-Shenker theory
\cite{Intriligator:1994}, and the other is the SO($N$) theory of
Brodie-Cho-Intriligator \cite{Brodie:1998}. Both of these are
believed to exhibit ``misleading'' anomaly matchings, and both have
$\mathrm{Tr}R>0$ (or alternatively, $c<a$ for the putative IR fixed
points). Interestingly, we find that in both cases the correction
term coming from $V^{\mathrm{eff}}_{min}<0$ makes the RHS of
(\ref{eq:dkCorrected}) positive.\\

A possible explanation for why the result of Di~Pietro and
Komargodski does not apply when $V^{\mathrm{eff}}_{min}<0$ is as
follows. As stated in subsection~4.3 of \cite{DiPietro:2014}, an
assumption made in that work is that the $S_b^3$ partition function
$Z_{S^3}(b)$ of the 4d theory reduced on $S_\beta^1$ does not
diverge. Indeed, in all the theories with finite $Z_{S^3}$ that we
have studied, $V^{\mathrm{eff}}_{min}$ vanishes and consequently
(\ref{eq:dk}) is satisfied. [We have not been able to show that the
finiteness of $Z_{S^3}$ always implies $V^{\mathrm{eff}}_{min}=0$,
although we suspect that is the case; we will comment on this point
further in section~\ref{sec:Dis}.]

There exist theories with $V^{\mathrm{eff}}_{min}=0$, but in which
$V^{\mathrm{eff}}$ has flat directions and the locus of the
high-temperature localization is extended:
$\mathrm{dim}\mathfrak{h}_{qu}>0$. In such cases $Z_{S^3}$ diverges,
and therefore the arguments in \cite{DiPietro:2014} are not on solid
footing. The matrix-integral that computes $Z_{S^3}$ (via 3d
supersymmetric localization
\cite{Kapustin:2010,Jafferis:2012,Hama:2011sq})  must then be
regularized with a cut-off. Introducing a cut-off $\Lambda$, we
argue in section~\ref{sec:N=1Asy} that upon taking
$\Lambda\to\infty$, the $S_b^3$ partition function diverges in these
cases as $\Lambda^{\mathrm{dim}\mathfrak{h}_{qu}}$. The power-law
divergences in $Z_{S^3}$ were interpreted in \cite{DiPietro:2014} as
coming from the ``unlifted Coulomb branch'' of the reduced theory on
$S_b^3$. Di~Pietro and Komargodski presented intuitive arguments
suggesting that for theories with such unlifted Coulomb branches,
the relation (\ref{eq:dk}) remains valid at the leading order, but
there will be subleading corrections to it of the form
$\ln(1/\beta)$. We will show in section~\ref{sec:N=1Asy} that, when
$V^{\mathrm{eff}}_{min}=0$, the Di~Pietro-Komargodski formula for
the leading asymptotics indeed remains valid, and the subleading
correction to it is of the form $\mathrm{dim}\mathfrak{h}_{qu}\cdot
\ln(1/\beta)$. If one interprets $\mathfrak{h}_{qu}$ as the
``quantum Coulomb branch'' of the reduced theory on $S_b^3$, this
subleading correction is in accord with the prescription of
Di~Pietro and Komargodski. Furthermore, in section~\ref{sec:Dis} we
will argue intuitively that, when $V^{\mathrm{eff}}_{min}=0$, the
space $\mathfrak{h}_{qu}$ should resemble the unlifted (or quantum)
Coulomb branch of the 3d theory obtained by reducing the gauge
theory on the circle of $R^3\times S^1$.

In the examples where $V^{\mathrm{eff}}_{min}<0$, however, we find
that $Z_{S^3}$ diverges exponentially in $\Lambda$, as
$\Lambda\to\infty$. This severe divergence seems to undermine
the---three-dimensional---assumption of Di~Pietro and Komargodski.
As a result, the formula (\ref{eq:dk}) no longer holds, and the
correct asymptotics of the SUSY partition function is given by
(\ref{eq:dkCorrected}).\\

A refinement of the SUSY partition function is available for
Lagrangian $\mathcal{N}=2$ SCFTs. These have extended R-symmetry
group
SU($2$)$_{R_{\mathcal{N}=2}}\times$U($1$)$_{r_{\mathcal{N}=2}}$. We
can then consider the \emph{$\mathcal{N}=2$ partition function}
$Z^{\mathcal{N}=2}(b,\beta,m_v)$, where $m_v$ is a background
U($1$)$_v$ gauge field along $S^1_\beta$, that couples to a specific
linear combination of U($1$)$_{r_{\mathcal{N}=2}}$ and the Cartan of
SU($2$)$_{R_{\mathcal{N}=2}}$. We will analyze the asymptotics of
this partition function in section~\ref{sec:N=2andSchur}. The Schur
limit \cite{Gadde:2011MC} of $Z^{\mathcal{N}=2}(b,\beta,m_v)$,
defined by setting $b=1$ and $m_v=i/3$, has been the subject of much
recent work. We will show in section~\ref{sec:N=2andSchur} that the
high-temperature asymptotics of the \emph{Schur partition function}
is given by
\begin{equation}
\ln Z^{\mathrm{Schur}}(\beta)\approx
-\frac{\pi^2}{2\beta}\mathrm{Tr}R\
-\frac{3}{2}V^{\mathrm{eff}}_{min}(b=1,\beta)\quad \quad (\text{as }
\beta\to 0). \label{eq:bnCorrected}
\end{equation}
In particular, when $V^{\mathrm{eff}}_{min}=0$---which is when the
Di~Pietro-Komargodski formula for $Z^{\mathrm{SUSY}}(b,\beta)$
applies---we find
\begin{equation}
\ln Z^{\mathrm{Schur}}(\beta)\approx
-\frac{\pi^2}{2\beta}\mathrm{Tr}R\quad \quad (\text{as } \beta\to
0,\text{ when } V^{\mathrm{eff}}_{min}=0). \label{eq:bn}
\end{equation}
This relation was recently observed by Buican and Nishinaka to hold
in a large set of Lagrangian and non-Lagrangian examples
\cite{Buican:2015a}.\\

Dual gauge theories must have identical partition functions.
Comparison of the SUSY partition functions of supersymmetric gauge
theories with a U($1$) R-symmetry provides one of the strongest
tests of any proposed duality between such theories
\cite{Dolan:2008,Spirido:2009}. The full comparison of the
matrix-integrals computing such partition functions is, however,
extremely challenging, except for the few cases (corresponding to
various SQCD-type theories
\cite{Dolan:2008,Spirido:2009,Spirido:2011or}) already established
in the mathematics literature (e.g. \cite{Rains:2005}). Rather,
known dualities are frequently used to conjecture new identities
between multi-variable matrix-integrals of elliptic hypergeometric
type \cite{Dolan:2008,Spirido:2009,Spirido:2011or,Kutasov:2014}.

We propose comparison of the high-temperature asymptotics of the
SUSY partition functions. This comparison provides two new simple
tests for dualities between SUSY gauge theories with a U($1$)
R-symmetry. The first test is the comparison of
$V^{\mathrm{eff}}_{min}$, which according to (\ref{eq:dkCorrected})
determines the leading high-temperature asymptotics of the SUSY
partition functions. The second test is the comparison of the
dimension of the locus of minima of $V^{\mathrm{eff}}$---i.e.
$\mathrm{dim}\mathfrak{h}_{qu}$; this is an integer which, as we
briefly mentioned above, determines the subleading $\ln(1/\beta)$
term in the high-temperature asymptotics of $\ln Z^{\mathrm{SUSY}}$.
These two high-temperature tests are independent of `t~Hooft anomaly
matchings (which in turn can be thought of as arising from
comparison of the \emph{low-temperature asymptotics} of an
equivariant generalization of $Z^{\mathrm{SUSY}}$
\cite{Bobev:2015}). They may thus help to diagnose situations with
misleading anomaly matchings. A few concrete applications of these
two duality tests can be found in subsection \ref{sec:tests}.\\

The rest of this paper is organized as follows. In the remaining of
the present section we first summarize our notation and terminology,
and then proceed to mention the relation of our findings to previous
work.

In section \ref{sec:back} we present the mathematical background
required for the quantitative analysis in the body of the paper. The
main result of section~\ref{sec:back} is the uniform estimate
(\ref{eq:GammaOffCenter2}) for the high-temperature asymptotics of
the elliptic gamma function.

Section \ref{sec:N=1Asy} contains our main findings. There we show
the high-temperature localization of the SUSY partition function
$Z^{\mathrm{SUSY}}(b,\beta)$, obtain the effective potential
$V^{\mathrm{eff}}$ that determines the locus of the high-temperature
localization, establish the validity of the Di~Pietro-Komargodski
formula (\ref{eq:dk}) when $V^{\mathrm{eff}}$ is positive
semi-definite, and demonstrate its modified version
(\ref{eq:dkCorrected}) for theories with $V^{\mathrm{eff}}_{min}<0$.
Section \ref{sec:N=1Asy} is the lengthiest section of this paper,
partly because it includes several examples that are analyzed quite
explicitly.

In section \ref{sec:N=2andSchur} we analyze the high-temperature
asymptotics of the $\mathcal{N}=2$ partition function
$Z^{\mathcal{N}=2}(b,\beta,m_v)$, and establish the formula
(\ref{eq:bnCorrected}) for the asymptotics of its Schur limit.
Section~\ref{sec:N=2andSchur} includes also the high-temperature
analysis of the superconformal index of the $E_6$ SCFT, which is the
only non-Lagrangian theory studied in this paper.

Our concluding remarks are made in section \ref{sec:Dis}, and the
appendices contain some technical details that are not essential for
following the discussion in the main text.

\subsection{Notation and terminology}

\subsubsection*{Partition functions and indices}

The \textbf{SUSY partition function} $Z^{\mathrm{SUSY}}(b,\beta)$ is
the path-integral of the 4d Lagrangian supersymmetric R-symmetric
theory on Euclidean $S_b^3\times S_1^\beta$, in presence of a
specific (as in \cite{Romelsberger:2005eg,Festuccia:2011})
background U($1$)$_R$ gauge field along $S_1^\beta$, and with
periodic boundary conditions around the circle. This is the object
computed by supersymmetric localization in
\cite{Closset:2014,Assel:2014}, and their result (with a minor
correction of a regularization procedure, as explained in
\cite{Ardehali:2015b,Assel:2015s}) is our starting point. For
superconformal theories, $Z^{\mathrm{SUSY}}(b,\beta)$ coincides, up
to a Casimir energy factor (see Eq.~(\ref{eq:LagEquivZ}) below),
with the \textbf{superconformal index} $\mathcal{I}(b,\beta)$, which
we sometimes refer to as the \textbf{Romelsberger index}, or simply
as \textbf{the index}. More commonly, the index is written as a
function of $p$,$q$, which are related to $b$,$\beta$ via
$p=e^{-\beta b}$, $q=e^{-\beta b^{-1}}$. Alternatively, we can
express the partition function or the index, in terms of the complex
structure moduli $\sigma,\tau$ of the space $S_b^3\times S_1^\beta$;
these are related to $p,q$ via $p=e^{2\pi i\sigma}$, $q=e^{2\pi
i\tau}$. We always assume $b,\beta$ to be positive real numbers, and
thus $\sigma,\tau$ to be pure imaginary in the upper half plane;
$p,q$ are then real numbers in $]0,1[$. The high-temperature limit
corresponds to $\beta\to0$ with $b$ fixed.

For non-conformal supersymmetric gauge theories with well-defined
$Z^{\mathrm{SUSY}}(b,\beta)$, we take Eq.~(\ref{eq:LagEquivZ}) below
as the definition of the \emph{Romelsberger index} (or \emph{the
index}) $\mathcal{I}(b,\beta)$. This way we avoid the awkward use of
the term ``superconformal index'' for non-conformal theories.

A further background gauge field $m_v$, which we take to be pure
imaginary and in the upper half plane, can serve to refine the
partition functions of $\mathcal{N}=2$ SCFTs with R-symmetry group
SU($2$)$_{R_{\mathcal{N}=2}}\times$U($1$)$_{r_{\mathcal{N}=2}}$. The
charge $Q_v$ that $m_v$ couples to is
\begin{equation}
Q_v=-(r_{\mathcal{N}=2}+R_{\mathcal{N}=2}).\label{eq:backLinVterm}
\end{equation}
We denote the resulting partition function by
$Z^{\mathcal{N}=2}(b,\beta,m_v)$, and refer to it as the
$\mathbf{\mathcal{N}=2}$ \textbf{partition function}. This partition
function coincides, up to a Casimir energy factor (see
Eq.~(\ref{eq:LagEquivN=2Z}) below), with the
$\mathbf{\mathcal{N}=2}$ \textbf{superconformal index}
$\mathcal{I}(b,\beta,m_v)$, which we frequently refer to as
\textbf{the} $\mathbf{\mathcal{N}=2}$ \textbf{index}. The
high-temperature limit corresponds to $\beta\to0$ with $b,m_v$
fixed.

\subsubsection*{Special functions}

The $q$-Pochhammer symbol, often written in the mathematics
literature as $(a;q)_\infty$, will be denoted below by $(a;q)$, and
will be called the Pochhammer symbol.

The elliptic gamma function, commonly written as $\Gamma_e(z;p,q)$,
will be denoted below by $\Gamma(z;p,q)$. We sometimes write
$\Gamma(z;p,q)$ as $\Gamma(x;\sigma,\tau)$, or simply as
$\Gamma(z)$. Also, the arguments of elliptic gamma functions are
frequently written with ``ambiguous'' signs (as in $\Gamma(\pm
x;\sigma,\tau)$); by that one means a multiplication of several
gamma functions each with a ``possible'' sign of the argument (as in
$\Gamma(+x;\sigma,\tau)\times \Gamma(-x;\sigma,\tau)$). Similarly
$\Gamma(z^{\pm 1}):=\Gamma(z;p,q)\times \Gamma(z^{-1};p,q)$.

The hyperbolic gamma function will be denoted by the standard
$\Gamma_h(x;\omega_1,\omega_2)$, with $\omega_1=i b$ and $\omega_2=i
b^{-1}$. For convenience, we will frequently write $\Gamma_h(x)$
instead of $\Gamma_h(x;\omega_1,\omega_2)$, and $\Gamma_h(x\pm y)$
instead of $\Gamma_h(x+y)\Gamma_h(x-y)$.

\subsubsection*{Asymptotic analysis}

When writing asymptotic relations, we use the symbol $\sim$ to
indicate all-orders asymptotic equalities. For example, we write
$f(\beta)\sim g(\beta)$, if the small-$\beta$ asymptotic expansions
of $f(\beta)$ and $g(\beta)$ coincide to all orders in $\beta$. This
notation is standard, and appears, for instance, in
\cite{Zagier:2006}.

We will also use the non-standard notation $f(\beta)\simeq
g(\beta)$, whenever $\ln f(\beta)\sim \ln g(\beta)$.

Finally, we use the symbol $\approx$ to indicate ``approximate
asymptotic equality''. We will not make this statement more precise,
and instead explicitly mention the error involved whenever using
$\approx$ below.

\subsubsection*{Convex polytopes}

By a $j$-face we mean an element of dimension $j$ in a convex
polytope. We define the unique $d$-face of a $d$-dimensional
polytope to be the polytope itself.

We call a $d$-dimensional polytope a prismatoid if all its vertices
(i.e. $0$-faces) lie in either of two parallel codimension one
hyperplanes. A prismatoid with only one vertex in one of the two
hyperplanes will be referred to as a pyramid.

\subsection{Relation to previous work}

Our discussion of the high-temperature asymptotics of the SUSY
partition function of Lagrangian gauge theories relies heavily on
the machinery developed by Rains \cite{Rains:2009}. In fact Rains's
results are immediately applicable to SU($N$) and Sp($N$) SQCD-type
theories, and yield asymptotics of the form (\ref{eq:dk}).

The fact that Rains's method gives the leading high-temperature
asymptotics of the Romelsberger index $\mathcal{I}(b,\beta)$ in
accord with the formula (\ref{eq:dk}) was identified and pointed out
for SU($N$) and Sp($N$) SQCD-type theories in
\cite{Dolan:2011sv,Spiridonov:2012sv} and \cite{Niarchos:2012}. The
relation (\ref{eq:LagEquivZ}), which was obtained later in
\cite{Ardehali:2015b,Assel:2015s}, would then imply the formula
(\ref{eq:dk}) for $Z^{\mathrm{SUSY}}$. (Other pioneering works on
the high-temperature limit of the 4d superconformal index include
\cite{Gadde:2011ibn,Imamura:2011ibn}, which clarified the relation
between the 4d index and the $S^3$ partition function, but did not
address the Cardy-like asymptotics of the index.)

An argument for the general validity of the formula (\ref{eq:dk})
appeared first in the work of Di~Pietro and Komargodski
\cite{DiPietro:2014}, who used methods completely different from
those of Rains. Importantly, Di~Pietro and Komargodski improved the
qualitative understanding of the role of unlifted Coulomb branches
in the high-temperature asymptotics of the index
\cite{Aharony:2013a,Aharony:2013b}, to a quantitative discussion and
argued that such unlifted Coulomb branches would only introduce
subleading logarithmic corrections to the formula (\ref{eq:dk}), but
would not modify the leading behavior.

In the present paper we show that Rains's rigorous approach can be
adapted, with minor modifications, for analyzing any SUSY partition
function (or Romelsberger index) given as an elliptic hypergeometric
integral. We are thus able to find the conditions under which the
Di~Pietro-Komargodski formula applies. In particular, we find that
the formula (\ref{eq:dk}) does not apply in certain interacting
SCFTs with $c<a$.

In \cite{Ardehali:2014,Ardehali:2015a} certain results from
holography were derived, over which we do not present any
improvement here. However, the holographic---large-$N$---results
were extrapolated there to conjecture prescriptions for extracting
the central charges of any finite-$N$ SCFT from its superconformal
index. In \cite{Ardehali:2015b} it was shown that those
prescriptions are equivalent to the statement that there is no
$O(\beta)$ term in the high-temperature asymptotics of $\ln
Z^{\mathrm{SUSY}}(\beta)$. In the present paper we \emph{almost}
establish that this statement is correct whenever
$\mathrm{dim}\mathfrak{h}_q=0$ (see the comments below
(\ref{eq:ISSindexAsy3})). On the other hand, we find a
counterexample which has $\mathrm{dim}\mathfrak{h}_q=1$: for the
SO($3$) SQCD with two flavors, $\ln Z^{\mathrm{SUSY}}(\beta)$
\emph{does have} an $O(\beta)$ term in its high-temperature
expansion (see (\ref{eq:SO3indexAsy3}) below). Our results thus
indicate that the finite-$N$ conjectures of
\cite{Ardehali:2014,Ardehali:2015a} are not necessarily true if
$\mathrm{dim}\mathfrak{h}_{qu}>0$.

In \cite{Ardehali:2015b} the SUSY partition function of free U($1$)
vector and free chiral multiplets were studied in the
high-temperature limit. The corresponding expressions were then
conjectured to be true for all SCFTs (with finite $N$). Here we rule
out that possibility, although we find that, except for the
$\ln(1/\beta)$ term in the asymptotics of $\ln
Z^{\mathrm{SUSY}}(\beta)$ conjectured in that work, the conjecture
in \cite{Ardehali:2015b} is correct for theories whose
$V^{\mathrm{eff}}$ has a unique minimum at the origin of
$\mathfrak{h}_{cl}$.

We have organized our discussion in section~\ref{sec:N=1Asy}
according to the degree of divergence of $Z_{S^3}$. For certain 3d
$\mathcal{N}=4$ theories, the criteria for finiteness of $Z_{S^3}$
have already been analyzed in three-dimensional terms
\cite{Kapustin:2010f} (see also \cite{Willett:2011m,Safdi:2012c} for
related discussions in the context of a particular 3d
$\mathcal{N}=2$ model). Our perspective on this problem is a bit
different, as we consider 3d $\mathcal{N}=2$ theories obtained from
dimensional reduction of 4d $\mathcal{N}=1$ gauge theories whose
index we would like to study.

Finally, Buican and Nishinaka \cite{Buican:2015a} recently noted a
Cardy-like behavior in the Schur index of a variety of Lagrangian
and non-Lagrangian $\mathcal{N}=2$ theories. We show in the present
paper that for all Lagrangian theories with a semi-simple gauge
group the Cardy-like behavior noted in \cite{Buican:2015a} is valid
in theories where the Di~Pietro-Komargodski formula for the SUSY
partition function is satisfied---i.e. when $V^{\mathrm{eff}}$ is
positive semi-definite.\\

\section{Mathematical background}\label{sec:back}

In subsection \ref{sec:special} below, we define the Pochhammer
symbol, the elliptic gamma function, and the hyperbolic gamma
function.

In subsection \ref{sec:someAsy} we review the asymptotic estimates
of the special functions discussed in subsection \ref{sec:special}.
These estimates form the mathematical basis of our high-temperature
analysis of SUSY partition functions. The only new estimate, and the
main result of the present section, is the relation
(\ref{eq:GammaOffCenter2}) for the asymptotics of the elliptic gamma
function. All the other estimates for the elliptic and hyperbolic
gamma functions have appeared (sometimes in slightly different
forms) already in the work of Rains \cite{Rains:2009}; we only
present them in a way more suited for the physical application. Even
the estimate (\ref{eq:GammaOffCenter2}) is only a minor modification
of the results in Proposition~2.12 and Corollary~3.1 of
\cite{Rains:2009}.

The important estimates are the asymptotics of the Pochhammer symbol
in Eq.~(\ref{eq:PochAsy}), the ``leading estimate''
(\ref{eq:GammaOffCenter2}) and the ``central estimate''
(\ref{eq:GammaToHypGamma}) for the elliptic gamma function, and the
asymptotics of the hyperbolic gamma function in
Eq.~(\ref{eq:hyperbolicGammaAsy}).

Subsection \ref{sec:gta} contains generalized triangle inequalities
due to Rains \cite{Rains:2009}, that we will need in the next
section when determining the locus of minima of certain effective
potentials.

\subsection{Useful special functions}\label{sec:special}

The \textbf{Pochhammer symbol} ($|q|\in ]0,1[$)
\begin{equation}
(a;q):=\prod_{k=0}^{\infty}(1-a q^k),\label{eq:PochDef}
\end{equation}
is related to the more familiar Dedekind eta function via
\begin{equation}
\eta(\tau)=q^{1/24}(q;q), \label{eq:etaPoc}
\end{equation}
with $q=e^{2\pi i\tau}.$

The eta function has an SL$(2,\mathbb{Z})$ modular property that
will be useful for us: $\eta(-1/\tau)=\sqrt{-i\tau}\eta(\tau)$.

The Pochhammer symbol $(q;q)$ equals the inverse of the generating
function of integer partitions. It also appears in the index of 4d
SUSY gauge theories that contain vector multiplets.\\

The \textbf{elliptic gamma function} is defined as
($\mathrm{Im}(\tau),\mathrm{Im}(\sigma) >0$)
\begin{equation}
\Gamma(x;\sigma,\tau):=\prod_{j,k\ge
0}\frac{1-z^{-1}p^{j+1}q^{k+1}}{1-z p^{j}q^{k}},\label{eq:GammaDef}
\end{equation}
with $z:=e^{2\pi i x}$, $p:=e^{2\pi i \sigma}=e^{-\beta b}$, and
$q:=e^{2\pi i \tau}=e^{-\beta b^{-1}}$. The above expression gives a
meromorphic function of $x\in\mathbb{C}$. For generic choice of
$\tau$ and $\sigma$, the elliptic gamma has simple poles at
$x=l-m\sigma-n\tau$, with $m,n\in\mathbb{Z}^{\ge 0}$,
$l\in\mathbb{Z}$.

The elliptic gamma function appears in the exact solution of some
important 2d integrable lattice models. It also features in the
index of 4d Lagrangian SUSY QFTs that contain chiral multiplets.\\

Following Rains \cite{Rains:2009}, we define the \textbf{hyperbolic
gamma function} by
\begin{equation}
\Gamma_h(x;\omega_1,\omega_2):=\exp
\left(\mathrm{PV}\int_{\mathbb{R}}\frac{e^{2\pi i x w}}{(e^{2\pi
i\omega_1 w}-1)(e^{2\pi i\omega_2
w}-1)}\frac{\mathrm{d}w}{w}\right).\label{eq:hyperbolicGamma}
\end{equation}
The above expression makes sense only for
$0<\mathrm{Im}(x)<2\mathrm{Im}(\omega)$, with
$\omega:=(\omega_1+\omega_2)/2$. In that domain, the function
defined by (\ref{eq:hyperbolicGamma}) satisfies
\begin{equation}
\Gamma_h(x+\omega_2;\omega_1,\omega_2)=2\sin (\frac{\pi
x}{\omega_1})\Gamma_h(x;\omega_1,\omega_2).\label{eq:hyperbolicGammaRecursion}
\end{equation}
This relation can then be used for an inductive meromorphic
continuation of the hyperbolic gamma function to all
$x\in\mathbb{C}$. For generic $\omega_1,\omega_2$ in the upper half
plane, the resulting meromorphic function
$\Gamma_h(x;\omega_1,\omega_2)$ has simple zeros at
$x=\omega_1\mathbb{Z}^{\ge1}+\omega_2\mathbb{Z}^{\ge1}$ and simple
poles at $x=\omega_1\mathbb{Z}^{\le0}+\omega_2\mathbb{Z}^{\le0}$.

We will encounter the hyperbolic gamma function in the $S_b^3$
partition function of 3d supersymmetric gauge theories which we will
obtain from reducing 4d gauge theories on the $S^1$ of $S_b^3\times
S^1$.

\subsection{Some asymptotic analysis}\label{sec:someAsy}

\indent We say $f(\beta)=O(g(\beta))$ as $\beta\to0$, if there exist
positive real numbers $C,\beta_0$ such that for all $\beta<\beta_0$
we have $|f(\beta)|< C|g(\beta)|$. We say $f(x,\beta)=O(g(x,\beta))$
\emph{uniformly} over $S$ as $\beta\to0$, if there exist positive
real numbers $C,\beta_0$ such that for all $\beta<\beta_0$ and all
$x\in S$ we have $|f(x,\beta)|< C|g(x,\beta)|$.

We will write $f(\beta)=o(g(\beta))$, if $f(\beta)/g(\beta)\to 0$ as
$\beta\to 0$.

We use the symbol $\sim$ when writing the all-orders asymptotics of
a function. For example, we have
\begin{equation}
\ln(\beta+e^{-1/\beta})\sim \ln\beta,\quad\quad(\text{as $\beta\to
0$})
\end{equation}
because we can write the LHS as the sum of $\ln\beta$ and
$\ln(1+e^{-1/\beta}/\beta)$, and the latter is beyond all-orders in
$\beta$.

More precisely, we say $f(\beta)\sim g(\beta)$ as $\beta\to 0$, if
we have $f(\beta)- g(\beta)=O(\beta^n)$ for any (arbitrarily large)
natural $n$.

We will write $f(\beta)\simeq g(\beta)$ if $\ln f(\beta)\sim \ln
g(\beta)$ (with an appropriate choice of branch for the logarithms).
By writing $f(x,\beta)\simeq g(x,\beta)$ we mean that $\ln
f(x,\beta)\sim \ln g(x,\beta)$ for all $x$ on which
$f(x,\beta),g(x,\beta)\neq 0$, and that $f(x,\beta)= g(x,\beta)=0$
for all $x$ on which
either $f(x,\beta)=0$ or $g(x,\beta)=0$.\\

With the above notations at hand, we can asymptotically analyze the
Pochhammer symbol as follows. The low-temperature ($T\to 0$, with
$q=e^{-1/T}$) behavior is trivial:
\begin{equation}
(q;q)\simeq 1\quad\quad(\text{as $1/\beta\to
0$}).\label{eq:PochLowT}
\end{equation}

The high-temperature ($\beta\to 0$, with $q=e^{-\beta}$) asymptotics
is nontrivial. It can be obtained using the SL($2,\mathbb{Z}$)
modular property of the eta function, which yields
\begin{equation}
\ln \eta(\tau=\frac{i\beta}{2\pi})\sim
-\frac{\pi^2}{6\beta}+\frac{1}{2}\ln(\frac{2\pi}{\beta})\quad\quad(\text{as
$\beta\to 0$}).
\end{equation}
The above relation, when combined with (\ref{eq:etaPoc}), implies
\begin{equation}
\ln(q;q)\sim
-\frac{\pi^2}{6\beta}+\frac{1}{2}\ln(\frac{2\pi}{\beta})+\frac{\beta}{24}\quad\quad(\text{as
$\beta\to
0$}).\label{eq:PochAsy}\\
\end{equation}

Next, we write estimates for the elliptic gamma function. Although
its low-temperature asymptotics is not needed for our main purposes
below, we encourage the reader without prior familiarity with the
elliptic gamma to convince herself that for fixed $r\in]0,2[$
\begin{equation}
\begin{split}
\frac{1}{\Gamma(x;\sigma,\tau)}\simeq  1-z,\quad\text{and}\quad
\Gamma((pq)^{r/2}z)\simeq 1,\quad\quad(\text{as $1/\beta\to 0$, for
$x\in\mathbb{R}$}) \label{eq:GammaLowT}
\end{split}
\end{equation}
both valid uniformly over ($x\in$) $\mathbb{R}$.

\begin{figure}[t]
\centering
    \includegraphics[scale=1]{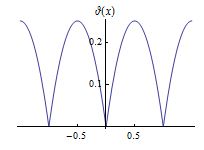}
    \hspace{2cm}
    \includegraphics[scale=1]{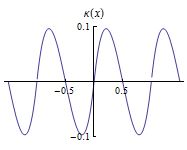}
\caption{The even function $\vartheta(x)$ (on the left) and the odd
function $\kappa(x)$ (on the right). Both are continuous and
periodic, and their fundamental domain can be taken to be
$[-1/2,1/2]$. \label{fig:thetakappa}}
\end{figure}

The high-temperature asymptotics of the elliptic gamma function is
quite nontrivial. From Proposition~2.11 of \cite{Rains:2009} we
obtain the following \emph{uniform} estimate over ($x\in$)
$\mathbb{R}$ (c.f. Proposition~2.12 of \cite{Rains:2009}, with
$v_{\mathrm{there}}=\beta_{\mathrm{here}}/2\pi$; see also
appendix~\ref{app:rains}):
\begin{equation}
\boxed{\begin{split}
\ln\Gamma(x+r(\frac{\sigma+\tau}{2});\sigma,\tau)= 2\pi
i(-\frac{\kappa(x)}{12\tau\sigma}
+(r-1)\frac{\tau+\sigma}{4\tau\sigma}\vartheta(x)-(r-1)\frac{\tau+\sigma}{24\tau\sigma})+O(\beta^0)&\\
(\text{for fixed $r\in]0,2[$})&.\label{eq:GammaOffCenter2}
\end{split}}
\end{equation}
Following Rains \cite{Rains:2009}, we have defined the continuous,
positive, even, periodic function\footnote{This function is closely
related to the functions $1-[x]_+^2$ (in Appendix~D of
\cite{Gross:1981}) and $g(x)$ (in Appendix~A of \cite{Unsal:2006})
appearing in the context of \emph{perturbative} corrections to
low-energy effective actions on $R^3\times S^1$.}
\begin{equation}
\begin{split}
\vartheta(x)&:=\{x\}(1-\{x\})\\
&\left(=|x|-x^2 \quad\quad\text{for
$x\in[-1,1]$}\right),\label{eq:varthetaDef}
\end{split}
\end{equation}
with $\{x\}:=x-\lfloor x\rfloor$. We have also introduced the
continuous, odd, periodic function
\begin{equation}
\begin{split}
\kappa(x)&:=\{x\}(1-\{x\})(1-2\{x\})\\
&\left(=2x^3-3x|x|+x \quad\quad\text{for
$x\in[-1,1]$}\right).\label{eq:kappaDef}
\end{split}
\end{equation}
These functions are displayed in Figure~\ref{fig:thetakappa}.

The real number $r$ in (\ref{eq:GammaOffCenter2}) will be
interpreted in the next section as the R-charge of a chiral
multiplet.

To analyze the SUSY partition function of theories that contain
vector multiplets, we will need an estimate similar to
(\ref{eq:GammaOffCenter2}) that would apply when $r=0$. The
following asymptotic relation, valid uniformly over compact subsets
of $\mathbb{R}$ avoiding an $O(\beta)$ neighborhood of $\mathbb{Z}$,
gives the desired estimate (c.f. Proposition~2.12 of
\cite{Rains:2009}):
\begin{equation}
\begin{split}
\ln\left(\frac{1}{\Gamma(x;\sigma,\tau)\Gamma(-x;\sigma,\tau)}\right)=
2\pi
i(\frac{\tau+\sigma}{2\tau\sigma}\vartheta(x)-\frac{\tau+\sigma}{12\tau\sigma})+O(\beta^0).\label{eq:GammaOffCenter2Vec}
\end{split}
\end{equation}
Note that the above relation would follow from a (sloppy) use of
(\ref{eq:GammaOffCenter2}) with $r=0$. But unlike
(\ref{eq:GammaOffCenter2}), the above estimate is not valid
uniformly over $\mathbb{R}$. For real $x$ in an $O(\beta)$
neighborhood of $\mathbb{Z}$, the following slightly weaker version
of (\ref{eq:GammaOffCenter2Vec}) applies (c.f. Corollary~3.1 of
\cite{Rains:2009}):
\begin{equation}
\begin{split}
\frac{1}{\Gamma(x;\sigma,\tau)\Gamma(-x;\sigma,\tau)}=O(\exp [2\pi
i(\frac{\tau+\sigma}{2\tau\sigma}\vartheta(x)-\frac{\tau+\sigma}{12\tau\sigma})])\quad\quad(\text{as
$\beta\to 0$}).\label{eq:GammaOffCenter2VecB}
\end{split}
\end{equation}
(A stronger estimate in this region can be obtained by relating the
product on the LHS to a product of theta functions, and then using
the modular property of the theta function.)

The reason (\ref{eq:GammaOffCenter2VecB}) is weaker than
(\ref{eq:GammaOffCenter2Vec}) is roughly that the argument of $O$ on
the RHS of (\ref{eq:GammaOffCenter2VecB}) is an overestimate of the
LHS, in particular (as the reader can easily check) for
$x\in\mathbb{Z}$.

Since the estimates (\ref{eq:GammaOffCenter2}),
(\ref{eq:GammaOffCenter2Vec}), and (\ref{eq:GammaOffCenter2VecB})
encode only the leading asymptotics of the elliptic gamma function,
we refer to them as the ``leading estimates''. These estimates alone
will suffice for obtaining the effective potential
on the moduli space of the holonomies in the next section.\\

A much more precise estimate for the elliptic gamma function is
(c.f. Proposition~2.10 in \cite{Rains:2009})
\begin{equation}
\begin{split}
\ln\Gamma((pq)^{r/2}z;p,q)\sim 2\pi i
R_{0}(x+r(\frac{\tau+\sigma}{2});\sigma,\tau)+\ln\Gamma_h(\frac{2\pi
x}{\beta}+\omega r;\omega_1,\omega_2)&,\\
(\text{as } \beta\to 0,  \text{ for }
x\in]-1,1[)&\label{eq:GammaToHypGamma}
\end{split}
\end{equation}
valid uniformly over (fixed, $\beta$-independent) compact subsets of
the domain $]-1,1[$, assuming $r\in]0,2[$ is fixed, and with
\begin{equation}
\begin{split}
R_{0}(x;\sigma,\tau):=-\frac{x^3}{6\tau\sigma}+\frac{\tau+\sigma}{4\tau\sigma}x^2
-\frac{\tau^2+\sigma^2+3\tau\sigma+1}{12\tau\sigma}x
+\frac{1}{24}\frac{\tau+\sigma}{\tau\sigma}+\frac{1}{24}(\tau+\sigma).\label{eq:RhDef}
\end{split}
\end{equation}
The restriction we imposed on the range of $r$ ensures that the (log
of the) hyperbolic gamma function on the RHS of
(\ref{eq:GammaToHypGamma}) is well-defined at $x=0$.

The domain of validity of (\ref{eq:GammaToHypGamma}) can be easily
extended from $x\in]-1,1[$ to $x\in\mathbb{R}$, by replacing every
$x$ on the RHS with $x_\mathbb{Z}:=x-\mathrm{nint}(x)$, where nint is the nearest-integer function. This is because the LHS of
(\ref{eq:GammaToHypGamma}) is a function of $z$, and $z=e^{2\pi i
x}$ is invariant under $x\to x+1$.

At $x=0$, the physical content of the relation
(\ref{eq:GammaToHypGamma}) is the well-known fact that the
superconformal index of a free 4d chiral multiplet (the elliptic
gamma on the LHS) reduces at high temperatures to the
squashed-three-sphere partition function of the 3d chiral multiplet
(the hyperbolic gamma on the RHS) obtained by reducing the 4d
multiplet on $S^1$.

The vector-multiplet analog of (\ref{eq:GammaToHypGamma}) reads
\begin{equation}
\begin{split}
\frac{1}{\Gamma(z^{\pm 1})}\simeq \frac{e^{-2\pi i[
R_{0}(x;\sigma,\tau)+R_{0}(-x;\sigma,\tau)]}}{\Gamma_h(\pm\frac{2\pi
x}{\beta})},\quad\quad (\text{as } \beta\to 0,  \text{ for }
x\in]-1,1[)\label{eq:GammaToHypGammaVec}
\end{split}
\end{equation}
which is valid uniformly over (fixed, $\beta$-independent) compact
subsets of the domain $]-1,1[$. Note that a (sloppy) use of
(\ref{eq:GammaToHypGamma}) for $r=0$ would have yielded
(\ref{eq:GammaToHypGammaVec}) correctly.

We refer to (\ref{eq:GammaToHypGamma}) and
(\ref{eq:GammaToHypGammaVec}) as the ``central estimates'' for the
elliptic gamma function.\\

Finally, we will need the asymptotics of the hyperbolic gamma
function. Corollary~2.3 of \cite{Rains:2009} implies that for
$x\in\mathbb{R}$
\begin{equation}
\begin{split}
\ln\Gamma_h(x+r\omega;\omega_1,\omega_2)= -\frac{i\pi}{2} x|x|-i\pi
(r-1)\omega|x|+O(1), \quad\quad(\text{as
$|x|\to\infty$})\label{eq:hyperbolicGammaAsy}
\end{split}
\end{equation}
for any fixed real $r$, and fixed $b>0$. The above relation allows
us to determine, among other things, whether or not the three-sphere
partition functions we obtain from the high-temperature limit of
SUSY partition functions are finite.

\subsection{Generalized triangle inequalities}\label{sec:gta}

After using the estimates (\ref{eq:GammaOffCenter2}) and
(\ref{eq:hyperbolicGammaAsy}), we will need some relations that the
functions $\vartheta(x)$ and $|x|$ satisfy.

The most important relation, which yields several others as
corollaries, is Rains's generalized triangle inequality. The
Lemma~3.2 of \cite{Rains:2009} says that for any sequence of real
numbers $c_1,\dots, c_n$, $d_1,\dots, d_n$, the following inequality
holds:
\begin{equation}
\sum_{1\le i,j\le n}\vartheta(c_i-d_j)-\sum_{1\le i<j\le
n}\vartheta(c_i-c_j)-\sum_{1\le i<j\le n}\vartheta(d_i-d_j)\ge
\vartheta(\sum_{1\le i\le n}(c_i-d_i)),\label{eq:RainsGTI}
\end{equation}
with equality iff the sequence can be permuted so that either
\begin{equation}
\{c_1\}\le\{d_1\}\le\{c_2\}\le\cdots\le
\{d_{n-1}\}\le\{c_n\}\le\{d_n\},\label{cond:gti1}
\end{equation}
or
\begin{equation}
\{d_1\}\le\{c_1\}\le\{d_2\}\le\cdots\le
\{c_{n-1}\}\le\{d_n\}\le\{c_n\}.\label{cond:gti2}
\end{equation}
The proof can be found in \cite{Rains:2009}.

Re-scaling with $c_i,d_i\mapsto vc_i,vd_i$, taking $v\to 0^+$, and
using the relation $\vartheta(vx)=v|x|-v^2 x^2$ (which holds for
small enough $v$), Rains obtains the following corollary of
(\ref{eq:RainsGTI}):
\begin{equation}
\sum_{1\le i,j\le n}|c_i-d_j|-\sum_{1\le i<j\le
n}|c_i-c_j|-\sum_{1\le i<j\le n}|d_i-d_j|\ge |\sum_{1\le i\le
n}(c_i-d_i)|,\label{eq:RainsGTIav}
\end{equation}
with equality iff the sequence can be permuted so that either
\begin{equation}
c_1\le d_1 \le c_2 \le\cdots\le  d_{n-1} \le c_n \le d_n,
\end{equation}
or
\begin{equation}
d_1 \le c_1 \le d_2 \le\cdots\le  c_{n-1} \le d_n \le c_n.
\end{equation}\\

\section{Asymptotics of the SUSY partition function}\label{sec:N=1Asy}

The SUSY partition function of a supersymmetric gauge theory with a
semi-simple gauge group $G$ (which we think of as a compact matrix
Lie group) is given by \cite{Assel:2014,Ardehali:2015b,Assel:2015s}
(see also \cite{Closset:2014,Peelaers:2014})
\begin{equation}
\begin{split}
Z^{\mathrm{SUSY}}(b,\beta)=e^{-\beta E_{\mathrm{susy}}(b)}
\mathcal{I}(b,\beta),\label{eq:LagEquivZ}
\end{split}
\end{equation}
with the \emph{Romelsberger index} of the SUSY gauge theory obtained
from
\begin{equation}
\begin{split}
\mathcal{I}(b,\beta)=\frac{(p;p)^{r_G}(q;q)^{r_G}}{|W|}\int
\mathrm{d}^{r_G}x \frac{\prod_\chi \prod_{\rho^{\chi}
\in\Delta_\chi}\Gamma((pq)^{r_\chi/2}
z^{\rho^{\chi}})}{\prod_{\alpha_+}\Gamma(
z^{\pm\alpha_+})},\label{eq:LagEquivIndex}
\end{split}
\end{equation}
and the \emph{SUSY Casimir energy} \cite{Assel:2015s} given by
\begin{equation}
\begin{split}
E_{\mathrm{susy}}(b)=\frac{i}{6} \mathrm{Tr} [R\omega]^3
+i\left(\frac{b^2+b^{-2}}{24}\right)\mathrm{Tr}[R\omega].\label{eq:EcBequiv}
\end{split}
\end{equation}

In (\ref{eq:LagEquivIndex}), $p=e^{-\beta b},\ q=e^{-\beta b^{-1}}$,
with $\beta,b\in]0,\infty[$. The rank of the gauge group is denoted
by $r_G$. The $r_\chi$, which we assume to be in the interval
$]0,2[$, are the R-charges of chiral multiplets $\chi$ in the
theory. The chiral multiplets sit in representations
$\mathcal{R}_\chi$ of the gauge group, whose set of weights we have
denoted by $\Delta_\chi$. The set $\Delta_\chi$ consists of as many
weights $\rho^{\chi}$ as the dimension of the representation
$\mathcal{R}_\chi$. Our symbolic notation $z^{\rho^{\chi}}$ should
be understood as $z_1^{\rho^{\chi}_1}\times\dots\times
z_{r_G}^{\rho^{\chi}_{r_G}}$, where $\rho^{\chi}\equiv
(\rho^{\chi}_1,\dots,\rho^{\chi}_{r_G})$. The $\alpha_+$ are the
positive roots of $G$, and $|W|$ is the order of the Weyl group of
$G$. The integral is over $x_i\in [-1/2,1/2]$ (or alternatively,
over the maximal torus of $G$ in the space of $z_i=e^{2\pi i x_i}$).
Note that a given positive root is determined by $r_G$ numbers
$(\alpha_1,\dots,\alpha_{r_G})$; by $z^{\alpha_+}$ we mean
$z_1^{\alpha_1}\times\dots\times z_{r_G}^{\alpha_{r_G}}$.

The numerator of the integrand of (\ref{eq:LagEquivIndex}) comes
from the chiral multiplets. The denominator of the integrand
together with the prefactor of the integral can be thought of as the
contribution of the vector multiplet(s).\\

Eq.~(\ref{eq:LagEquivZ}) describes the way to arrive at the index
via the ``Lagrangian'' path-integral that defines
$Z^{\mathrm{SUSY}}(b,\beta)$. There is an alternative
``Hamiltonian'' route to the index via
\begin{equation}
\mathcal
I(b,\beta)=\mathrm{Tr}\left[(-1)^Fe^{-\hat\beta(\Delta-2j_2-\fft32r)}
p^{j_1+j_2+\fft12r}q^{-j_1+j_2+\fft12r}\right].
\label{eq:equivIndexDef}
\end{equation}
The trace in the above relation is over the Hilbert space of the
theory on $S^3\times \mathbb{R}$, with $S^3$ the unit round
three-sphere, and $\mathbb{R}$ the time direction. The quantum
numbers $(j_1,j_2)$ label the charges of a state under the Cartan of
the $\mathrm{SU}(2)_1\times\mathrm{SU}(2)_2$ isometry group of
$S^3$, while the R-charge is denoted $r$, and $\Delta$---which
coincides with the conformal dimension in a superconformal
theory---is as in \cite{Festuccia:2011}. The index is independent of
$\hat\beta$, because it only receives contributions from states with
$\Delta-2j_2-\fft32r=0$. In a superconformal theory, these states
correspond to operators that sit in short representations of the
superconformal algebra. The index---or alternatively the SUSY
partition function---of an SCFT thus encodes exact
(non-perturbative) information about the operator spectrum of the
underlying theory.\\

Since the expression in Eq.~(\ref{eq:LagEquivIndex}) might seem a
bit complicated, let us specialize it to a very simple case: the
SU($2$) SQCD with three flavors. The gauge group SU($2$) has rank
$r_G=1$. The Weyl group of SU($N$) is the permutation group of $N$
elements, so it has order $N!$, which for SU($2$) becomes $2$. We
have three chiral quark multiplets with
$\rho^{\chi_1}_1,\rho^{\chi_2}_1,\rho^{\chi_3}_1=\pm 1$, and three
chiral anti-quark multiplets with
$\rho^{\chi_{4}}_1,\rho^{\chi_{5}}_1,\rho^{\chi_{6}}_1=\mp1$ (each
of the chiral multiplets has two weights ($\pm 1$), because they sit
in two-dimensional representations of the gauge group). All the
chiral multiplets have R-charge $r_\chi=1/3$. Finally, the group
SU($2$) has two roots, corresponding to the raising and lowering
operators of the 3d angular momentum, and the positive root (the
raising operator) has $\alpha_+=2$. All in all, we get for this
simple example
\begin{equation}
\begin{split}
Z^{\mathrm{SUSY}}_{N_c=2,N_f=3}(b,\beta)=e^{-\beta
E_{\mathrm{susy}}(b)}\frac{(p;p)(q;q)}{2}\int_{-1/2}^{1/2}
\mathrm{d}x \frac{\Gamma^6((pq)^{1/6} z^{\pm 1} )}{\Gamma( z^{\pm
2})},\label{eq:LagIndexSU2SQCD}
\end{split}
\end{equation}
where
$E_{\mathrm{susy}}(b)=i\frac{\omega^3}{6}\mathrm{Tr}R^3+i(\frac{b^2+b^{-2}}{24})\omega\mathrm{Tr}R$,
with $\mathrm{Tr}R^3=-5/9$ and $\mathrm{Tr}R=-5$. The interested
reader is invited to show, using the low-temperature estimates
(\ref{eq:PochLowT}) and (\ref{eq:GammaLowT}), that as $1/\beta\to 0$
the partition function is dominated by the vacuum energy:
$Z^{\mathrm{SUSY}}_{N_c=2,N_f=3}(b,\beta)\simeq e^{-\beta
E_{\mathrm{susy}}(b)}$.

We will spell out the SUSY partition function of several other
theories below. The reader can also consult
\cite{Dolan:2008,Spirido:2009,Spirido:2011or,Kutasov:2014} wherein
explicit expressions are given for the indices of many more
physically interesting supersymmetric QFTs.\\

As a warm-up for our high-temperature analysis, let's study the
low-temperature asymptotics of the SUSY partition function
(\ref{eq:LagEquivZ}). The estimates (\ref{eq:PochLowT}) and
(\ref{eq:GammaLowT}) (the latter being valid assuming
$r_\chi\in]0,2[$) simplify the index (\ref{eq:LagEquivIndex}) at low
temperatures as
\begin{equation}
\begin{split}
\mathcal{I}(b,\beta)\simeq\frac{1}{|W|}\int \mathrm{d}^{r_G}x
\prod_{\alpha_+}\left((1- z^{\alpha_+})(1-
z^{-\alpha_+})\right)=1,\quad\quad (\text{as $1/\beta\to
0$})\label{eq:WeylIntegral}
\end{split}
\end{equation}
with the equality on the RHS resulting from the Weyl integral
formula. The relation (\ref{eq:LagEquivZ}) then yields the universal
low-temperature asymptotics
\begin{equation}
Z^{\mathrm{SUSY}}(b,\beta)\simeq e^{-\beta
E_{\mathrm{susy}}(b)}\quad\quad (\text{as $1/\beta\to 0$, with $b$
fixed}).\label{eq:equivZlowT}
\end{equation}
Recall that the symbol $\simeq$ indicates equality to all
orders---in $1/\beta$---after taking the logarithm of both sides.

A relation similar to (\ref{eq:equivZlowT}) holds for an equivariant
generalization of $Z^{\mathrm{SUSY}}(b,\beta)$, which we denote by
$Z^{\mathrm{SUSY}}(b,\beta;m_a)$. The latter is computed in presence
of real background gauge fields $m_a$ along $S^1_\beta$, each
coupling a conserved U($1$)$_a$ current in the theory; setting all
$m_a$ to zero, we recover $Z^{\mathrm{SUSY}}(b,\beta)$. In that
relation, $E_{\mathrm{susy}}(b)$ is replaced with
$E_{\mathrm{susy}}(b;m_a)$ which can be obtained from
(\ref{eq:EcBequiv}) by shifting every $R\omega$ in it to
$R\omega+Q_a m_a$, with $Q_a$ the U($1$)$_a$ charge of the chiral
fermions in the theory. As emphasized in \cite{Bobev:2015},
$E_{\mathrm{susy}}(b;m_a)$ contains complete information about the
linear and cubic `t~Hooft anomalies of the SUSY QFT. Therefore (if
all $r_\chi$ are in $]0,2[$) the `t~Hooft anomaly matching
conditions correspond to matching the \emph{low-temperature}
asymptotics of the equivariant SUSY partition functions of dual SUSY
QFTs. [As we will discuss in
subsections~\ref{sec:Z3finite}--\ref{sec:c<a} below, various
`t~Hooft anomalies appear also in the high-temperature asymptotics
of $Z^{\mathrm{SUSY}}(b,\beta;m_a)$, but their
appearance is not as universal as in $E_{\mathrm{susy}}(b;m_a)$.]\\

We now get to the main subject of the present section. The
high-temperature asymptotics of the partition function in
(\ref{eq:LagEquivZ}) is found as follows. The Casimir energy factor
is of course negligible at the leading order. The Pochhammer symbols
in the prefactor of (\ref{eq:LagEquivIndex}) can be immediately
replaced with their asymptotic expressions obtainable from
(\ref{eq:PochAsy}). Focusing on the divergent asymptotics we have
\begin{equation}
(p;p)^{r_G}(q;q)^{r_G}\approx e^{-\pi^2(b+b^{-1})r_G/6\beta}\times
\left(\frac{2\pi}{\beta}\right)^{r_G},\label{eq:prePochAsyLead}
\end{equation}
which is accurate to within a multiplicative $O(\beta^0)$ factor.

The leading asymptotics of the integrand of (\ref{eq:LagEquivIndex})
can be obtained from the leading estimates
(\ref{eq:GammaOffCenter2}) and (\ref{eq:GammaOffCenter2Vec}).
Combining these two estimates with (\ref{eq:prePochAsyLead}), we
find that $Z^{\mathrm{SUSY}}$ and $\mathcal{I}$ simplify at high
temperatures to
\begin{equation}
\begin{split}
Z^{\mathrm{SUSY}}(b,\beta)\approx\mathcal I(b,\beta)\approx
\left(\frac{2\pi}{\beta}\right)^{r_G} \int_{\mathfrak{h}_{cl}}
\mathrm{d}^{r_G}x\
e^{-[\mathcal{E}^{DK}_0(b,\beta)+V^{\mathrm{eff}}(\mathbf{x};b,\beta)]+i\Theta(\mathbf{x};\beta)},\label{eq:LagIndexSimp1}
\end{split}
\end{equation}
with $\mathfrak{h}_{cl}$ the unit hypercube $x_i\in [-1/2,1/2]$, and
with
\begin{equation}
\begin{split}
\mathcal{E}^{DK}_0(b,\beta)=-i\frac{\pi^2}{3\beta}\mathrm{Tr}[R\omega],\label{eq:EdkEquiv}
\end{split}
\end{equation}
\begin{equation}
\begin{split}
V^{\mathrm{eff}}(\mathbf{x};b,\beta)=\frac{4\pi^2}{\beta}(\frac{b+b^{-1}}{2})L_h(\mathbf{x}),\label{eq:VeffEquiv1}
\end{split}
\end{equation}
\begin{equation}
\begin{split}
\Theta(\mathbf{x};\beta)=\frac{8\pi^3
}{\beta^2}Q_h(\mathbf{x}).\label{eq:ThetaDef}
\end{split}
\end{equation}
For convenience we have introduced
$\mathbf{x}:=(x_1,\dots,x_{r_G})$. The \emph{real} functions
$Q_h(\mathbf{x})$ and $L_h(\mathbf{x})$ are defined by
\begin{equation}
\begin{split}
Q_h(\mathbf{x}):=\frac{1}{12}\sum_{\chi}\sum_{\rho^{\chi}
\in\Delta_\chi}\kappa(\langle\rho^{\chi}\cdot
\mathbf{x}\rangle),\label{eq:QhDef}
\end{split}
\end{equation}
\begin{equation}
\begin{split}
L_h(\mathbf{x}):= \frac{1}{2}\sum_{\chi}(1-r_\chi)\sum_{\rho^{\chi}
\in\Delta_\chi}\vartheta(\langle\rho^{\chi}\cdot
\mathbf{x}\rangle)-\sum_{\alpha_+}\vartheta(\langle\alpha_+\cdot
\mathbf{x}\rangle).\label{eq:LhDef}
\end{split}
\end{equation}\\

Note that in (\ref{eq:LagIndexSimp1}) we are claiming that \emph{the
matrix-integral is approximated well with the integral of its
approximate integrand}. This is not entirely obvious. First of all,
while the estimate (\ref{eq:GammaOffCenter2}) for the
chiral-multiplet gamma functions is valid uniformly over the domain
of integration, the estimate (\ref{eq:GammaOffCenter2Vec}) for the
vector-multiplet gamma functions is uniform only over compact
subsets of $\mathfrak{h}_{cl}$ that avoid an $O(\beta)$ neighborhood
of the \emph{Stiefel diagram}
\begin{equation}
\mathcal{S}_g:=\bigcup_{\alpha_+}\{\mathbf{x}\in
\mathfrak{h}_{cl}|\langle\alpha_+\cdot
\mathbf{x}\rangle\in\mathbb{Z}\}.\label{eq:SgDef}
\end{equation}
Let's denote this neighborhood by $\mathcal{S}^{(\beta)}_g$.
Intuitively speaking, we expect the estimate
(\ref{eq:GammaOffCenter2VecB}), which applies also on
$\mathcal{S}^{(\beta)}_g$, to guarantee that our unreliable use of
(\ref{eq:GammaOffCenter2Vec}) over this small region modifies the
asymptotics\footnote{A stronger version of
(\ref{eq:GammaOffCenter2VecB}) implies that the expression
(\ref{eq:LhDef}) for $L_h$ should be corrected on
$\mathcal{S}^{(\beta)}_g$. The correction is negligible
($o(\beta^0)$) though, except in an $O(e^{-1/\beta})$ neighborhood
of $\mathcal{S}_g$. In particular, the corrected $L_h$ diverges on
$\mathcal{S}_g$, as the integrand of (\ref{eq:LagEquivIndex})
vanishes there.} at most by a multiplicative $O(1)$ factor. This is
also the error of the estimates used in deriving
(\ref{eq:LagIndexSimp1}) from (\ref{eq:LagEquivIndex}). Therefore
the logarithms of the two sides of the symbols $\approx$ in
(\ref{eq:LagIndexSimp1}) are equal up to an $O(\beta^0)$ error. When
$\Theta=Q_h=0$, the claim in the previous sentence can be justified
more carefully, as we outline below (\ref{eq:LagIndexSimp6noTheta}).

Aside from the issue of non-uniform estimates discussed in the
previous paragraph, a second subtlety may arise in going from an
estimate of the integrand to an estimate for the integral:
cancelations may occur in the actual integral, that do not occur
when integrating the estimated integrand; if this happens, the RHS
of (\ref{eq:LagIndexSimp1}) would overestimate the LHS. Such an
overestimation would be symptomized by divergent corrections that
would arise when trying to improve (\ref{eq:LagIndexSimp1}) to
higher accuracy. When $\Theta=Q_h=0$, the absence of such subtleties
is equivalent to the finiteness of the $O(\beta^0)$ term on the RHS
of (\ref{eq:LagIndexSimp6noTheta}) (see the discussion below
(\ref{eq:ISSindexAsy3}) for instance); we check this finiteness in
some of our explicit examples below, but it can be more generally
demonstrated for non-chiral theories (c.f. \cite{Ardehali:thesis}).

In short, (\ref{eq:LagIndexSimp1}) is demonstrably valid---up to an
$O(\beta^0)$ error upon taking the logarithm of the two sides---in
non-chiral theories (which have $Q_h=0$); we leave its validity---up
to the said error---for chiral theories (which may have $Q_h\neq0$)
as a conjecture.\\

Studying the small-$\beta$ behavior of the multiple-integral on the
RHS of (\ref{eq:LagIndexSimp1}) is now an exercise (albeit a quite
nontrivial one) in asymptotic analysis. Before explaining the
result, we comment on some important properties of the functions
$Q_h$ and $L_h$ introduced above.

The real function $Q_h$ appearing in the phase
$\Theta(\mathbf{x};\beta)$ is \emph{piecewise quadratic}, because
the cubic terms in it cancel thanks to the gauge-gauge-gauge anomaly
cancelation condition:
\begin{equation}
\frac{\partial^3 Q_h(\mathbf{x})}{\partial x_i\partial x_j\partial
x_k}=\sum_\chi\sum_{\rho^\chi\in\Delta_\chi}\rho^\chi_i\rho^\chi_j\rho^\chi_k=0.
\end{equation}
Moreover, as a consequence of the vanishing of the
gauge-gravitational-gravitational anomaly, $Q_h$ is stationary at
the origin:
\begin{equation}
\frac{\partial Q_h(\mathbf{x})}{\partial
x_i}|_{\mathbf{x}=0}=\frac{1}{12}\sum_\chi\sum_{\rho^\chi\in\Delta_\chi}\rho^\chi_i=0.\label{eq:Qstationary}
\end{equation}
We leave it to the interested reader to verify that
$Q_h(\mathbf{x})$ has a continuous first derivative. Also,
$Q_h(\mathbf{x})$ is odd under $\mathbf{x}\to -\mathbf{x}$, and
vanishes at $\mathbf{x}=0$; these properties follow from the fact
that the function $\kappa(x)$ defined in (\ref{eq:kappaDef}) is a
continuous odd function of its argument.

As a result of its oddity, $Q_h(\mathbf{x})$ identically vanishes if
the nonzero $\rho^\chi$ come in pairs with opposite signs; we refer
to theories with such matter content as \emph{non-chiral}; most of
the specific examples that we study in the present paper are of this
kind.

When all $x_i$ are small enough, so that the absolute value of all
the arguments of the $\kappa$ functions in $Q_h$ are less than $1$,
we can use $\kappa(x)=2x^3-3x|x|+x$ to simplify $Q_h$. The resulting
expression---which equals $Q_h$ for $x_i$ small enough---can then be
considered as defining a function $\tilde{Q}_{S^3}(x)$ for any
$x_i\in\mathbb{R}$. Explicitly, we have
\begin{equation}
\begin{split}
\tilde{Q}_{S^3}(\mathbf{x})=-\frac{1}{4}\sum_{\chi}\sum_{\rho^{\chi}
\in\Delta_\chi}\langle\rho^{\chi}\cdot
\mathbf{x}\rangle|\langle\rho^{\chi}\cdot
\mathbf{x}\rangle|,\label{eq:hypS3q}
\end{split}
\end{equation}
with no linear or cubic terms thanks to the cancelation of the
gauge-gravitational-gravitational and gauge-gauge-gauge anomalies.
The homogeneity of $\tilde{Q}_{S^3}$ will be important for us below.
The reason for the subscript $S^3$ will become clear shortly.

The star of our show, the real function $L_h$, determines the
effective potential\footnote{Somewhat surprisingly, $L_h$ also
appears in the $n\to1$ limit of the zero-point energy associated to
nonzero spatial holonomies on $S^1\times S^3/\mathbb{Z}_n$; c.f.
Eq.~(29) of the arXiv preprint of \cite{Benini:2011et} (with $\nu,a$
in there set to zero). It might be possible to clarify this
coincidence by analytically continuing the results of
\cite{Benini:2011et} (see also \cite{Razamat:2013bw,Nieri:2015}) to
non-integer $n$, and then using modular properties of the
generalized elliptic gamma functions employed in that work.}
$V^{\mathrm{eff}}(\mathbf{x};b,\beta)$. It is \emph{piecewise
linear}; the quadratic terms in it cancel because of the ABJ
U($1$)$_R$-gauge-gauge anomaly cancelation:
\begin{equation}
\frac{\partial^2 L_h(\mathbf{x})}{\partial x_i\partial
x_j}=\sum_\chi(r_\chi-1)\sum_{\rho^\chi\in\Delta_\chi}\rho^\chi_i\rho^\chi_j+\sum_{\alpha}\alpha_{i}\alpha_{j}=0.
\end{equation}
Also, $L_h$ is continuous, is even under $\mathbf{x}\to
-\mathbf{x}$, and vanishes at $\mathbf{x}=0$; these properties
follow from the properties of the function $\vartheta(x)$ defined in
(\ref{eq:varthetaDef}). We refer to $L_h(\mathbf{x})$ as the
\emph{Rains function} of the gauge theory. This function has been
analyzed in \cite{Rains:2009} in the context of the elliptic
hypergeometric integrals associated to SU($N$) and Sp($N$) SQCD
theories.

When all $x_i$ are small enough, such that the absolute value of the
argument of every $\vartheta$ function in $L_h$ is smaller than $1$,
we can use $\vartheta(x)=|x|-x^2$ to simplify the Rains function.
The resulting expression---which equals $L_h$ for small $x_i$---can
then be considered as defining a function
$\tilde{L}_{S^3}(\mathbf{x})$ for any $x_i\in\mathbb{R}$.
Explicitly, we have\footnote{Interestingly, on a discrete subset of
its domain (corresponding to the cocharacter lattice of the gauge
group $G$), the function $\tilde{L}_{S^3}$ coincides (up to
normalization) with the $S^2\times S^1$ Casimir energy $\epsilon_0$
\cite{Imamura:2011mp} associated to monopole sectors of the 3d
$\mathcal{N}=2$ theory obtained from dimensional reduction of the 4d
gauge theory. Similarly, $b_0(a)$ in \cite{Imamura:2011mp} is
related to $\tilde{Q}_{S^3}$ above. Also, an analog of the $q_{0i}$
of that work would appear in our analysis if we turn on equivariant
parameters. The observation in the previous footnote might provide a
clue for understanding this set of coincidences. In the context of
3d $\mathcal{N}=4$ theories, a different connection between
$\tilde{L}_{S^3}$ and 3d monopoles was discussed in
\cite{Kapustin:2010f}.}
\begin{equation}
\begin{split}
\tilde{L}_{S^3}(\mathbf{x})=\frac{1}{2}\sum_{\chi}(1-r_\chi)\sum_{\rho^{\chi}
\in\Delta_\chi}|\langle\rho^{\chi}\cdot
\mathbf{x}\rangle|-\sum_{\alpha_+}|\langle\alpha_+\cdot
\mathbf{x}\rangle|.\label{eq:hypURains}
\end{split}
\end{equation}
Note that there is no quadratic term in $\tilde{L}_{S^3}$, thanks to
the cancelation of the U($1$)$_R$-gauge-gauge anomaly. The
homogeneity of $\tilde{L}_{S^3}$ will be important for us below. We
have added a subscript $S^3$ in $\tilde{L}_{S^3}$, because this
function plays an important role in determining whether or not the
$S^3$ partition function of the gauge theory reduced on $S^1_\beta$
is finite.\\

The high-temperature analysis of the integral
(\ref{eq:LagIndexSimp1}) proceeds as follows. First take the factor
$e^{-\mathcal{E}^{DK}_0(b,\beta)}$ outside the integral. Then, since
the real part of the exponent of the integrand is proportional to
$-L_h(\mathbf{x})/\beta$, the $\beta\to 0$ limit exponentially
suppresses the integrand away from the locus of the minima of
$L_h(\mathbf{x})$. This argument suggests, though does not prove,
the high-temperature localization of $Z^{\mathrm{SUSY}}(b,\beta)$.
(A rigorous analysis must first resolve the tension between
minimizing $V^{\mathrm{eff}}$ and making $\Theta$ stationary.)

To make more precise statements, we now focus on the cases where
$\Theta=0$; more specifically, we will keep non-chiral theories in
mind. The cases with $\Theta\neq 0$ (hence with chiral matter
content) require more care, and will not be treated in generality
here; we will study a couple of such examples below.

Setting $\Theta=0$, and writing $V^{\mathrm{eff}}$ in terms of the
Rains function $L_h$, (\ref{eq:LagIndexSimp1}) simplifies to
\begin{equation}
\begin{split}
Z^{\mathrm{SUSY}}(b,\beta)\approx
\left(\frac{2\pi}{\beta}\right)^{r_G}
e^{-\mathcal{E}^{DK}_0(b,\beta)}\int_{\mathfrak{h}_{cl}}
\mathrm{d}^{r_G}x\
e^{-\frac{4\pi^2}{\beta}(\frac{b+b^{-1}}{2})L_h(\mathbf{x})}\quad
(\text{as $\beta\to 0$, when
$Q_h=0$}).\label{eq:LagIndexSimp1noTheta}
\end{split}
\end{equation}

The asymptotic small-$\beta$ analysis of the above integral is
straightforward, but somewhat detailed. Therefore we first give a
brief outline of how the analysis proceeds and what the final result
looks like. The integral localizes, as $\beta\to 0$, around the
locus of the minima of $L_h$. This locus is a subset of
$\mathfrak{h}_{cl}$ that we denote by $\mathfrak{h}_{qu}$, and write
its dimension as $\mathrm{dim}\mathfrak{h}_{qu}$. The integration
goes over $\mathrm{dim}\mathfrak{h}_{qu}$ directions along
$\mathfrak{h}_{qu}$, and $r_G-\mathrm{dim}\mathfrak{h}_{qu}$
directions perpendicular to it. Along the directions perpendicular
to $\mathfrak{h}_{qu}$, the integrand decays exponentially; to get
an order one (instead of
$O(\beta^{r_G-\mathrm{dim}\mathfrak{h}_{qu}})$) result from
integrating along them, it turns out that one has to absorb
$r_G-\mathrm{dim}\mathfrak{h}_{qu}$ factors of $2\pi/\beta$ into the
integral. This leaves $\mathrm{dim}\mathfrak{h}_{qu}$ factors of
$2\pi/\beta$, besides the exponential factors that we have already
seen in Eq.~(\ref{eq:betaLoc}) of the introduction. The end result
is displayed in Eq.~(\ref{eq:LagIndexSimp6noTheta}). The reader not
interested in a more careful derivation of that result is invited to
continue reading from Eq.~(\ref{eq:LagIndexSimp6noTheta}), and skip
the detailed analysis below.

To analyze the integral in (\ref{eq:LagIndexSimp1noTheta}) more
carefully, first note that the integrand is not smooth over
$\mathfrak{h}_{cl}$. We therefore break $\mathfrak{h}_{cl}$ into
sets on which $L_h$ is linear. These sets can be obtained as
follows. Define
\begin{equation}
\mathcal{S}_\chi:=\bigcup_{\rho^\chi\in\Delta^{\neq
0}_\chi}\{\mathbf{x}\in\mathfrak{h}_{cl}|\langle\rho^{\chi}\cdot
\mathbf{x}\rangle\in\mathbb{Z}\},\quad\quad
\mathcal{S}:=\bigcup_{\chi}\mathcal{S}_\chi\cup
\mathcal{S}_g,\label{eq:SSsDef}
\end{equation}
with $\Delta^{\neq 0}_\chi$ ($\subset\Delta_\chi$) the set of
nonzero weights of $\mathcal{R}_\chi$. Note that everywhere in
$\mathfrak{h}_{cl}$, except on $\mathcal{S}$, the function $L_h$ is
guaranteed to be linear---and therefore smooth.

The set $\mathcal{S}$ consists of a union of codimension one affine
hyperplanes inside the space of the $x_i$. These hyperplanes chop
$\mathfrak{h}_{cl}$ into (finitely many, convex) polytopes
$\mathcal{P}_n$. The integral in (\ref{eq:LagIndexSimp1noTheta})
then decomposes to
\begin{equation}
\begin{split}
Z^{\mathrm{SUSY}}(b,\beta)\approx\sum_n
\left(\frac{2\pi}{\beta}\right)^{r_G}
e^{-\mathcal{E}^{DK}_0(b,\beta)}\int_{\mathcal{P}_n}
\mathrm{d}^{r_G}x\
e^{-\frac{4\pi^2}{\beta}(\frac{b+b^{-1}}{2})L_h(\mathbf{x})}.\label{eq:LagIndexSimp2noTheta}
\end{split}
\end{equation}
Since $L_h$ is linear on each $\mathcal{P}_n$, its minimum over
$\mathcal{P}_n$ is guaranteed to be realized on
$\partial\mathcal{P}_n$. Let us assume that this minimum occurs on
the $k$th $j$-face of $\mathcal{P}_n$, which we denote by
$j_n$-$\mathcal{F}^k_n$. We denote the value of $L_h$ on this
$j$-face by $L_{h\ \mathrm{min}}^n$. Equipped with this notation, we
can write (\ref{eq:LagIndexSimp2noTheta}) as
\begin{equation}
\begin{split}
Z^{\mathrm{SUSY}}(b,\beta)\approx\sum_n
\left(\frac{2\pi}{\beta}\right)^{r_G}
e^{-\mathcal{E}^{DK}_0(b,\beta)-\frac{4\pi^2}{\beta}(\frac{b+b^{-1}}{2})L_{h\
\mathrm{min}}^n}\int_{\mathcal{P}_n} \mathrm{d}^{r_G}x\
e^{-\frac{4\pi^2}{\beta}(\frac{b+b^{-1}}{2})\Delta
L^n_h(\mathbf{x})},\label{eq:LagIndexSimp3noTheta}
\end{split}
\end{equation}
where $\Delta L^n_h(\mathbf{x}):=L_h(\mathbf{x})-L_{h\
\mathrm{min}}^n$ is a linear function on $\mathcal{P}_n$. Note that
$\Delta L^n_h(\mathbf{x})$ vanishes on $j_n$-$\mathcal{F}^k_n$, and
it increases as we go away from $j_n$-$\mathcal{F}^k_n$ and into the
interior of $\mathcal{P}_n$. [The last sentence, as well as the rest
of the discussion leading to (\ref{eq:LagIndexSimp6noTheta}), would
receive a trivial modification if $j_n=r_G$ (corresponding to
constant $L_h$ over $\mathcal{P}_n$).] Therefore as $\beta\to 0$,
the integrals in (\ref{eq:LagIndexSimp3noTheta}) localize around
$j_n$-$\mathcal{F}^k_n$.

To further simplify the $n$th integral in
(\ref{eq:LagIndexSimp3noTheta}), we now adopt a set of new
coordinates---affinely related to $x_i$ and with unit
Jacobian---that are convenient on $\mathcal{P}_n$. We pick a point
on $j_n$-$\mathcal{F}^k_n$ as the new origin, and parameterize
$j_n$-$\mathcal{F}^k_n$ with $\bar{x}_1,...,\bar{x}_{j_n}$. We take
$x_{\mathrm{in}}$ to parameterize a direction perpendicular to all
the $\bar{x}$s, and to increase as we go away from
$j_n$-$\mathcal{F}^k_n$ and into the interior of $\mathcal{P}_n$.
Finally, we pick $\tilde{x}_1,...,\tilde{x}_{r_G-j_n-1}$ to
parameterize the perpendicular directions to $x_{\mathrm{in}}$ and
the $\bar{x}$s. Note that, because $\Delta L^n_h$ is linear on
$\mathcal{P}_n$, it does not depend on the $\bar{x}$s; they
parameterize its flat directions. By re-scaling
$\bar{x},x_{\mathrm{in}},\tilde{x}\mapsto
\frac{\beta}{2\pi}\bar{x},\frac{\beta}{2\pi}x_{\mathrm{in}},\frac{\beta}{2\pi}\tilde{x}$,
we can absorb the $(\frac{2\pi}{\beta})^{r_G}$ factors in
(\ref{eq:LagIndexSimp3noTheta}) into the integrals, and write the
$n$th resulting integral as
\begin{equation}
\begin{split}
\int_{\frac{2\pi}{\beta}\mathcal{P}_n} \mathrm{d}^{j_n}\bar{x}\
\mathrm{d}x_{\mathrm{in}}\ \mathrm{d}^{r_G-j_n-1}\tilde{x}\
e^{-2\pi(\frac{b+b^{-1}}{2})\Delta
L^n_h(x_{\mathrm{in}},\mathbf{\tilde{x}})}.\label{eq:ithIntegralSimpl1}
\end{split}
\end{equation}
To eliminate $\beta$ from the exponent, we have used the fact that
$\Delta L^n_h$ depends homogenously on the new coordinates. We are
also denoting the re-scaled polytope schematically by
$\frac{2\pi}{\beta}\mathcal{P}_n$. Instead of integrating over all
of $\frac{2\pi}{\beta}\mathcal{P}_n$ though, we can restrict to
$x_{\mathrm{in}}<\epsilon/\beta$ with some (small) $\epsilon>0$. The
reason is that the integrand of (\ref{eq:ithIntegralSimpl1}) is
exponentially suppressed (as $\beta\to 0$) for
$x_{\mathrm{in}}>\epsilon/\beta$. We take $\epsilon>0$ to be small
enough such that a hyperplane at $x_{\mathrm{in}}=\epsilon/\beta$,
and parallel to $j_n$-$\mathcal{F}^k_n$, cuts off a prismatoid
$P^n_{\epsilon/\beta}$ from $\frac{2\pi}{\beta}\mathcal{P}_n$. After
restricting the integral in (\ref{eq:ithIntegralSimpl1}) to
$P^n_{\epsilon/\beta}$, the integration over the $\bar{x}$s is easy
to perform. The only potential difficulty is that the range of the
$\bar{x}$ coordinates may depend on $x_{\mathrm{in}}$ and the
$\tilde{x}s$. But since we are dealing with a prismatoid, the
dependence is linear, and by the time the range is modified
significantly (compared to its $O(1/\beta)$ size on the re-scaled
$j$-face $\frac{2\pi}{\beta}(j_n$-$\mathcal{F}^k_n)$), the integrand
is exponentially suppressed. Therefore we can neglect the dependence
of the range of the $\bar{x}$s on the other coordinates in
(\ref{eq:ithIntegralSimpl1}). The integral then simplifies to
\begin{equation}
\begin{split}
\left(\frac{2\pi}{\beta}\right)^{j_n}\
\mathrm{vol}(j_n\text{-}\mathcal{F}^k_n)\int_{\hat{P}^n_{\epsilon/\beta}}
\mathrm{d}x_{\mathrm{in}}\ \mathrm{d}^{r_G-j_n-1}\tilde{x}\
e^{-2\pi(\frac{b+b^{-1}}{2})\Delta
L^n_h(x_{\mathrm{in}},\mathbf{\tilde{x}})},\label{eq:ithIntegralSimpl2}
\end{split}
\end{equation}
where $\hat{P}^n_{\epsilon/\beta}$ is the pyramid obtained by
restricting $P^n_{\epsilon/\beta}$ to
$\bar{x}_1=...=\bar{x}_{j_n}=0$.

We now take $\epsilon\to\infty$ in (\ref{eq:ithIntegralSimpl2}).
This introduces exponentially small error, as the integrand is
exponentially suppressed (as $\beta\to 0$) for
$x_{\mathrm{in}}>\epsilon/\beta$. The resulting integral is strictly
positive, because it is the integral of a strictly positive
function. We denote by $I_n$ the result of the integral multiplied
by $\mathrm{vol}(j_n$-$\mathcal{F}^k_n)$. Putting everything
together, we can simplify (\ref{eq:LagIndexSimp3noTheta}) as
\begin{equation}
\begin{split}
Z^{\mathrm{SUSY}}(b,\beta)\approx\sum_n
e^{-\mathcal{E}^{DK}_0(b,\beta)-\frac{4\pi^2}{\beta}(\frac{b+b^{-1}}{2})L_{h\
\mathrm{min}}^n}\left(\frac{2\pi}{\beta}\right)^{j_n}
I_n.\label{eq:LagIndexSimp4noTheta}
\end{split}
\end{equation}
The dominant contribution comes, of course, from the terms/polytopes
whose $L_{h\ \mathrm{min}}^n$ is smallest. If these terms are
labeled by $n=n_\ast^1,n_\ast^2,...$, what we referred to as
$\mathfrak{h}_{qu}$ and $\mathrm{dim}\mathfrak{h}_{qu}$ above can
now be precisely defined via
\begin{equation}
\mathfrak{h}_{qu}:=\bigcup_{n_\ast}
j_{n_\ast}\text{-}\mathcal{F}^k_{n_\ast},\quad
\mathrm{dim}\mathfrak{h}_{qu}:=\mathrm{max}(j_{n_\ast}).
\end{equation}
Put colloquially, if $\mathfrak{h}_{qu}$ has multiple connected
components, by $\mathrm{dim}\mathfrak{h}_{qu}$ we mean the dimension
of the component(s) with greatest dimension, while if a connected
component consists of several intersecting flat elements inside
$\mathfrak{h}_{cl}$, by its dimension we mean the dimension of the
flat element(s) of maximal dimension.

The final result for the high-temperature asymptotics of
$Z^{\mathrm{SUSY}}(b,\beta)$ is then
\begin{equation}
\begin{split}
Z^{\mathrm{SUSY}}(b,\beta)\approx
e^{-\mathcal{E}^{DK}_0(b,\beta)-\frac{4\pi^2}{\beta}(\frac{b+b^{-1}}{2})L_{h\
\mathrm{min}}}\left(\frac{2\pi}{\beta}\right)^{\mathrm{dim}\mathfrak{h}_{qu}},
\label{eq:LagIndexSimp5noTheta}
\end{split}
\end{equation}
where $L_{h\ \mathrm{min}}:=L_{h\ \mathrm{min}}^{n_\ast}$.

Using the explicit expression (\ref{eq:EdkEquiv}) for
$\mathcal{E}^{DK}_0(b,\beta)$, and noting that
(\ref{eq:LagIndexSimp5noTheta}) is accurate up to a multiplicative
factor of order $\beta^0$, we arrive at
\begin{equation}
\boxed{\begin{split} \ln Z^{\mathrm{SUSY}}(b,\beta)=
-\frac{\pi^2}{3\beta}(\frac{b+b^{-1}}{2})(\mathrm{Tr}R+12L_{h\
\mathrm{min}})
+\mathrm{dim}\mathfrak{h}_{qu}\ln(\frac{2\pi}{\beta})+O(\beta^0)&\\
(\text{when }Q_h=0)&. \label{eq:LagIndexSimp6noTheta}
\end{split}}
\end{equation}\\

As the last step in deriving (\ref{eq:LagIndexSimp6noTheta}), we now
outline the argument justifying, when $Q_h=0$, our use of the
estimate (\ref{eq:GammaOffCenter2Vec}) on $\mathcal{S}_g^{(\beta)}$.
First, an analysis similar to the one that took us from
(\ref{eq:LagIndexSimp1noTheta}) to (\ref{eq:LagIndexSimp6noTheta}),
shows that (\ref{eq:LagIndexSimp6noTheta}) is not modified if the
region $\mathcal{S}_g^{(\beta)}$ is excised from the integral
(\ref{eq:LagIndexSimp1noTheta}) (in particular, after excising
$\mathcal{S}_g^{(\beta)}$, since $L_h$ is continuous and piecewise
linear, $L_{h\ \mathrm{min}}$ moves up by an $O(\beta)$ amount,
leaving (\ref{eq:LagIndexSimp6noTheta}) unchanged). The effect of
the integral over $\mathcal{S}_g^{(\beta)}$ is thus negligible
(additive $O(\beta^0)$) on the asymptotics
(\ref{eq:LagIndexSimp6noTheta}). On the other hand, the estimate
(\ref{eq:GammaOffCenter2VecB}) guarantees that the actual
contribution to $Z^{\mathrm{SUSY}}(b,\beta)$ coming from
$\mathcal{S}_g^{(\beta)}$ is of the same order as the integral we
just found negligible. Thus our previously unjustified use of
(\ref{eq:GammaOffCenter2Vec}) on $\mathcal{S}_g^{(\beta)}$
introduces a negligible ($O(\beta^0)$)
error in (\ref{eq:LagIndexSimp6noTheta}).\\

Let us now discuss the obstruction to performing a similar analysis
for chiral theories with $Q_h\neq 0$. Assuming
(\ref{eq:LagIndexSimp1}) [see the comments below (\ref{eq:LhDef})],
for such theories (\ref{eq:LagIndexSimp3noTheta}) would apply,
except that the exponent of the integrand would contain an
$i\Theta$. However, the resulting integrals can not be written in a
form similar to (\ref{eq:ithIntegralSimpl1}) in any obvious way. In
subsections~\ref{sec:Z3finite} and \ref{sec:Coul} below, we consider
special cases with $Q_h\neq 0$, where this obstruction can be
bypassed.\\

In the remaining of this section we consider several specific
examples. We will have the following two goals in mind:
\begin{itemize}
\item \textbf{Deriving more precise asymptotics} than
(\ref{eq:LagIndexSimp6noTheta}). We will find that improving
(\ref{eq:LagIndexSimp6noTheta}) to include the $O(\beta^0)$ term is
generally straightforward, but further obtaining the $O(\beta)$ term
is difficult for theories with $\mathrm{dim}\mathfrak{h}_{qu}>0$.
The only example with $\mathrm{dim}\mathfrak{h}_{qu}>0$ for which we
will improve (\ref{eq:LagIndexSimp6noTheta}) to $O(\beta)$---and in
fact to all-orders---accuracy is the SO($3$) SQCD with two flavors
in the special case $b=1$. What enables us to obtain such a precise
result for this theory is a remarkable equality between its SUSY
partition function and the Schur partition function of the SU($2$)
$\mathcal{N}=4$ SYM. The latter is known to coincide with the
partition function of a free-fermion system on a circle
\cite{Bourdier:2015}, and is thus exceptionally well under control.
The Schur partition function of the SU($N$) $\mathcal{N}=4$ SYM is
studied in appendix~\ref{app:drukker}.
\item \textbf{Analyzing cases with nonzero} $\mathbf{Q_h}$ (and hence
nonzero $\Theta$). If $\mathbf{x}=0$ is the unique minimum of the
Rains function, nonzero $Q_h$ can in fact be easily accommodated,
and as we will show in subsection~\ref{sec:Z3finite} the relation
(\ref{eq:LagIndexSimp6noTheta}) remains valid. A theory exemplifying
this scenario is the magnetic Pouliot theory \cite{Pouliot:1995}
with $N_f=7$. When the Rains function is minimized away from the
origin, or when $\mathrm{dim}\mathfrak{h}_{qu}>0$, nonzero $Q_h$
makes it difficult (as already mentioned above) to obtain precise
general statements similar to (\ref{eq:LagIndexSimp6noTheta}). The
$\mathbb{Z}_3$ orbifold theory studied in subsection~\ref{sec:Coul}
below, exemplifies the scenario with $Q_h\neq 0$ and
$\mathrm{dim}\mathfrak{h}_{qu}>0$. However, it appears that $Q_h$
vanishes on $\mathfrak{h}_{qu}$ in that case, suggesting that
(\ref{eq:LagIndexSimp6noTheta}) is not modified for the
$\mathbb{Z}_3$ orbifold theory either.\\
\end{itemize}

\subsection{$Z_{S^3}$ finite}\label{sec:Z3finite}

Consider the cases where $\mathbf{x}=0$ is the unique minimum of the
Rains function $L_h(\mathbf{x})$; in particular, it is isolated.
Then it follows from (\ref{eq:LagIndexSimp1}) that the
matrix-integral of $Z^{\mathrm{SUSY}}$ localizes around
$\mathbf{x}=0$, and receives exponentially small contribution from
everywhere else. (Note that when $\Theta\neq 0$, to obtain the
asymptotics of (\ref{eq:LagIndexSimp1}) one must in general first
resolve the tension between minimizing $V^{\mathrm{eff}}$ and making
$\Theta$ stationary. But, firstly, since according to
Eq.~(\ref{eq:Qstationary}) $Q_h$ is stationary at the origin, we
have the best of both worlds in the present subsection, and need not
worry about stationarity of the phase $\Theta$. Secondly, in
deriving (\ref{eq:LagIndexSimp2}) we will not even use the
stationarity of $Q_h$ below; the result in (\ref{eq:LagIndexSimp2})
then justifies focusing on a neighborhood of the origin, as the
positivity of the Rains function everywhere else guarantees that the
correction to (\ref{eq:LagIndexSimp2}) coming from the rest of
$\mathfrak{h}_{cl}$ is exponentially suppressed.) We can thus
restrict the domain of integration in (\ref{eq:LagEquivIndex}) to a
small neighborhood of $\mathbf{x}=0$, say a hypercube
$\mathfrak{h}_{cl}^{\epsilon}$ defined by $|x_i|<\epsilon$, in which
the central estimates (\ref{eq:GammaToHypGamma}) and
(\ref{eq:GammaToHypGammaVec}) apply. [In asymptotic analysis, the
procedure of cutting down the range of integration to some
manageable size is sometimes called \emph{tails pruning}.] Using the
central estimates (\ref{eq:GammaToHypGamma}) and
(\ref{eq:GammaToHypGammaVec}) for every elliptic gamma function
inside the integrand of (\ref{eq:LagEquivIndex}), we obtain a
product of several $e^{2\pi i R_0}$ factors the result of which we
denote by $e^{2\pi i R^{\mathrm{integrand}}_0(\mathbf{x};b,\beta)}$,
and also one hyperbolic gamma function for each elliptic gamma
function. On the other hand, according to (\ref{eq:PochAsy}), we
have the following estimate for the Pochhammer symbols in the
prefactor of (\ref{eq:LagEquivIndex}):
\begin{equation}
(p;p)^{r_G}(q;q)^{r_G}\simeq e^{2\pi i\cdot r_G\cdot
R_0^{U(1)}(b,\beta)}\times
\left(\frac{2\pi}{\beta}\right)^{r_G},\quad\quad (\text{as $\beta\to
0$})\label{eq:prePochAsy}
\end{equation}
where we have defined
\begin{equation}
e^{2\pi i
R_0^{U(1)}(b,\beta)}:=e^{-\pi^2(b+b^{-1})/6\beta}e^{(b+b^{-1})\beta/24}.\label{eq:R0U1}
\end{equation}
We now define $e^{2\pi i R^{\mathrm{total}}_0}$ as follows
\begin{equation}
e^{2\pi i R^{\mathrm{total}}_0(\mathbf{x};b,\beta)}:=e^{2\pi i\cdot
r_G\cdot R_0^{U(1)}(b,\beta)}\cdot e^{2\pi i
R^{\mathrm{integrand}}_0(\mathbf{x};b,\beta)}.\label{eq:RtotDef}
\end{equation}

The small-$\beta$ asymptotics of the partition function
(\ref{eq:LagEquivZ}) can then be written as
\begin{equation}
\begin{split}
Z^{\mathrm{SUSY}}(b,\beta)\simeq e^{-\beta
E_{\mathrm{susy}}(b)}\frac{(\frac{2\pi}{\beta})^{r_G}}{|W|}\int_{\mathfrak{h}_{cl}^{\epsilon}}
\mathrm{d}^{r_G}x\ e^{2\pi i
R^{\mathrm{total}}_0(\mathbf{x};b,\beta)} \frac{\prod_\chi
\prod_{\rho^{\chi} \in\Delta_\chi}\Gamma_h(r_\chi\omega
+\frac{2\pi}{\beta}\langle\rho^{\chi}\cdot
\mathbf{x}\rangle)}{\prod_{\alpha_+}\Gamma_h(\pm\frac{2\pi}{\beta}\langle\alpha_+\cdot
\mathbf{x}\rangle)}.\label{eq:LagIndexSimp2fourD}
\end{split}
\end{equation}

Since the function $R_0(x;\sigma,\tau)$ defined in (\ref{eq:RhDef})
is a cubic polynomial in $x$, one might expect
$R^{\mathrm{total}}_0(\mathbf{x};b,\beta)$ to be also cubic in
$x_i$. But because of the gauge-gauge-gauge anomaly cancelation, the
cubic terms in $R^{\mathrm{total}}_0$ cancel. Because of the
vanishing of the ABJ anomaly for the U($1$)$_R$ current, the
quadratic terms in $R^{\mathrm{total}}_0$ also cancel. In fact
$R^{\mathrm{total}}_0$ is completely $x_i$-independent, because the
linear terms in it also cancel due to the vanishing of the mixed
gauge-U($1$)$_R^2$ and gauge-gravity-gravity anomalies (see the
related discussion in section~5 of \cite{Assel:2014}). The end
result is
\begin{equation}
e^{2\pi i
R^{\mathrm{total}}_0(\mathbf{x};b,\beta)}=e^{-\mathcal{E}_0^{DK}(b,\beta)+\beta
E_{\mathrm{susy}}(b)}.\label{eq:RhTot}
\end{equation}

Using the above simplification for $R^{\mathrm{total}}_0$, and
re-scaling the integration variables in (\ref{eq:LagIndexSimp2fourD}) via
$x\mapsto (\frac{2\pi}{\beta})x$, we arrive at
\begin{equation}
\begin{split}
Z^{\mathrm{SUSY}}(b,\beta)\simeq e^{-\mathcal{E}_0^{DK}(b,\beta)}
Z_{S^3}(b;2\pi\epsilon/\beta),\quad \quad (\text{as $\beta\to
0$})\label{eq:LagIndexSimp2}
\end{split}
\end{equation}
where
\begin{equation}
Z_{S^3}(b;\Lambda):=\frac{1}{|W|}\int_{\Lambda} \mathrm{d}^{r_G}x\
\frac{\prod_\chi \prod_{\rho^{\chi}
\in\Delta_\chi}\Gamma_h(r_\chi\omega +\langle\rho^{\chi}\cdot
\mathbf{x}\rangle)}{\prod_{\alpha_+}\Gamma_h(\pm\langle\alpha_+\cdot
\mathbf{x}\rangle)}, \label{eq:LagZ3d}
\end{equation}
is the matrix-integral computing the squashed-three-sphere partition
function of the theory reduced on $S_\beta^1$ (c.f. Eq.~(5.23) of
\cite{Aharony:2013a}), assuming the same R-charge assignments in the
reduced theory as those directly descending from the parent 4d
theory. We are keeping the cut-off $\Lambda$ explicit, emphasizing
that the integration is over the hypercube $|x_i|<\Lambda$.

The RHS of (\ref{eq:LagIndexSimp2}) still has an intricate
temperature-dependence through the $\beta$-dependent cut-off for
$Z_{S^3}$. Our final step in analyzing the high-temperature
asymptotics of $Z^{\mathrm{SUSY}}$ for theories with finite
$Z_{S^3}$ is to argue that taking $\epsilon\to\infty$ in
(\ref{eq:LagIndexSimp2}) introduces exponentially small error. [In
asymptotic analysis, the procedure of extending the range of
integration to an infinitely large set, over which computations are
simplified, is sometimes called \emph{tails completion}.] Upon using
the asymptotics of the hyperbolic gamma function in
(\ref{eq:hyperbolicGammaAsy}), we find that the integrand of
$Z_{S^3}(b;\Lambda)$ can be estimated, as $x\to\infty$, by
\begin{equation}
\frac{\prod_\chi \prod_{\rho^{\chi}
\in\Delta_\chi}\Gamma_h(r_\chi\omega +\langle\rho^{\chi}\cdot
\mathbf{x}\rangle)}{\prod_{\alpha_+}\Gamma_h(\pm\langle\alpha_+\cdot
\mathbf{x}\rangle)}\approx e^{-2\pi
(\frac{b+b^{-1}}{2})\tilde{L}_{S^3}(\mathbf{x})+2\pi
i\tilde{Q}_{S^3}(\mathbf{x})}, \label{eq:LagZ3dIntegrandAsy}
\end{equation}
with $\tilde{L}_{S^3}$ and $\tilde{Q}_{S^3}$ the functions defined
in (\ref{eq:hypURains}) and (\ref{eq:hypS3q}). Our assumption that
$\mathbf{x}=0$ is an isolated minimum of $L_h(\mathbf{x})$ now
implies that for $\mathbf{x}$ small enough,
$\tilde{L}_{S^3}(\mathbf{x})$ is strictly positive. But since
$\tilde{L}_{S^3}(\mathbf{x})$ is a homogenous function of
$\mathbf{x}$, its strict positivity for small enough $\mathbf{x}$
implies its strict positivity for all $x$. As a result, for
$x_i\propto\Lambda$ the integrand of $Z_{S^3}(b;\Lambda)$ is
exponentially small as $\Lambda\to\infty$, and tails completion
introduces an error that is exponentially small in the cut-off. Thus
taking $\epsilon\to\infty$ in (\ref{eq:LagIndexSimp2}) introduces an
error of the type $e^{-\epsilon/\beta}$, and we can write
\begin{equation}
\begin{split}
\ln Z^{\mathrm{SUSY}}(b,\beta)\sim -\mathcal{E}_0^{DK}(b,\beta)+ \ln
Z_{S^3}(b),\quad\quad (\text{as $\beta\to
0$})\label{eq:LagIndexSimp3}
\end{split}
\end{equation}
where $Z_{S^3}(b):=Z_{S^3}(b;\infty)$ is the squashed-three-sphere
partition function with the cut-off removed. The symbol $\sim$
indicates that the error is beyond all-orders, but our arguments
above imply the stronger result that the error is exponentially
small, of the type $e^{-1/\beta}$.

We have demonstrated that if $\mathbf{x}=0$ is the unique global
minimum of the Rains function, then $Z_{S^3}(b)$ is finite and
(\ref{eq:LagIndexSimp3}) holds. Our arguments show in fact that if
$\mathbf{x}=0$ is an isolated \emph{local} minimum of $L_h$, then
$Z_{S^3}(b)$ is finite (although (\ref{eq:LagIndexSimp3}) does not
hold if $\mathbf{x}=0$ is not a global minimum). Conversely, if
$Z_{S^3}(b)$ is finite, then $\mathbf{x}=0$ is an isolated local
minimum of the Rains function. This is because for $Z_{S^3}(b)$ to
be finite, its integrand in (\ref{eq:LagZ3d}) must decay at large
$\mathbf{x}$. Hence $\tilde{L}_{S^3}(\mathbf{x})$ must be positive
for large $\mathbf{x}$, and because of its homogeneity, also for
small nonzero $\mathbf{x}$. But for small $\mathbf{x}$ the two
functions $\tilde{L}_{S^3}$ and $L_h$ coincide. Therefore $L_h$ is
strictly positive for small enough but nonzero $\mathbf{x}$. Since
$L_h(\mathbf{x}=0)=0$, the desired result follows.\\

Combining (\ref{eq:LagIndexSimp3}) and (\ref{eq:LagEquivZ}), we
obtain
\begin{equation}
\begin{split}
\ln \mathcal{I}(b,\beta)\sim -\mathcal{E}_0^{DK}(b,\beta)+ \ln
Z_{S^3}(b)+\beta E_{\mathrm{susy}}(b)\quad\quad (\text{as $\beta\to
0$}).\label{eq:LagIndexSimp3Romelsberger}
\end{split}
\end{equation}
A relation similar to (\ref{eq:LagIndexSimp3Romelsberger}) holds for
an equivariant generalization of $\mathcal{I}(b,\beta)$, which we
denote by $\mathcal{I}(b,\beta;m_a)$. The latter contains fugacities
$u_a=e^{i\beta m_a}$ associated to conserved U($1$)$_a$ charges of
the theory (the U($1$)s may reside in the Cartan torus of a
non-abelian group); setting all $u_a$ to $1$, we recover
$\mathcal{I}(b,\beta)$. In that generalized relation,
$E_{\mathrm{susy}}(b)$ is replaced with $E_{\mathrm{susy}}(b;m_a)$.
Therefore all the `t~Hooft anomalies of a SUSY gauge theory with a
U($1$) R-symmetry, with a semi-simple gauge group, and with a Rains
function that is minimized only at the origin, can be extracted from
the high-temperature asymptotics of the equivariant Romelsberger
index of the theory. This statement is related (but not equivalent)
to some of the claims in \cite{Spiridonov:2012sv}, which were made
there in the context of SU($N$) SQCD.

\subsubsection{$A_k$ SQCD theories with $N_f>\frac{2N}{k+1}$}

Take now the example of $A_k$ SQCD with SU($N$) gauge group. This
theory has a chiral multiplet with R-charge $r_a=\frac{2}{k+1}$ in
the adjoint, $N_f$ flavors in the fundamental with R-charge
$r_f=1-\frac{2}{k+1}\frac{N}{N_f}$, and $N_f$ flavors in the
anti-fundamental with R-charge $r_{\bar{f}}=r_f$. For $r_f$ to be
positive we must have $N_f>2N/(k+1)$.

We will not bother commenting on the IR phase of the theory on flat
space for various ranges of parameters. What matters for us is that
the supersymmetric partition function of the theory on $S_b^3\times
S^1_\beta$ is well-defined if $N_f>2N/(k+1)$.

The SUSY partition function is (c.f. \cite{Dolan:2008})
\begin{equation}
\begin{split}
Z^{\mathrm{SUSY}}_{A_k}(b,\beta)=e^{-\beta
E_{\mathrm{susy}}(b)}&\frac{(p;p)^{N-1}(q;q)^{N-1}}{N!}\Gamma^{N-1}((pq)^{r_a/2})\int
\mathrm{d}^{N-1}x\\ &\left(\prod_{1\le i<j\le
N}\frac{\Gamma((pq)^{r_a/2}(z_i/z_j)^{\pm 1})}{\Gamma((z_i/z_j)^{\pm
1})}\right) \prod_{i=1}^{N}\Gamma^{N_f}((pq)^{r_f/2} z_i^{\pm
1}),\label{eq:SQCDindex}
\end{split}
\end{equation}
with $\prod_{i=1}^{N} z_i=1$.

\begin{figure}[t]
\centering
    \includegraphics[scale=.7]{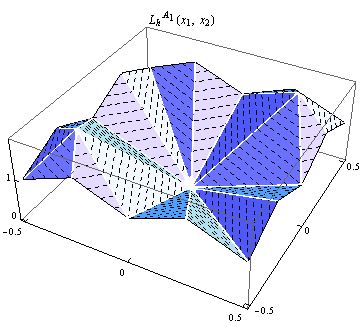}
\caption{The Rains function of the $A_1$ SU($3$) theory---also known
as SU($3$) SQCD---for $N_f>3$. Note that the minimum lies at
$x_1=x_2=0$. \label{fig:A1}}
\end{figure}

The Rains function of the theory is
\begin{equation}
\begin{split}
L_h^{A_k}(x_1,\dots,x_{N-1})&=N_f
(1-r_f)\sum_{i=1}^{N}\vartheta(x_i)+(1-r_a)\sum_{1\le i<j\le
N}\vartheta(x_i-x_j)-\sum_{1\le i<j\le N}\vartheta(x_i-x_j)\\
&=\frac{2}{k+1}(N\sum_{i}\vartheta(x_i)-\sum_{1\le i<j\le
N}\vartheta(x_i-x_j)).\label{eq:AkRains}
\end{split}
\end{equation}
The $x_N$ in the above expression is constrained by $\sum_{i=1}^{N}
x_i\in\mathbb{Z}$, although since $\vartheta(x)$ is periodic with
period one we can simply replace $x_N\to -x_1-\dots-x_{N-1}$. For
$k=1$ and $N=3$, the resulting function is illustrated in
Figure~\ref{fig:A1}.

We recommend that the reader convince herself that the Rains
function in (\ref{eq:AkRains}) can be easily written down by
examining the integrand of (\ref{eq:SQCDindex}). Whenever the index
(or the SUSY partition function) of a theory is available in the
literature, a similar examination of the integrand quickly yields
the theory's $L_h$ and $Q_h$ functions.

Using Rains's generalized triangle inequality (\ref{eq:RainsGTI}),
in the special case where $d_i=0$, we find that the above function
is minimized when all $x_i$ are zero. This establishes that the
integrand of (\ref{eq:SQCDindex}) is localized around $x_i=0$, and
is exponentially suppressed everywhere else, as $\beta\to 0$. The
asymptotic relation (\ref{eq:LagIndexSimp3}) then applies with
\begin{equation}
\begin{split}
Z^{A_k}_{S^3}(b)=\frac{\Gamma_h^{N-1}(r_a \omega)}{N!} \int
\mathrm{d}^{N-1}x \left(\prod_{1\le i<j\le N}\frac{\Gamma_h(r_a
\omega\pm (x_i-x_j) )}{\Gamma_h(\pm (x_i-x_j) )}\right)
\prod_{i=1}^{N}\Gamma_h^{N_f}(r_f \omega\pm x_i).\label{eq:SQCD3dZ}
\end{split}
\end{equation}
The convergence of the above integral (over
$x_1,\dots,x_{N-1}\in]-\infty,\infty[$) follows from our general
discussion above (\ref{eq:LagIndexSimp3}), but it can also be
explicitly verified using the estimate (\ref{eq:hyperbolicGammaAsy})
and the generalized triangle inequality (\ref{eq:RainsGTIav}).\\

\begin{figure}[t]
\centering
    \includegraphics[scale=.7]{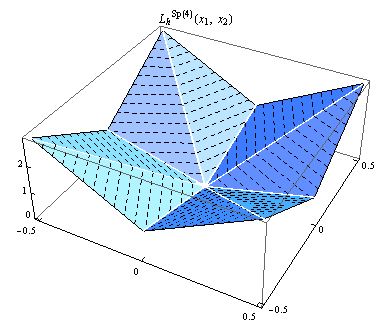}
\caption{The Rains function of the Sp($4$) SQCD theory for $N_f>3$.
Note that the minimum lies at $x_1=x_2=0$. \label{fig:Sp4}}
\end{figure}

A similar story applies to the $D$ and $E$ type SU($N$) SQCD
theories \cite{Kutasov:2014}, and also to the Sp($2N$) SQCD
theories. We leave it as an exercise for the interested reader to
reproduce the plot of the Rains function of the Sp($4$) SQCD for
$N_f>3$ shown in Figure~\ref{fig:Sp4}. (The Romelsberger index of
the Sp($2N$) SQCD theories can be found in \cite{Dolan:2008}.
Lemma~3.3 of Rains \cite{Rains:2009} establishes that
$L_h^{Sp(2N)}(\mathbf{x})$ is minimized only at $\mathbf{x}=0$, for
any $N_f>N+1$.)

\subsubsection{The magnetic Pouliot theory with $N_f=7$}

We now consider a chiral theory with $Q_h\neq 0$. This is the
SU($3$) theory \cite{Pouliot:1995} with seven chiral multiplets of
R-charge $r_{\bar{f}}=13/21$ in the anti-fundamental representation,
a chiral multiplet of R-charge $r_{w}=2/3$ in the symmetric tensor
representation, and 28 chiral gauge-singlets with R-charge
$r_{M}=4/7$.

The SUSY partition function is (c.f. \cite{Spirido:2011or})
\begin{equation}
\begin{split}
Z^{\mathrm{SUSY}}_{P_{m7}}(b,\beta)=e^{-\beta
E_{\mathrm{susy}}(b)}&\frac{(p;p)^{2}(q;q)^{2}}{3!}\Gamma^{28}((pq)^{r_M/2})\int
\mathrm{d}^{2}x\\ &\left(\prod_{1\le i<j\le
3}\frac{\Gamma((pq)^{r_w/2}z_i z_j)}{\Gamma((z_i/z_j)^{\pm
1})}\right) \prod_{i=1}^{3}\Gamma((pq)^{r_w/2}
z_i^{2})\Gamma^{7}((pq)^{r_{\bar{f}}/2} z_i^{-
1}),\label{eq:PouliotIndex}
\end{split}
\end{equation}
with $\prod_{i=1}^{3} z_i=1$.

\begin{figure}[t]
\centering
    \includegraphics[scale=.7]{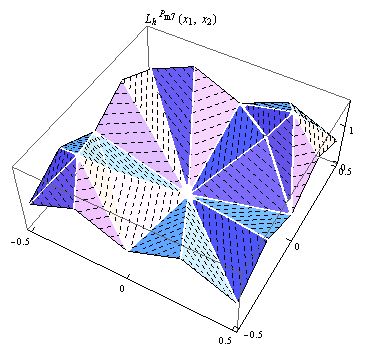}
\caption{The Rains function of the SU($3$) magnetic Pouliot theory
with $N_f=7$. Note that the minimum lies at $x_1=x_2=0$.
\label{fig:Pm7}}
\end{figure}

The Rains function of the theory is
\begin{equation}
\begin{split}
L_h^{P_{m7}}(x_1,x_2)&=\frac{1}{2}(1-r_w)\sum_{1\le i<j\le
3}\vartheta(x_i+x_j)+\frac{1}{2}(1-r_w)\sum_{i=1}^{3}\vartheta(2x_i)\\
&\ \ +7\cdot\frac{1}{2}(1-r_{\bar{f}})\sum_{i=1}^{3}\vartheta(x_i)-\sum_{1\le i<j\le 3}\vartheta(x_i-x_j)\\
&=\frac{1}{6}\sum_{1\le i<j\le
3}\vartheta(x_i+x_j)+\frac{1}{6}\sum_{i=1}^{3}\vartheta(2x_i)+\frac{13}{6}\sum_{i=1}^{3}\vartheta(x_i)-\sum_{1\le
i<j\le 3}\vartheta(x_i-x_j),
\end{split}
\end{equation}
with $x_3=-x_1-x_2$. This function is illustrated in
Figure~\ref{fig:Pm7}.

\begin{figure}[h]
\centering
    \includegraphics[scale=.7]{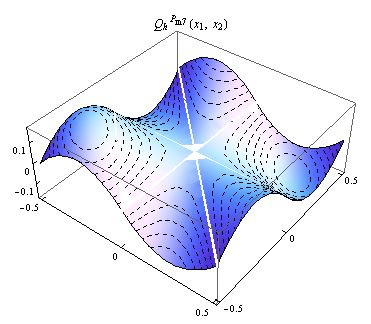}
\caption{The $Q_h$ function of the SU($3$) magnetic Pouliot theory
with $N_f=7$. Note the stationarity at $x_1=x_2=0$.
\label{fig:Pm7Qh}}
\end{figure}

For this theory $Q_h$ is given by
\begin{equation}
\begin{split}
Q_h^{P_{m7}}(x_1,x_2)=\frac{1}{12}\sum_{1\le i<j\le
3}\kappa(x_i+x_j)+\frac{1}{12}\sum_{i=1}^{3}\kappa(2x_i)+7\cdot\frac{1}{12}\sum_{i=1}^{3}\kappa(-x_i),
\end{split}
\end{equation}
again with $x_3=-x_1-x_2$. This function is illustrated in
Figure~\ref{fig:Pm7Qh}.

As Figure~\ref{fig:Pm7} demonstrates, the Rains function has a
unique minimum at the origin of $\mathfrak{h}_{cl}$. Therefore the
asymptotics (\ref{eq:LagIndexSimp3}) applies, with some $Z_{S^3}(b)$
whose derivation we leave to the interested reader.\\

\subsection{$Z_{S^3}$ power-law divergent (or: The effect of an unlifted Coulomb branch)}\label{sec:Coul}

Assume now that $\mathbf{x}=0$ is a global minimum of $L_h$, and
that the zero-set of $L_h$ is a connected subset of
$\mathfrak{h}_{cl}$, which we refer to as $\mathfrak{h}_{qu}$. In
such cases the Rains function has ``flat directions'' along
$\mathfrak{h}_{qu}$. These flat directions present an obstruction to
the tails pruning of the previous subsection. Therefore we can not
obtain asymptotic expressions as precise as
(\ref{eq:LagIndexSimp3}). Nonzero $Q_h$ would present another
difficulty in the general analysis. We thus assume for now that
$Q_h=0$; the $\mathbb{Z}_3$ orbifold studied below provides an
example with $Q_h\neq 0$. Equation (\ref{eq:LagIndexSimp6noTheta})
then reads
\begin{equation}
\begin{split}
\ln Z^{\mathrm{SUSY}}(b,\beta)=
-\mathcal{E}^{DK}_0(b,\beta)+\mathrm{dim}\mathfrak{h}_{qu}\cdot
\ln\left(\frac{2\pi}{\beta}\right)+O(1)\quad\quad (\text{as
$\beta\to 0$}).\label{eq:LagIndexSimp1Coul2}
\end{split}
\end{equation}

We will derive more precise asymptotic expressions in our detailed
case-studies below. But before that, some comments on the relation
between the logarithmic term in (\ref{eq:LagIndexSimp1Coul2}) and
the ``unlifted Coulomb branches'' are in order.

The SUSY partition function is computed by a path-integral on
$S_b^3\times S_\beta^1$. If the theory on $S_b^3\times S_\beta^1$
contains unlifted (or quantum) zero-modes with a compact target
manifold, the path-integral receives a multiplicative contribution
from the volume of the target space. The parameters $x_i$ that we
have used above, are related via $\sigma_i=\frac{2\pi}{\beta}x_i$ to
the scalar zero-modes $\sigma_i$ associated with the holonomies
(c.f. section~2 of \cite{Aharony:2013a}). Since $x_i$ are periodic
with period one, the volume of the target space of the $\sigma_i$ is
proportional to $(\frac{2\pi}{\beta})^{r_G}$; we say proportional,
because one must mod out the product space by the large gauge
transformations associated to the Weyl group of $G$; this introduces
an $O(1)$ factor though, and can be neglected for our current
discussion. Not all of the $r_G$ scalars $\sigma_i$ are quantum zero
modes; some of them are lifted by quantum mechanically generated
potentials. We interpret $V^{\mathrm{eff}}$ as (the high-temperature
asymptotics of) such a potential, and thus conclude that the
unlifted zero modes are those $\sigma_i$ that correspond to the
$x_i$ parameterizing $\mathfrak{h}_{qu}$. Because the target space
of these unlifted $\sigma_i$ decompactifies as $\beta\to 0$, we say
that theories with $\mathrm{dim}\mathfrak{h}_{qu}>0$ experience
\emph{Coulomb branch decompactification at high temperatures} on
$S_b^3\times S_\beta^1$. The word ``Coulomb branch'' is used because
the $\sigma_i$ parameterize (part of) the Coulomb branch of the 3d
$\mathcal{N}=2$ gauge theory obtained by reducing the 4d
$\mathcal{N}=1$ gauge theory on $S_\beta^1$.

With the same assumptions that $\mathbf{x}=0$ is a global minimum of
$L_h$ and that the zero-set of $L_h$ is a connected subspace of
$\mathfrak{h}_{cl}$ denoted $\mathfrak{h}_{qu}$, we can demonstrate
that $Z_{S^3}(b;\Lambda)$ defined in (\ref{eq:LagZ3d}) must be
power-law divergent in $\Lambda.$ Combining (\ref{eq:LagZ3d}) and
(\ref{eq:LagZ3dIntegrandAsy}) we can find the leading
$\Lambda$-dependence of $Z_{S^3}(b;\Lambda)$ from
\begin{equation}
Z_{S^3}(b;\Lambda)\approx\frac{1}{|W|}\int_{\Lambda}
\mathrm{d}^{r_G}x\ e^{-2\pi
(\frac{b+b^{-1}}{2})\tilde{L}_{S^3}(\mathbf{x})}.
\label{eq:LagZ3dCoul}
\end{equation}

Now since $L_h$ has flat directions near $\mathbf{x}=0$, so does
$\tilde{L}_{S^3}$. Assume that the flat directions of
$\tilde{L}_{S^3}$ parameterize a space of dimension
$\mathrm{dim}\mathfrak{h}_{qu}$ (this is the case in the SO($2N+1$)
SQCD and the $\mathcal{N}=4$ SYM examples below, and we conjecture
that it is the case also in the $\mathbb{Z}_2$ and $\mathbb{Z}_3$
orbifold theories; more generally, we suspect---but have not been
able to show---that whenever $L_{h\ \mathrm{min}}=0$, the flat
directions of $\tilde{L}_{S^3}$ parameterize a space of dimension
$\mathrm{dim}\mathfrak{h}_{qu}$). Along these flat directions the
integrand of $Z_{S^3}(b;\Lambda)$ does not decay at large $\Lambda$,
and thus upon taking the cut-off to infinity $Z_{S^3}(b;\Lambda)$
diverges as $\Lambda^{\mathrm{dim}\mathfrak{h}_{qu}}$. Such
power-law divergences are expected for 3d theories with an unlifted
Coulomb branch \cite{DiPietro:2014}.

\subsubsection{SO($2N+1$) SQCD with $N_f>2N-1$}

Consider the SO($n$) SQCD theories with $N_f$ chiral matter
multiplets of R-charge $r=1-\frac{n-2}{N_f}$ in the vector
representation. For the R-charges to be greater than zero, and the
gauge group to be semi-simple, we must have $0<n-2<N_f$.

We perform the analysis for odd $n$; the analysis for even $n$ is
completely analogous, and the result is similar. The SUSY partition
function of SO($2N+1$) SQCD is given by (c.f. \cite{Dolan:2008})
\begin{equation}
\begin{split}
Z^{\mathrm{SUSY}}_{SO(2N+1)}(b,\beta)=e^{-\beta
E_{\mathrm{susy}}(b)}&\frac{(p;p)^N(q;q)^N}{2^N
N!}\Gamma^{N_f}((pq)^{r/2})\\
&\times\int \mathrm{d}^N x
\frac{\prod_{j=1}^{N}\Gamma^{N_f}((pq)^{r/2}z_j^{\pm1})}{\prod_{j=1}^{N}
\Gamma(z_j^{\pm1})\prod_{i<j}(\Gamma((z_i z_j)^{\pm1})\Gamma((z_i/
z_j)^{\pm1}))}.\label{eq:SONindex}
\end{split}
\end{equation}

\begin{figure}[t]
\centering
    \includegraphics[scale=.7]{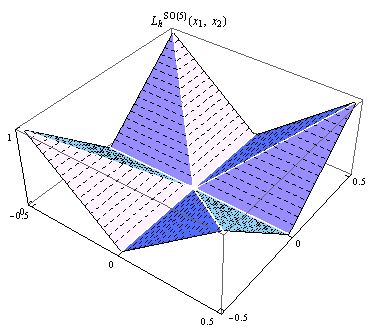}
\caption{The Rains function of the SO($5$) SQCD theory with $N_f>3$.
Note the flat directions along the axes. \label{fig:SO5}}
\end{figure}

The Rains function of the theory is
\begin{equation}
\begin{split}
L_h^{SO(2N+1)}(\mathbf{x})=(2N-2)\sum_{j=1}^{N}\vartheta(x_j)
-\sum_{1\le i<j\le N}\vartheta(x_i+x_j)-\sum_{1\le i<j\le
N}\vartheta(x_i-x_j).\label{eq:soNRains}
\end{split}
\end{equation}
For the case $N=2$, corresponding to the SO($5$) theory, this
function is illustrated in Figure~\ref{fig:SO5}.

To find the minima of the above function, we need the following
result, valid for $-1/2\le x_i\le 1/2$:
\begin{equation}
\begin{split}
(2N-2)\sum_{1\le j\le N}\vartheta(x_j)&-\sum_{1\le i<j\le
N}\vartheta(x_i+x_j)-\sum_{1\le i<j\le
N}\vartheta(x_i-x_j)=2\sum_{1\le i<j\le
N}\mathrm{min}(|x_i|,|x_j|)\\
&=2(N-1)\mathrm{min}(|x_i|)+2(N-2)\mathrm{min}_2(|x_i|)+\cdots
+2\mathrm{min}_{N-1}(|x_i|),\label{eq:SONvarthetaId}
\end{split}
\end{equation}
where $\mathrm{min}(|x_i|)$ stands for the smallest of
$|x_1|,\dots,|x_N|$, while $\mathrm{min}_2(|x_i|)$ stands for the
next to smallest element, and so on. To prove
(\ref{eq:SONvarthetaId}), one can first verify it for $N=2$, and
then use induction for $N>2$.

Applying (\ref{eq:SONvarthetaId}) we find that the Rains function in
(\ref{eq:soNRains}) is minimized to zero when one (and only one) of
the $x_j$ is nonzero, and the rest are zero. This follows from the
fact that $\mathrm{max}(|x_i|)$ does not show up on the RHS of
(\ref{eq:SONvarthetaId}). Therefore, unlike for the theories of the
previous subsection, here the matrix-integral is not localized
around the origin of the $x_i$ space, but \emph{localized around the
axes}. Equation (\ref{eq:LagIndexSimp1Coul2}) thus simplifies to
\begin{equation}
\begin{split}
\ln Z^{\mathrm{SUSY}}_{SO(2N+1)}(b,\beta)=
-\mathcal{E}^{DK}_0(b,\beta)+
\ln\left(\frac{2\pi}{\beta}\right)+O(1)\quad\quad (\text{as
$\beta\to 0$}).\label{eq:LagIndexSimp1Coul2SON}
\end{split}
\end{equation}

\subsubsection*{More precise asymptotics}

Below we improve the asymptotic relation
(\ref{eq:LagIndexSimp1Coul2SON}) by obtaining the $O(1)$ term in it.
The following discussion is somewhat detailed, and the reader not
interested in the technical nuances is invited to skip to
Eq.~(\ref{eq:SONindexAsy2}) and continue reading from there.

Symmetry implies that we can compute the contribution from around
the $x_1$ axis, and multiply the result by $N$ to get the final
result. Since away from the axes the integrand is exponentially
small, to compute the contribution coming from around the $x_1$ axis
we can assume $|x_2|,\dots,|x_N| < \epsilon$; this is the
\emph{tails pruning}. Unlike in the previous subsection though, now
there is one direction (namely $x_1$) in which we can not prune.

Neglect for the moment the region where $|x_1|$ is smaller than or
equal to some small fixed $\varepsilon_1>0$. Then for all the gamma
functions that contain $x_1$ in their argument we can use the
following estimate, valid uniformly over (fixed,
$\beta$-independent) compact subsets of $\mathbb{R}$ avoiding
$\mathbb{Z}$ (c.f. Proposition~2.11 in \cite{Rains:2009}):
\begin{equation}
\begin{split}
\ln\Gamma(x+r(\frac{\sigma+\tau}{2});\sigma,\tau)\sim 2\pi
iQ_{+}(\{x\}+r(\frac{\sigma+\tau}{2});\sigma,\tau),\quad (\text{as
$\beta\to 0$, for $x\in\mathbb{R}\backslash\mathbb{Z}$})
\label{eq:GammaAsyPosExt}
\end{split}
\end{equation}
where $r$ can be any (fixed) real number, and
\begin{equation}
\begin{split}
Q_{+}(x;\sigma,\tau)=&-\frac{x^3}{6\tau\sigma}+\frac{\tau+\sigma+1}{4\tau\sigma}x^2-\frac{\tau^2+\sigma^2+3\tau\sigma+3\tau+3\sigma+1}{12\tau\sigma}x\\
&+\frac{1}{24}(\tau+\sigma+1)(1+\tau^{-1}+\sigma^{-1}).\label{eq:MmaDef}
\end{split}
\end{equation}
Using the estimate (\ref{eq:GammaToHypGamma}) for all the rest of
the elliptic gamma functions (namely, those that do not contain
$x_1$ in their argument), we obtain various factors of $e^{2\pi i
R_0}$, as well as one hyperbolic gamma for each of the elliptic
gammas. It turns out that after using the aforementioned estimates
the dependence of the integrand on $x_1$ completely drops out (this
is essentially because (\ref{eq:SONvarthetaId}) is independent of
$\mathrm{max}(|x_i|)$). Therefore the integral over $x_1$ can be
performed. The result is ($R^{\mathrm{integrand}}_h$ stands for the
sum of the various $Q_+$es and $R_0$s)
\begin{equation}
\begin{split}
Z^{\mathrm{SUSY}}_{SO(2N+1)}(b,\beta)\approx e^{-\beta
E_{\mathrm{susy}}(b)} &\frac{(p;p)^N(q;q)^N}{2^N
(N-1)!}\Gamma^{N_f}((pq)^{r/2})\times e^{2\pi i R^{\mathrm{integrand}}_h}\times\int \mathrm{d}^{N-1} x\\
& \frac{\prod_{j=2}^{N}\Gamma_h^{N_f}(\omega r \pm \frac{2\pi
x_j}{\beta})}{\prod_{j=2}^{N} \Gamma_h(\pm \frac{2\pi
x_j}{\beta})\prod_{2\le i<j\le N}(\Gamma_h(\pm \frac{2\pi
(x_i+x_j)}{\beta})\Gamma_h(\pm \frac{2\pi
(x_i-x_j)}{\beta}))},\label{eq:SONindexAsy1}
\end{split}
\end{equation}
with all the $N-1$ integrals over size $\epsilon$ neighborhoods of
the axes $x_{i\neq 1}=0$. Note that the denominator of the prefactor
is now $(N-1)!$, because we multiplied by $N$ to take into account
the contribution to the matrix-integral coming from all the other
axes $x_{i\neq 1}$.

In the matrix-integral of (\ref{eq:SONindexAsy1}), there remains
temperature-dependence through hyperbolic gamma functions whose
argument contains $x_{j\neq 1}/\beta$. To remove this dependence, we
re-scale the $N-1$ variables $x_{j\neq 1}$, by using $N-1$ out of
the $N$ factors of $(\frac{\beta}{2\pi})$ that the asymptotics of
the Pochhammer symbols in the prefactor provide. After the
re-scaling, the resulting integrals range over size $\epsilon/\beta$
neighborhoods of $x_{i\neq 1}=0$. Thus, although there is no
temperature-dependence in the integrand, now the ranges of the
integrals depend on $\beta$. To remove this latter dependence we
need to argue that at large $x_{i\neq 1}$ ($\propto1/\beta$) the
integrand of (\ref{eq:SONindexAsy1}) is exponentially small (of the
type $e^{-1/\beta}$), and therefore \emph{tails completion}
introduces negligible error. Here the asymptotics of the hyperbolic
gamma (\ref{eq:hyperbolicGammaAsy}) can be used to show that for
large $x_{i\neq 1}$ the asymptotics of the integrand of
(\ref{eq:SONindexAsy1}) is
\begin{equation}
\frac{\prod_{j=2}^{N}\Gamma_h^{N_f}(\omega r \pm \frac{2\pi
x_j}{\beta})}{\prod_{j=2}^{N} \Gamma_h(\pm \frac{2\pi
x_j}{\beta})\prod_{2\le i<j\le N}(\Gamma_h(\pm \frac{2\pi
(x_i+x_j)}{\beta})\Gamma_h(\pm \frac{2\pi
(x_i-x_j)}{\beta}))}\approx\exp(-2\pi(\frac{b+b^{-1}}{2})\tilde{u}(\mathbf{x})),
\end{equation}
with
\begin{equation}
\tilde{u}(\mathbf{x}):=(2N-2)\sum_{j=2}^{N}|x_j| -\sum_{2\le i<j\le
N}|x_i+x_j|-\sum_{2\le i<j\le N}|x_i-x_j|
\end{equation}

We now need the following corollary of (\ref{eq:SONvarthetaId}):
\begin{equation}
\begin{split}
2N\sum_{1\le j\le N}|x_j|&-\sum_{1\le i<j\le N}|x_i+x_j|-\sum_{1\le
i<j\le N}|x_i-x_j|=2\sum_{1\le i<j\le
N}\mathrm{min}(|x_i|,|x_j|)+2\sum_{1\le j\le N}|x_j|\\
&=2N\mathrm{min}(|x_i|)+2(N-1)\mathrm{min}_2(|x_i|)+\cdots
+4\mathrm{min}_{N-1}(|x_i|)+2\mathrm{max}(|x_i|).\label{eq:SONabsValId}
\end{split}
\end{equation}

The relation (\ref{eq:SONabsValId}) guarantees that
$\tilde{u}(\mathbf{x})$ is strictly positive for nonzero
$\mathbf{x}$, and that it is proportional to $\Lambda$ when
$x_i=\pm\Lambda$. The tails completion of the matrix-integral in
(\ref{eq:SONindexAsy1}) is thus (exponentially) safe.

The asymptotic analysis is straightforward from here: since $x_1$
did not need re-scaling, one out of the $N$ factors of
$(\frac{\beta}{2\pi})$ coming from the prefactor remains. This one
factor of $(\frac{\beta}{2\pi})$, along with the exponential pieces
of the asymptotics of the Pochhammer symbols in (\ref{eq:SONindex}),
and all the $e^{2\pi i R_0}$ factors coming from the elliptic gamma
functions in (\ref{eq:SONindex}), provide the leading asymptotics of
the SO($N$) SQCD partition function.

One of the assumptions we used to arrive at the above conclusion was
that the contribution of the matrix-integral from the region
$|x_1|<\varepsilon_1$ was negligible. To justify this assumption we
argue as follows. Since $\varepsilon_1$ can be taken to be
arbitrarily small, and since the error we have introduced by using
(\ref{eq:GammaAsyPosExt}) in the integrand of (\ref{eq:SONindex}) is
uniformly bounded as $\beta\to 0$, the contribution to the
matrix-integral from the region of size $\varepsilon_1$ is an $o(1)$
factor of the contribution we have computed so far. Thus our
assumption, that the ``small region'' can be neglected when
computing the asymptotics of the partition function, is valid with
relative error which is $o(1)$.

All in all, we find the following asymptotic relation:
\begin{equation}
\begin{split}
\ln Z^{\mathrm{SUSY}}_{SO(2N+1)}(b,\beta)=
-\mathcal{E}_0^{DK}(b,\beta)+\ln(\frac{2\pi}{\beta})+\ln
Y_{3d}(b)+o(1),\quad\quad (\text{as $\beta\to
0$})\label{eq:SONindexAsy2}
\end{split}
\end{equation}
where
\begin{equation}
Y_{3d}(b)=\frac{\Gamma_h^{N_f}(\omega r)}{2^N (N-1)!} \int
\mathrm{d}^{N-1} x \frac{\prod_{j=1}^{N-1}\Gamma_h^{N_f}(\omega r
\pm x_j)}{\prod_{j=1}^{N-1} \Gamma_h(\pm x_j)\prod_{1\le i<j\le
N-1}(\Gamma_h(\pm (x_i+x_j))\Gamma_h(\pm (x_i-x_j)))},
\end{equation}
with the $x_j$ integrals going over the whole real line.\\

Take now the special case of the SO($3$) theory with two flavors;
i.e. $N=1$, $N_f=2$. Set moreover $b=1$. The asymptotic expression
(\ref{eq:SONindexAsy2}) simplifies in this case to
\begin{equation}
\begin{split}
\ln Z^{\mathrm{SUSY}}_{SO(3)}(\beta)= \ln(\frac{2\pi}{\beta})+\ln
(\frac{\Gamma_h^{2}(i r;i,i)}{2})+o(1),\quad\quad (\text{as
$\beta\to 0$})\label{eq:SO3indexAsy1}
\end{split}
\end{equation}
with $r=1/2$. Note that there is no $1/\beta$ term on the RHS,
because the SO($3$) theory has $\mathrm{Tr}R=0$ (and also $L_{h\
min}=0$). Employing
\begin{equation}
\begin{split}
\ln \Gamma_h(ix;i,i)=(x-1)\ln (1-e^{-2\pi i x})-\frac{1}{2\pi
i}Li_2(e^{-2\pi i
x})+\frac{i\pi}{2}(x-1)^2-\frac{i\pi}{12},\label{eq:hypGammab=1}
\end{split}
\end{equation}
and noting $Li_2(-1)=-\pi^2/12$, we find that
$\Gamma_h(i/2;i,i)=1/\sqrt{2}$. Therefore (\ref{eq:SO3indexAsy1})
can be further simplified to
\begin{equation}
\begin{split}
\ln Z^{\mathrm{SUSY}}_{SO(3)}(\beta)= \ln(\frac{2\pi}{\beta})-2\ln
2+o(1)\quad\quad (\text{as $\beta\to 0$}).\label{eq:SO3indexAsy2}
\end{split}
\end{equation}

\subsubsection*{Much more precise asymptotics for the SO($3$) theory with $N_f=2$ when $b=1$}

Luckily, the asymptotic expansion in (\ref{eq:SO3indexAsy2}) can be
completed to all orders, with the result
\begin{equation}
\begin{split}
\ln Z^{\mathrm{SUSY}}_{SO(3)}(\beta)\sim
\ln(\frac{\pi}{2\beta}-\frac{1}{2\pi})\quad\quad (\text{as $\beta\to
0$}).\label{eq:SO3indexAsy3}
\end{split}
\end{equation}

To derive the above all-orders asymptotics, we first note the
following remarkable coincidence: the SUSY partition function of the
SO($3$) theory with two flavors precisely matches the $v=(p
q)^{1/6}$ specialization of the $\mathcal{N}=2$ partition function
of the SU($2$) $\mathcal{N}=4$ theory, to be described in the next
section. In particular, when $b=1$, the said $\mathcal{N}=2$
partition function becomes the Schur partition function of the
SU($2$) $\mathcal{N}=4$ theory, and the latter is exceptionally well
under control. The result in (\ref{eq:SO3indexAsy3}) is what one
gets for $\ln Z^{Schur}_{SU(2)\ \mathcal{N}=4}$, as demonstrated in
appendix~\ref{app:drukker}.

\subsubsection{SU($N$) $\mathcal{N}=4$ SYM}

The $\mathcal{N}=4$ theory is another important example with
high-temperature Coulomb branch decompactification on $S^3\times
S^1$.

The SU($N$) theory has the following SUSY partition function
\cite{Spiridonov:2010sv}:
\begin{equation}
\begin{split}
Z^{\mathrm{SUSY}}_{\mathcal{N}=4}(b,\beta)=e^{-\beta
E_{\mathrm{susy}}(b)}&\frac{(p;p)^{N-1}(q;q)^{N-1}}{
N!}\Gamma^{3(N-1)}((pq)^{1/3})\\
&\times \int \mathrm{d}^{N-1} x \prod_{1\le i<j\le
N}\frac{\Gamma^{3}((pq)^{1/3}(z_i/z_j)^{\pm1})}{\Gamma((z_i/z_j)^{\pm1})},\label{eq:N=4index}
\end{split}
\end{equation}
with $\prod_{i=1}^{N}z_i=1$.

Recall that for the theories of the previous subsection, the
integrand of the matrix-integral was everywhere exponentially
smaller than in the origin of the $x_i$ space; in other words, the
integral localized at a point. In the SO($N$) SQCD case, we saw that
the integral localizes around the (one-real-dimensional) axes of the
$x_i$ space. We will shortly find that for the $\mathcal{N}=4$
theory the matrix-integral does not localize at all.

The Rains function of the theory is
\begin{equation}
\begin{split}
L_h^{\mathcal{N}=4}=3(1-\frac{2}{3})\sum_{1\le i<j\le
N}\vartheta(x_i-x_j)-\sum_{1\le i<j\le N}\vartheta(x_i-x_j)=0.
\end{split}
\end{equation}
In other words, there is no effective potential for the interaction
of the holonomies, and the matrix-integral does not localize:
$\mathfrak{h}_{qu}=\mathfrak{h}_{cl}$.
Eq.~(\ref{eq:LagIndexSimp1Coul2}) thus dictates
\begin{equation}
\begin{split}
\ln Z_{\mathcal{N}=4}^{\mathrm{SUSY}}(b,\beta)=
(N-1)\ln(\frac{2\pi}{\beta})+O(\beta^0).\label{eq:N=4indexAsy1}
\end{split}
\end{equation}
There is no $O(1/\beta)$ term on the RHS, because $\mathrm{Tr}R=0$
for the $\mathcal{N}=4$ theory (and also $L_{h\ \mathrm{min}}=0$).

\subsubsection*{More precise asymptotics}

Neglecting the contribution to the integral coming from a small
(size $\varepsilon_1$) neighborhood of $\mathcal{S}_g$ (see the
similar discussion for the SO($N$) SQCD theory above), we can use
the estimate (\ref{eq:GammaAsyPosExt}) for all the gamma functions
in the integrand of (\ref{eq:N=4index}), and obtain that the
integrand is in fact approximately equal to one. Therefore the
integral is asymptotically equal to
$\mathrm{vol}(\mathfrak{h}_{cl})=1$, and the asymptotic analysis of
$Z_{\mathcal{N}=4}^{\mathrm{SUSY}}$ becomes trivial: the only
contributions that are $O(1)$ or larger come from the integral's
prefactor. These can be estimated using (\ref{eq:PochAsy}) and
(\ref{eq:GammaToHypGamma}). All in all, we find
\begin{equation}
\begin{split}
\ln Z_{\mathcal{N}=4}^{\mathrm{SUSY}}(b,\beta)=
(N-1)\ln(\frac{2\pi}{\beta})+3(N-1)\ln\Gamma_h(\frac{2}{3}\omega)-\ln
N!+o(1)\quad (\text{as $\beta\to 0$}).\label{eq:N=4indexAsy}
\end{split}
\end{equation}

\subsubsection{The $\mathbb{Z}_2$ orbifold theory}

We now study a quiver gauge theory, to illustrate how easily Rains's
method generalizes to theories with more than one simple factor in
their gauge group.

Consider the $\mathbb{Z}_2$ orbifold of the $\mathcal{N}=4$ SYM with
SU($N$) gauge group. The theory consists of two SU($N$) gauge
groups, with one chiral multiplet in the adjoint of each, and one
doublet of bifundamental chiral multiplets from each gauge group to
the other. All the chiral multiplets have R-charge $r=2/3$.

The SUSY partition function is given by (c.f. \cite{Gadde:2010holo})
\begin{equation}
\begin{split}
Z^{\mathrm{SUSY}}_{\mathbb{Z}_2}(b,\beta)=\ &e^{-\beta
E_{\mathrm{susy}}(b)} (\prod_{k=1,2}[
\frac{(p;p)^{N-1}(q;q)^{N-1}}{N!}\Gamma^{N-1}((pq)^{1/3})\int
\mathrm{d}^{N-1}x^{(k)}\\ &\left(\prod_{1\le i<j\le
N}\frac{\Gamma((pq)^{1/3}(z^{(k)}_i/z^{(k)}_j)^{\pm
1})}{\Gamma((z^{(k)}_i/z^{(k)}_j)^{\pm 1})}\right)]) \times
\prod_{i,j=1}^{N}\Gamma^2((pq)^{1/3} (z^{(1)}_i/z^{(2)}_j)^{\pm
1}),\label{eq:Z2orbZ}
\end{split}
\end{equation}
with $\prod_{i=1}^{N} z^{(1)}_i=\prod_{i=1}^{N} z^{(2)}_i=1$.

\begin{figure}[t]
\centering
    \includegraphics[scale=.7]{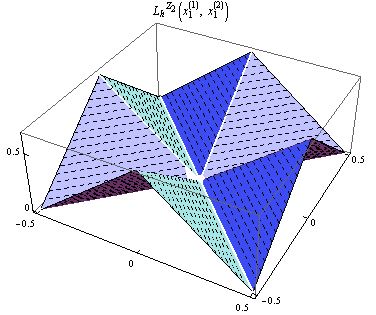}
\caption{The Rains function of the SU($2$)$\times$SU($2$) orbifold
theory. Note the flat directions along $|x^{(1)}_1|=|x^{(2)}_1|$.
\label{fig:z2Lh}}
\end{figure}

The Rains function of the theory is
\begin{equation}
\begin{split}
L_h^{\mathbb{Z}_2}(\mathbf{x}^{(1)},\mathbf{x}^{(2)})&=-\frac{2}{3}\sum_{1\le
i<j\le N}\vartheta(x^{(1)}_i-x^{(1)}_j)-\frac{2}{3}\sum_{1\le i<j\le
N}\vartheta(x^{(2)}_i-x^{(2)}_j)+\frac{2}{3}\sum_{i,j=1}^{N}\vartheta(x^{(1)}_i-x^{(2)}_j).
\end{split}
\end{equation}
For the case $N=2$, corresponding to the SU($2$)$\times$SU($2$)
theory, this function is illustrated in Figure~\ref{fig:z2Lh}.

The generalized triangle inequality (\ref{eq:RainsGTI}) applies with
$c=x^{(1)},d=x^{(2)}$, and implies that $L_h^{\mathbb{Z}_2}$ is
positive semi-definite. It moreover shows that $L_h^{\mathbb{Z}_2}$
vanishes if the $x^{(1)}_i,x^{(2)}_j$ can be permuted such that
either of (\ref{cond:gti1}) or (\ref{cond:gti2}) holds. For
simplicity we consider all $x^{(1)}_i$ to be positive and very
small, except for $x^{(1)}_N=-x^{(1)}_1-\dots-x^{(1)}_{N-1}$ being
negative and very small, and similarly for $x^{(2)}_j$. Assuming
either (\ref{cond:gti1})  or (\ref{cond:gti2}), we conclude that
$x^{(1)}_i=x^{(2)}_i$. Based on this result, and also the $N=2$ case
whose Rains function is displayed in Figure~\ref{fig:z2Lh}, we
conjecture that for the $\mathbb{Z}_2$ orbifold theory
$\mathrm{dim}\mathfrak{h}_{qu}=N-1$, and thereby
\begin{equation}
\begin{split}
\ln Z^{\mathrm{SUSY}}_{\mathbb{Z}_2}(b,\beta)=
-\mathcal{E}^{DK}_0(b,\beta)+(N-1)
\ln\left(\frac{2\pi}{\beta}\right)+O(1)\quad\quad (\text{as
$\beta\to 0$}).\label{eq:Z2indexAsy}
\end{split}
\end{equation}

\subsubsection{The $\mathbb{Z}_3$ orbifold theory}

The SU($N$)$^3$ quiver is our second (and last) example with
$Q_h\neq 0$. More precisely, it is for $N>2$ that the model is
chiral, and has nonzero $Q_h$, since the fundamental and
anti-fundamental representations of SU($2$) are equivalent. The
quiver has three chiral multiplets with R-charge $2/3$ going from
the first node to the second, a similar triplet going from the
second node to the third, and a last triplet going from to the third
node to the first.

Similarly to the case of the $\mathbb{Z}_2$ orbifold theory $i)$
Rains's generalized triangle inequality (\ref{eq:RainsGTI})
establishes that $L_{h}^{\mathbb{Z}_3}$ is positive semi-definite,
and $ii)$ based on an argument made in the region where
$x^{(1,2,3)}_i$ are small (and positive except for $i=N$) we
conjecture that also for this theory
$\mathrm{dim}\mathfrak{h}_{qu}=N-1$.

Although $Q_h^{\mathbb{Z}_3}$ does not identically vanish for $N>2$,
our numerical investigation for $N=3$ indicates that it vanishes on
$\mathfrak{h}_{qu}$, and we suspect $Q_h^{\mathbb{Z}_3}$ to keep
vanishing on $\mathfrak{h}_{qu}$ for all $N\ge3$. This suggests that
the nonzero $Q_h$ function of the $\mathbb{Z}_3$ orbifold theory
does not affect the leading high-temperature asymptotics of its SUSY
partition function. We are thus led to conjecture
\begin{equation}
\begin{split}
\ln Z^{\mathrm{SUSY}}_{\mathbb{Z}_3}(b,\beta)=
-\mathcal{E}^{DK}_0(b,\beta)+(N-1)
\ln\left(\frac{2\pi}{\beta}\right)+O(1)\quad\quad (\text{as
$\beta\to 0$}).\label{eq:Z3indexAsy}
\end{split}
\end{equation}\\

\subsection{$Z_{S^3}$ exponentially divergent (or: The curious case of the SCFTs with $c<a$)}\label{sec:c<a}

In this subsection we consider examples of Lagrangian SCFTs arising
as IR fix points of R-symmetric SUSY gauge theories with a
semi-simple gauge group, with $Q_h=0$, and with $\mathrm{Tr}R>0$.

We write the asymptotics in terms of the central charges $a$ and $c$
of our theories. Since $\mathrm{Tr}R=-16(c-a)$, the
Di~Pietro-Komargodski formula for SCFTs reads
\begin{equation}
\ln Z^{\mathrm{SUSY}}(b,\beta)\approx
\frac{16\pi^2}{3\beta}(\frac{b+b^{-1}}{2})(c-a)\quad\quad (\text{as
$\beta\to 0$}).\label{eq:dkSCFTs}
\end{equation}

On the other hand, the formula (\ref{eq:LagIndexSimp6noTheta})
becomes
\begin{equation}
\ln Z^{\mathrm{SUSY}}(b,\beta)=
\frac{16\pi^2}{3\beta}(\frac{b+b^{-1}}{2})(c-a-\frac{3}{4}L_{h\
min})+\mathrm{dim}\mathfrak{h}_{qu}\ln(\frac{2\pi}{\beta})+O(\beta^0),\label{eq:dkSCFTsCorrected}
\end{equation}
with $L_{h\ \mathrm{min}}$ the minimum of the Rains function over
$\mathfrak{h}_{cl}$. Note that the leading piece takes the same form
as the Di~Pietro-Komargodski formula, but with the ``shifted $c-a$''
defined as $(c-a)_{\mathrm{shifted}}:= c-a-\frac{3}{4}L_{h\
\mathrm{min}}$; this last relation appears to be analogous to the
equation $c_{\mathrm{eff}}= c-24h_{\mathrm{min}}$ frequently
discussed in the context of non-unitary 2d CFTs (see e.g.
\cite{Kutasov:1991ks}).

For SCFTs with $c<a$, the RHS of the Di~Pietro-Komargodski formula
(\ref{eq:dkSCFTs}) becomes negative. Interestingly, in the SCFTs
with $c<a$ studied below, the correction term $-\frac{3}{4}L_{h\
min}$ makes the RHS of (\ref{eq:dkSCFTsCorrected}) positive. In
other words $(c-a)_{\mathrm{shifted}}>0$.\\

In the theories studied in this subsection, $\mathbf{x}=0$ is not a
local minimum of $L_h$; it is in fact a local maximum. We now argue
that when $\mathbf{x}=0$ is not a local minimum of the Rains
function, $Z_{S^3}(b;\Lambda)$ defined in (\ref{eq:LagZ3d}) diverges
exponentially in $\Lambda$ as $\Lambda\to\infty$.

Our starting point for the argument is the relation (we are assuming
$Q_h=0$)
\begin{equation}
Z_{S^3}(b;\Lambda)\approx\frac{1}{|W|}\int_{\Lambda}
\mathrm{d}^{r_G}x\ e^{-2\pi
(\frac{b+b^{-1}}{2})\tilde{L}_{S^3}(\mathbf{x})},
\label{eq:LagZ3dExpDiv}
\end{equation}
which we obtained in subsection \ref{sec:Coul}. The assumption that
$\mathbf{x}=0$ is not a local minimum implies that there are
neighboring points of $\mathbf{x}=0$ where the effective potential,
and hence the Rains function, is negative. Since for small enough
$\mathbf{x}$ the Rains function and $\tilde{L}_{S^3}$ coincide, we
learn that there are $\mathbf{x}\neq 0$ points where
$\tilde{L}_{S^3}$ is negative. Since $\tilde{L}_{S^3}$ is a
homogenous function of $\mathbf{x}$, we conclude that there are
directions along which we can take $|\mathbf{x}|\propto\Lambda$ and
have $\tilde{L}_{S^3}(\mathbf{x})\propto -\Lambda$. The integrand of
(\ref{eq:LagZ3dExpDiv}) would become exponentially large if
$|\mathbf{x}|$ becomes large along those directions, and
$Z_{S^3}(b;\Lambda)$ would diverge exponentially in $\Lambda$.

\subsubsection{The SU($2$) ISS model}

There are two famous interacting Lagrangian $\mathcal{N}=1$ SCFTs
with $c<a$. The first is the Intriligator-Seiberg-Shenker (ISS)
model of dynamical SUSY breaking \cite{Intriligator:1994}. The
theory is formulated in the UV as an SU($2$) vector multiplet with a
single chiral multiplet in the four-dimensional representation of
the gauge group. Although originally suspected to confine (and to
break supersymmetry upon addition of a tree-level superpotential)
\cite{Intriligator:1994}, the theory is currently believed to flow
to an interacting SCFT in the IR
\cite{Intriligator:2005,Poppitz:2009}, where the chiral multiplet
has R-charge $3/5$. The IR SCFT would then have $c-a=-7/80$.

The SUSY partition function of this theory is (c.f.
\cite{Vartanov:2010})
\begin{equation}
\begin{split}
Z^{\mathrm{SUSY}}_{ISS}(b,\beta)=e^{-\beta
E_{\mathrm{susy}}(b)}\frac{(p;p)(q;q)}{2}\int\mathrm{d}x
\frac{\Gamma((pq)^{3/10}z^{\pm 1})\Gamma((pq)^{3/10}z^{\pm
3})}{\Gamma(z^{\pm 2})}.\label{eq:ISSindex}
\end{split}
\end{equation}

\begin{figure}[t]
\centering
    \includegraphics[scale=1]{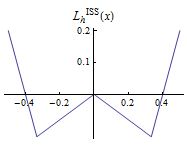}
\caption{The Rains function of the SU($2$) ISS theory. Note that the
minima lie at $x=\pm 1/3$. \label{fig:ISS}}
\end{figure}

The Rains function of the theory is
\begin{equation}
\begin{split}
L^{ISS}_h(x)=\frac{2}{5}\vartheta(x)
+\frac{2}{5}\vartheta(3x)-\vartheta(2x).
\end{split}
\end{equation}
This function is plotted in Figure~\ref{fig:ISS}.

A direct examination reveals that $L^{ISS}_h(x)$ is minimized at
$x=\pm 1/3$, and $L^{ISS}_h(\pm 1/3)=-2/15$. The asymptotics of
$Z^{\mathrm{SUSY}}_{ISS}$ is hence given according to
(\ref{eq:dkSCFTsCorrected}) by
\begin{equation}
\begin{split}
\ln Z^{\mathrm{SUSY}}_{ISS}(b,\beta)=
\frac{\pi^2}{15\beta}(\frac{b+b^{-1}}{2})+O(\beta^0).\label{eq:AsyISS1}
\end{split}
\end{equation}
In other words we have $(c-a)_{\mathrm{shifted}}=c-a+1/10=1/80$.

\subsubsection*{Much more precise asymptotics}

We now proceed to improve the asymptotic relation (\ref{eq:AsyISS1})
to all-orders accuracy. The reader not interested in the technical
details is invited to skip to the relation (\ref{eq:ISSindexAsy3})
and continue reading from there.

We know that the integral (\ref{eq:ISSindex}) localizes around
$x=\pm 1/3$ at high temperatures. Therefore we prune down the
integration range to two small neighborhoods of size $\epsilon$
around $x=1/3$ and $x=-1/3$. The $x\to-x$ symmetry then implies that
we can compute only the integral around $x=1/3$, and multiply the
result by two.

In an $O(\epsilon)$ neighborhood around $x=1/3$, the arguments of
the gamma functions $\Gamma((pq)^{3/10}z^{\pm 1})$ and
$\Gamma(z^{\pm 2})$ inside the integrand of (\ref{eq:ISSindex}) are
such that the estimate (\ref{eq:GammaAsyPosExt}) applies to them.
But the gamma functions $\Gamma((pq)^{3/10}z^{\pm 3})$ need special
care now: for $x\ge1/3$ their argument is such that we can not use
the central estimate (\ref{eq:GammaToHypGamma}) for them. To get
around this, as mentioned below (\ref{eq:RhDef}) we can replace
every $x$ on the RHS of (\ref{eq:GammaToHypGamma}) with $\{x\}$, to
obtain (after scaling $x\mapsto 3x$)
\begin{equation}
\begin{split}
\Gamma((pq)^{r/2} z^{\pm 3})\simeq e^{2\pi i
R_{0}(\frac{\beta}{2\pi}\omega r \pm
3x;\sigma,\tau)}\Gamma_h(r\omega \pm\frac{2\pi
(3x-1)}{\beta})e^{2\pi i (1-r)\frac{2\pi\omega}{\beta}(6x-1)}&.\\
(\text{For } 0<3x<2,\ \text{as $\beta\to
0$}.)&\label{eq:GammaToHypGammaShift1}
\end{split}
\end{equation}
The extended range of applicability of the above estimate allows us
to approximate $\Gamma((pq)^{3/10}z^{\pm 3})$ for $x\ge1/3$, and
also uniformly on the $O(\epsilon)$ neighborhood of $x= 1/3$, which
is where (half of) the dominant contribution to the integral
(\ref{eq:ISSindex}) comes from (the other half comes from an
$O(\epsilon)$ neighborhood of $x=-1/3$).

Using the estimates (\ref{eq:PochAsy}),
(\ref{eq:GammaToHypGammaShift1}), and (\ref{eq:GammaAsyPosExt}),
defining a new variable $x':=x-1/3$, and then re-scaling $x'\mapsto
x'/(\beta/2\pi)$, we find that $Z^{\mathrm{SUSY}}_{ISS}(b,\beta)$ in
(\ref{eq:ISSindex}) simplifies to
\begin{equation}
\begin{split}
Z^{\mathrm{SUSY}}_{ISS}(b,\beta)\simeq \
e^{\frac{16\pi^2}{3\beta}(c-a+\frac{1}{10})(\frac{b+b^{-1}}{2})}
\times Y^{ISS}_{S^3}(b;2\pi\epsilon/\beta),\quad\quad (\text{as
$\beta\to 0$})\label{eq:ISSindexAsy4}
\end{split}
\end{equation}
with
\begin{equation}
Y^{ISS}_{S^3}(b;\Lambda)=\int_{-\Lambda}^{\Lambda}\mathrm{d}x'
e^{-\frac{4\pi}{5}(b+b^{-1})x'}\times\Gamma_h(3x'+(3/5)\omega)\Gamma_h(-3x'+(3/5)\omega).\label{eq:Yiss}
\end{equation}
The asymptotics of the hyperbolic gamma in
(\ref{eq:hyperbolicGammaAsy}) guarantees that the integrand in the
above equation is exponentially small at large $|x|$, and hence we
can safely complete the tails to the whole real line.

Our final asymptotic estimate is obtained by taking the logarithm of
(\ref{eq:ISSindexAsy4}):
\begin{equation}
\begin{split}
\ln Z^{\mathrm{SUSY}}_{ISS}(b,\beta)\sim
\frac{16\pi^2}{3\beta}(c-a)_{\mathrm{shifted}}(\frac{b+b^{-1}}{2})+
\ln Y^{ISS}_{S^3}(b),\quad\quad (\text{as $\beta\to
0$})\label{eq:ISSindexAsy3}
\end{split}
\end{equation}
with $Y^{ISS}_{S^3}(b)=Y^{ISS}_{S^3}(b;\infty)$, and
$(c-a)_{\mathrm{shifted}}=(c-a)+1/10=1/80$.

Now, if $Y^{ISS}_{S^3}(b)$ were found to vanish, then the
$O(\beta^0)$ term on the RHS of (\ref{eq:ISSindexAsy3}) would
diverge, the relation (\ref{eq:ISSindexAsy3}) would not make sense,
and we would need to redo the asymptotic analysis of
$Z^{\mathrm{SUSY}}_{ISS}(b,\beta)$ more carefully; the careful
analysis would then presumably lead us to an asymptotics different
from the one dictated by (\ref{eq:LagIndexSimp6noTheta}); that would
be a scenario exemplifying the subtle cancelations discussed below
(\ref{eq:SgDef}) [in the present case, the cancelation would be seen
at the level of $Y^{ISS}_{S^3}(b)$], and their consequential failure
of (\ref{eq:LagIndexSimp6noTheta}). However, it follows from
(\ref{eq:hyperbolicGamma}) that
\begin{equation}
\Gamma_h(-\mathrm{Re}(x)+i\mathrm{Im}(x);\omega_1,\omega_2)=(\Gamma_h(\mathrm{Re}(x)+i\mathrm{Im}(x);\omega_1,\omega_2))^\ast,\label{eq:hyperbolicGammaConj}
\end{equation}
with $\ast$ denoting complex conjugation; as a result the product of
the hyperbolic gamma functions in the integrand of (\ref{eq:Yiss})
is (real and) positive, and thus $Y^{ISS}_{S^3}(b)> 0$. Therefore
the unexpected cancelations discussed below (\ref{eq:SgDef}) do not
occur here.\\

An analysis similar to the one above can be performed for any SUSY
gauge theory with a semi-simple gauge group and non-chiral matter
content\footnote{Non-chirality guarantees that the subtle
cancelations discussed below (\ref{eq:SgDef}) do not occur (c.f.
\cite{Ardehali:thesis}); note, for example, how below
(\ref{eq:ISSindexAsy3}) we argued for the positivity of the
integrand of $Y^{ISS}_{S^3}(b)$, and thus for $Y^{ISS}_{S^3}(b)\neq
0$.} (hence $Q_h=0$), whose Rains function is minimized on a set of
points not consisting only of the origin (i.e.
$\mathrm{dim}\mathfrak{h}_{qu}=0$ and
$\mathfrak{h}_{qu}\backslash\{\mathbf{x}=0\}\neq\varnothing$). All
such theories would display asymptotics similar to
(\ref{eq:ISSindexAsy3}). In particular,
$\mathrm{dim}\mathfrak{h}_{qu}=0$ implies that the high-temperature
expansion of $\ln Z^{\mathrm{SUSY}}(b,\beta)$ (and in fact also that
of $\ln Z^{\mathrm{SUSY}}(b,\beta;m_a)$) terminates at $O(\beta^0)$.
We showed the latter statement in subsection~\ref{sec:Z3finite} for
theories whose Rains function is minimized only at the origin,
irrespective of whether their $Q_h$ was zero or not. It would be
interesting to prove (or disprove) the same general statement for
theories with $\mathrm{dim}\mathfrak{h}_{qu}=0$,
$\mathfrak{h}_{qu}\backslash\{\mathbf{x}=0\}\neq\varnothing$, and
nonzero $Q_h$.

\subsubsection{The SO($2N+1$) BCI model with $1<N<5$}

The second famous example of interacting Lagrangian $\mathcal{N}=1$
SCFTs with $c<a$ is provided by the ``misleading'' SO($n$) theory of
Brodie, Cho, and Intriligator \cite{Brodie:1998}. This is an
$\mathcal{N}=1$ SO($n$) gauge theory with a single chiral multiplet
in the two-index symmetric traceless tensor representation of the
gauge group. The theory is asymptotically free if $n\ge 5$. For
$5\le n<11$ the corresponding interacting IR SCFT is believed to
have $c-a=-(n-1)/16$ (for greater values of $n$ the R-symmetry of
the IR fixed point is believed to mix with an emergent accidental
symmetry, and thus more care is called for; c.f.
\cite{Intriligator:2010}).

For the SO($2N+1$) theory (with $1<N<5$) we have  (c.f.
\cite{Vartanov:2010})
\begin{equation}
\begin{split}
Z^{\mathrm{SUSY}}_{BCI}(b,\beta)=e^{-\beta
E_{\mathrm{susy}}(b)}&\frac{(p;p)^N(q;q)^N}{2^N
N!}\Gamma^N((pq)^{2/(2N+3)})\int\mathrm{d}^N x\\
&\prod_{i<j}\frac{\Gamma((pq)^{2/(2N+3)}z_i^{\pm1}
z_j^{\pm1})}{\Gamma(z_i^{\pm1}
z_j^{\pm1})}\prod_{j=1}^N\frac{\Gamma((pq)^{2/(2N+3)}z_j^{\pm1}
,(pq)^{2/(2N+3)}z_j^{\pm2})}{\Gamma(
z_j^{\pm1})}.\label{eq:BCIindex}
\end{split}
\end{equation}

\begin{figure}[t]
\centering
    \includegraphics[scale=.7]{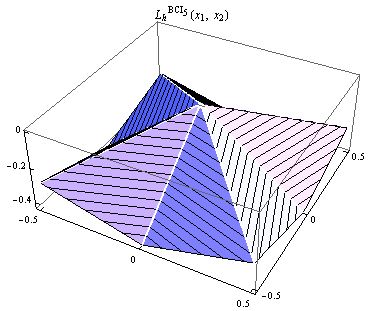}
\caption{The Rains function of the SO($5$) BCI theory. Note that the
function is maximized at the origin, and minimized at
$(x_1,x_2)=(0,\pm 1/2)$ and $(x_1,x_2)=(\pm 1/2,0)$.
\label{fig:BCI}}
\end{figure}

The Rains function of the theory is
\begin{equation}
\begin{split}
L_h^{BCI}(x)=\frac{4}{2N+3}\left((\frac{2N-1}{4})\sum_{j}\vartheta(2x_j)
-\sum_j\vartheta(x_j)-\sum_{i<j}\vartheta(x_i+x_j)-\sum_{i<j}\vartheta(x_i-x_j)\right).\label{eq:BCIRains}
\end{split}
\end{equation}
For $N=2$, corresponding to the SO($5$) theory, this function is
plotted in Figure~\ref{fig:BCI}.

To find the minima of the above function, we need the following
result, valid for $-1/2\le x_i\le 1/2$:
\begin{equation}
\begin{split}
&(\frac{2N-1}{4})\sum_{1\le j\le N}\vartheta(2x_j)-\sum_{1\le j\le
N}\vartheta(x_j)-\sum_{1\le i<j\le N}\vartheta(x_i+x_j)-\sum_{1\le
i<j\le N}\vartheta(x_i-x_j)=\\
&-\frac{3}{2}\sum_{ i<j}\mathrm{max}(|x_i|,|x_j|) +\frac{1}{2}\sum_{
i<j}\mathrm{min}(|x_i|,|x_j|)=\sum_{j}(-N+2j-\frac{3}{2})\mathrm{min}_{N-j+1}(|x_i|),\label{eq:BCIvarthetaId}
\end{split}
\end{equation}
with $\mathrm{min}_N(|x_i|):=\mathrm{max}(|x_i|)$. The proof of
(\ref{eq:BCIvarthetaId}) is similar to that of
(\ref{eq:SONvarthetaId}).

Note that the coefficient of the $j$th term on the RHS of
(\ref{eq:BCIvarthetaId}) is negative if $j\le\frac{N+1}{2}$, and
positive otherwise. This implies that the Rains function
(\ref{eq:BCIRains}) is minimized when $\lfloor
\frac{N+1}{2}\rfloor$ of the $|x_i|$ are maximized (i.e. $x_{i}=\pm
1/2$), and the rest of the $|x_i|$ are minimized (i.e. $x_{i}=0$).
Consequently, the minimum of the Rains function is
\begin{equation}
\begin{split}
L_{h\ \mathrm{min}}^{BCI}=-\frac{1}{2N+3}\sum_{1\le j\le \lfloor
\frac{N+1}{2}\rfloor}(2N+3-4j).\label{eq:BCIRainsMin}
\end{split}
\end{equation}
This is less than zero for any $N>1$. Therefore the
Di~Pietro-Komargodski formula needs to be modified in the SO($2N+1$)
BCI model with $1<N<5$.

For example, consider the SO($5$) theory corresponding to $N=2$.
This theory has $c-a=-1/4$. From Eq.~(\ref{eq:BCIRainsMin}) we have
in this case $L_{h\ min}^{BCI}(x)=-3/7$. The asymptotics of
$Z^{\mathrm{SUSY}}$ is therefore given according to
(\ref{eq:dkSCFTsCorrected}) by
\begin{equation}
\begin{split}
\ln Z^{\mathrm{SUSY}}_{BCI_5}(b,\beta)=
\frac{8\pi^2}{21\beta}(\frac{b+b^{-1}}{2})+O(\beta^0).\label{eq:AsyBCI1}
\end{split}
\end{equation}
In other words $(c-a)_{\mathrm{shifted}}=c-a+9/28=1/14$.

\subsubsection*{Much more precise asymptotics for the SO($5$) BCI theory}

We now proceed to improve (\ref{eq:AsyBCI1}) to all-orders accuracy.
The reader not interested in technical details is invited to skip to
the relation (\ref{eq:BCIindexAsy4}) and continue reading from
there.

The matrix-integral of the SO($5$) BCI theory localizes around two
points. We compute the contribution coming from around
$(x_1,x_2)=(\pm 1/2,0)$, and multiply the result by two to take into
account also the contribution coming from around $(x_1,x_2)=(0,\pm
1/2)$.

Analogously to (\ref{eq:GammaToHypGammaShift1}), this time we need
\begin{equation}
\begin{split}
\Gamma((pq)^{r/2} z_1^{\pm 2})\simeq e^{2\pi i
R_{0}(\frac{\beta}{2\pi}\omega r \pm
2x_1;\sigma,\tau)}\Gamma_h(r\omega \pm\frac{2\pi
(2x_1-1)}{\beta})e^{2\pi i (1-r)\frac{2\pi\omega}{\beta}(4x_1-1)}&.\\
(\text{For } 0<2x_1<2,\ \text{as $\beta\to
0$}.)&\label{eq:GammaToHypGammaShift1bci}
\end{split}
\end{equation}

Proceeding as in the case of the ISS model, this time defining
$x_1':=x_1-1/2$, pruning down to $|x'_1|,|x_2|<\epsilon$ and then
re-scaling $x_1',x_2\mapsto x_1'/(\beta/2\pi),x_2/(\beta/2\pi)$, we
arrive at
\begin{equation}
\begin{split}
Z^{\mathrm{SUSY}}_{BCI_5}(b,\beta)\simeq \
e^{\frac{16\pi^2}{3\beta}(c-a+\frac{9}{28})(\frac{b+b^{-1}}{2})}
\times Y^{BCI_5}_{S^3}(b;2\pi\epsilon/\beta),\quad\quad (\text{as
$\beta\to 0$})\label{eq:BCIindexAsy3}
\end{split}
\end{equation}
with
\begin{equation}
\begin{split}
Y^{BCI_5}_{S^3}(b;\Lambda)=&\frac{1}{2}\int_{-\Lambda}^{\Lambda}\mathrm{d}
x_1' \ \Gamma_h((4/7)\omega\pm
2x_1')\times\\
&\frac{\Gamma_h^2((4/7)\omega)}{2}\int_{-\Lambda}^{\Lambda}
\mathrm{d} x_2\ \frac{\Gamma_h((4/7)\omega\pm
x_2)\Gamma_h((4/7)\omega\pm 2x_2)} {\Gamma_h(\pm
x_2)}.\label{eq:Ybci}
\end{split}
\end{equation}
The asymptotics of the hyperbolic gamma
(\ref{eq:hyperbolicGammaAsy}) guarantees that the tails completion
is safe, and we obtain
\begin{equation}
\begin{split}
\ln Z^{\mathrm{SUSY}}_{BCI_5}(b,\beta)\sim
\frac{16\pi^2}{3\beta}(c-a)_{\mathrm{shifted}}(\frac{b+b^{-1}}{2})+
\ln Y^{BCI_5}_{S^3}(b),\quad\quad (\text{as $\beta\to
0$})\label{eq:BCIindexAsy4}
\end{split}
\end{equation}
with $Y^{BCI_5}_{S^3}(b)=Y^{BCI_5}_{S^3}(b;\infty)$, and
$(c-a)_{\mathrm{shifted}}=(c-a)+9/28=1/14$. From
(\ref{eq:hyperbolicGammaConj}) it follows that the integrands in
(\ref{eq:Ybci}) are (real and) positive; therefore
$Y^{BCI_5}_{S^3}(b)> 0$, assuring that the unexpected cancelations
discussed below (\ref{eq:SgDef}) do not occur here either.\\

\section{Asymptotics of the $\mathcal{N}=2$ partition function}
\label{sec:N=2andSchur}

In this section we focus on Lagrangian $\mathcal{N}=2$ SCFTs; these
have the extended R-symmetry group
SU($2$)$_{R_{\mathcal{N}=2}}\times$ U($1$)$_{r_{\mathcal{N}=2}}$.
The $\mathcal{N}=2$ theories are put on $S_b^3\times S_\beta^1$, and
their path-integral is computed in presence of a background gauge
field that couples to a specific linear combination of
U($1$)$_{r_{\mathcal{N}=2}}$ and the Cartan of
SU($2$)$_{R_{\mathcal{N}=2}}$. Denoting the latter by
U($1$)$_{R_{\mathcal{N}=2}}$, the said linear combination is
\begin{equation}
Q_v=-(r_{\mathcal{N}=2}+R_{\mathcal{N}=2}).\label{eq:backLinV1}
\end{equation}
The $\mathcal{N}=1$ R-symmetry is also a linear combination of
U($1$)$_{R_{\mathcal{N}=2}}$ and U($1$)$_{r_{\mathcal{N}=2}}$; it is
given by
\begin{equation}
r=\frac{2}{3}(2R_{\mathcal{N}=2}-r_{\mathcal{N}=2}).\label{eq:R1R2}
\end{equation}
The linear combination in (\ref{eq:backLinV1}) can hence be written
as
\begin{equation}
Q_v=\frac{3}{2}r-3R_{\mathcal{N}=2}.\label{eq:backLinV2}
\end{equation}

The path-integral computed in the presence of a background gauge
field (along the $S^1_\beta$, with value $m_v$, and with holonomy
$v=e^{i\beta m_v}$) coupling the linear combination
(\ref{eq:backLinV2}) defines the \emph{$\mathcal{N}=2$ partition
function} $Z^{\mathcal{N}=2}(b,\beta,m_v)$. When the gauge group $G$
is semi-simple, and when besides the $\mathcal{N}=2$ vector
multiplet the theory has chiral multiplets in the doublet of
SU($2$)$_{R_{\mathcal{N}=2}}$ forming hyper multiplets, a
localization computation yields \cite{Bobev:2015}
\begin{equation}
\begin{split}
Z^{\mathrm{SUSY}}(b,\beta,m_v)=e^{-\beta E_{\mathrm{susy}}(b,m_v)}
\mathcal{I}(b,\beta,m_v),\label{eq:LagEquivN=2Z}
\end{split}
\end{equation}
with
\begin{equation}
\begin{split}
\mathcal{I}(b,\beta,m_v)=\frac{(p;p)^{r_G}(q;q)^{r_G}}{|W|}\Gamma^{r_G}((pq)^{1/3}v)\int
\mathrm{d}^{r_G}x &\prod_{\alpha_+}\left(\frac{\Gamma((pq)^{1/3}v
z^{\pm\alpha_+})}{\Gamma( z^{\pm\alpha_+})}\right)\\
&\prod_\chi \prod_{\rho^{\chi}
\in\Delta_\chi}\Gamma((pq)^{r_\chi/2}v^{3(r_\chi-1)/2}
z^{\rho^{\chi}} ),\label{eq:LagEquivN=2Index}
\end{split}
\end{equation}
the \emph{$\mathcal{N}=2$ index} of the SUSY gauge theory, and
$E_{\mathrm{susy}}(b,m_v)$ the corresponding Casimir polynomial,
which can be obtained from (\ref{eq:EcBequiv}) by substituting on
its RHS
\begin{equation}
\omega\to\omega+\frac{3}{2}m_v.\label{eq:omegaShift}
\end{equation}
A quick way to see why this shift is expected is to note that the
argument of the chiral-multiplet gamma functions in
(\ref{eq:LagEquivN=2Index}) contain
$(pq)^{r_\chi/2}v^{3(r_\chi-1)/2}=e^{i\beta\omega r_\chi}\cdot
e^{i\beta m_v [3(r_\chi-1)/2]}$, whereas if $m_v$ were zero we would
only have $e^{i\beta\omega r_\chi}$. Consequently, to obtain the
dependence of various quantities on $m_v$, we can start with their
expression for when $m_v=0$, and replace in them every $\omega
r_\chi$ with $\omega r_\chi+m_v [3(r_\chi-1)/2]$. In particular,
this amounts to replacing every $(r_\chi-1)\omega$ with
$(r_\chi-1)[\omega+\frac{3}{2}m_v]$, which can be alternatively
realized as a shift in $\omega$, as the prescription
(\ref{eq:omegaShift}) indicates. A similar argument applies to the
$\mathcal{N}=2$ vector multiplets.\\

The Hamiltonian route to the $\mathcal{N}=2$ index is via
\begin{equation}
\mathcal
I(b,\beta,m_v)=\mathrm{Tr}\left[(-1)^Fe^{-\hat\beta(\Delta-2j_2-\fft32r)}
p^{j_1+j_2+\fft12r}q^{-j_1+j_2+\fft12r}v^{\frac{3}{2}r-3R_{\mathcal{N}=2}}\right].
\label{eq:IndexN=2Def}
\end{equation}

The case $m_v=0$ corresponds to the $\mathcal{N}=1$ index, which we
already know how to deal with. The new challenge is to find the
dependence of the asymptotics on $m_v$.\\

First of all, the low-temperature asymptotics is found as in the
previous section, and (assuming all $r_\chi$ are in $]0,2[$) reads
\begin{equation}
\mathcal I(b,\beta,m_v)\simeq 1 \Rightarrow
Z^{\mathrm{SUSY}}(b,\beta,m_v)\simeq e^{-\beta
E_{\mathrm{susy}}(b,m_v)}\quad (\text{as $1/\beta\to 0$, with
$b,m_v$ fixed}).\label{eq:equivZlowTN=2}
\end{equation}

To find the high-temperature asymptotics, we first use the estimates
(\ref{eq:prePochAsyLead}), (\ref{eq:GammaOffCenter2}), and
(\ref{eq:GammaOffCenter2Vec}) in the integrand of the
$\mathcal{N}=2$ index. Proceeding as in the previous section, we
find
\begin{equation}
\begin{split}
Z^{\mathcal{N}=2}(b,\beta,m_v)\approx\mathcal I(b,\beta,m_v)\approx
\left(\frac{2\pi}{\beta}\right)^{r_G} \int_{\mathfrak{h}_{cl}}
\mathrm{d}^{r_G}x\
e^{-[\mathcal{E}^{DK}_0(b,\beta,m_v)+V^{\mathrm{eff}}(\mathbf{x};b,\beta,m_v)]},\label{eq:LagN=2IndexSimp1}
\end{split}
\end{equation}
with $\mathcal{E}^{DK}_0(b,\beta,m_v)$ a function easily obtainable
from (\ref{eq:EdkEquiv}) by applying on its RHS the substitution
(\ref{eq:omegaShift}). The effective potential
$V^{\mathrm{eff}}(\mathbf{x};b,\beta,m_v)$ can be obtained
similarly, and reads
\begin{equation}
\begin{split}
V^{\mathrm{eff}}(\mathbf{x};b,\beta,m_v)=\frac{4\pi^2}{\beta}(\frac{b+b^{-1}}{2})L_h(\mathbf{x},m_v),\label{eq:VeffN=2Equiv1}
\end{split}
\end{equation}
where we have defined the \emph{$\mathcal{N}=2$ Rains function} as
\begin{equation}
\begin{split}
L_h(\mathbf{x},m_v)=(1+\frac{3m_v}{2\omega})L_h(\mathbf{x}).\label{eq:N=2Lh}
\end{split}
\end{equation}

Note that we have not included a phase $\Theta$ in
(\ref{eq:LagN=2IndexSimp1}), the way we did in the previous section.
The reason is that we are assuming the hyper multiplets consist of
pairs of chiral multiplets sitting in conjugate representations of
the gauge group. In other words, $Q_h=0$ for all the Lagrangian
$\mathcal{N}=2$ theories of our interest.

Remarkably, according to (\ref{eq:N=2Lh}), the effect of nonzero
$m_v$ in $L_h(\mathbf{x},m_v)$ is only a multiplicative overall
factor. Assuming that $m_v$ is---just like $\omega$---pure
imaginary, we conclude that nonzero $m_v$ does not modify the locus
of the high-temperature localization of the matrix-integral. We can
thus apply the shift (\ref{eq:omegaShift}) in the asymptotics of the
SUSY partition function in (\ref{eq:LagIndexSimp6noTheta}) to obtain
\begin{equation}
\begin{split}
\ln Z^{\mathcal{N}=2}(b,\beta,m_v)=
i\frac{\pi^2}{3\beta}\left(\omega+\frac{3}{2}m_v\right)(\mathrm{Tr}R+12L_{h\
\mathrm{min}})
+\mathrm{dim}\mathfrak{h}_{qu}\ln(\frac{2\pi}{\beta})+O(\beta^0).
\label{eq:LagIndexN=2noTheta}
\end{split}
\end{equation}\\

\subsection{Asymptotics of the Schur partition function and the Schur index}

An immediate corollary is the asymptotics of the \emph{Schur
partition function}, defined by setting in the $\mathcal{N}=2$
partition function $m_v=\frac{\omega}{3}=\frac{i}{3}$:
\begin{equation}
\begin{split}
\ln Z^{Schur}(\beta)= -\frac{\pi^2}{2\beta}(\mathrm{Tr}R+12L_{h\
\mathrm{min}})
+\mathrm{dim}\mathfrak{h}_{qu}\ln(\frac{2\pi}{\beta})+O(\beta^0)\quad\quad
(\text{as $\beta\to 0$}).\label{eq:bnCoul}
\end{split}
\end{equation}

The \emph{Schur index} $\mathcal{I}^{Schur}(\beta)$ is similarly
defined by setting $m_v=\frac{\omega}{3}=\frac{i}{3}$ in the
$\mathcal{N}=2$ index. The relation between $\mathcal{I}^{Schur}$
and $Z^{Schur}$ follows from (\ref{eq:LagEquivN=2Z}) to be
\begin{equation}
\begin{split}
Z^{Schur}(\beta)= e^{-\beta
c/2}\mathcal{I}^{Schur}(\beta),\label{eq:lagSchurZ}
\end{split}
\end{equation}
where
\begin{equation}
c:=\frac{1}{32}(9\mathrm{Tr}R^3-5\mathrm{Tr}R), \label{eq:cFromTrRs}
\end{equation}
is the $c$ central charge \cite{Anselmi:1997}.

To the order shown in (\ref{eq:bnCoul}), the asymptotics of
$\ln\mathcal{I}^{Schur}(\beta)$ and $\ln Z^{Schur}(\beta)$ match;
the difference is of course at order $\beta$.

When $L_{h\ \mathrm{min}}=0$, the leading asymptotics in
(\ref{eq:bnCoul}) gives the Cardy-like piece noted recently in some
examples by Buican and Nishinaka \cite{Buican:2015a}.\\

For theories whose Rains function is minimized only at the origin of
$\mathfrak{h}_{cl}$, we can apply the shift (\ref{eq:omegaShift}) to
(\ref{eq:LagIndexSimp3Romelsberger}), and obtain
\begin{equation}
\ln \mathcal{I}(b,\beta,m_v)\sim
i\frac{\pi^2}{3\beta}(\omega+\frac{3}{2}m_v)\mathrm{Tr}R+\ln
Z_{S^3}(b,m_v)+\beta E_{\mathrm{susy}}(b,m_v),\quad\quad (\text{as
}\beta\to 0)\label{eq:IasyFiniteZ3dN=2}
\end{equation}
with some $Z_{S^3}(b,m_v)$ which can be easily derived from
(\ref{eq:LagEquivN=2Index}). Upon setting
$m_v=\frac{\omega}{3}=\frac{i}{3}$ in (\ref{eq:IasyFiniteZ3dN=2}) we
find
\begin{equation}
\ln \mathcal{I}_{Schur}(\beta)\sim \frac{8\pi^2}{\beta}(c-a)+\ln
Z_{S^3}(b=1,m_v=i/3)+\beta c/2\quad\quad (\text{as }\beta\to
0),\label{eq:bnFiniteZ3d}
\end{equation}
where
\begin{equation}
a:=\frac{3}{32}(3\mathrm{Tr}R^3-\mathrm{Tr}R), \label{eq:aFromTrRs}
\end{equation}
is the $a$ central charge \cite{Anselmi:1997}.

Our preliminary (unpublished) results suggest that the asymptotic
relation (\ref{eq:bnFiniteZ3d}) also holds (with some
$Z_{S^3}(b=1,m_v=i/3)$) for all the non-Lagrangian $T_N$ SCFTs. The
Schur index of these theories is given in \cite{Gadde:2011rry2d4d}.

\subsubsection{The Schur partition function of SU($N$) $\mathcal{N}=4$ SYM}

As discussed in subsection~\ref{sec:Z3finite}, the Rains function of
the $\mathcal{N}=4$ theory vanishes. Therefore $L_{h\
\mathrm{min}}=0$ and $\mathrm{dim}\mathfrak{h}_{qu}=N-1$. Since for
this theory also $\mathrm{Tr}R=0$, (\ref{eq:bnCoul}) yields
\begin{equation}
\begin{split}
\ln Z^{Schur}_{SU(N)\ \mathcal{N}=4}(\beta)=
(N-1)\ln(\frac{2\pi}{\beta})+O(\beta^0)\quad\quad (\text{as
$\beta\to 0$}).\label{eq:bnCoulN=4}
\end{split}
\end{equation}

\subsubsection*{More precise asymptotics}

We now improve (\ref{eq:bnCoulN=4}) by reducing its error to $o(1)$.
The reader not interested in the technical details of the derivation
can skip to (\ref{eq:SUNschurIndexAsy}) and continue reading from
there.

The starting point is the matrix-integral computing the Schur
partition function (recall that
$\Gamma((pq)^{1/2};p,q)=\Gamma(q;q,q)=1$)
\begin{equation}
\begin{split}
Z^{Schur}_{SU(N)\
\mathcal{N}=4}(\beta)=e^{-\beta(N^2-1)/8}\frac{(q;q)^{2(N-1)}}{
N!}&\Gamma^{2(N-1)}(q^{1/2})\times\\
&\int \mathrm{d}^{N-1} x \prod_{1\le i<j \le
N}\frac{\Gamma(q(z_i/z_j)^{\pm 1})}{\Gamma((z_i/z_j)^{\pm
1})}\Gamma^{2}(q^{1/2}(z_i/z_j)^{\pm 1}),\label{eq:N=4indexSchurSU2}
\end{split}
\end{equation}
with the integral over $x_i\in[-1/2,1/2]$, and $\prod_{i=1}^N
z_i=1$. The estimate (\ref{eq:GammaAsyPosExt}) guarantees that,
outside an $\varepsilon_1$ neighborhood of $\mathcal{S}_g$, the
integrand is well approximated by unity. Therefore
\begin{equation}
\begin{split}
Z^{Schur}_{SU(N)\ \mathcal{N}=4}(\beta)=\frac{(q;q)^{2(N-1)}}{
N!}\Gamma^{2(N-1)}(q^{1/2})(1+o(1))\quad\quad (\text{as $\beta\to
0$}).\label{eq:n=4SU2SchurAsy1}
\end{split}
\end{equation}
The $o(1)$ error above comes from neglecting $i)$ the $e^{-\beta
c/2}$ prefactor; $ii)$ the contribution to the integral from the
$\varepsilon_1$ neighborhood of $\mathcal{S}_g$, where the estimate
(\ref{eq:GammaAsyPosExt}) does not apply, and the integrand differs
from unity by some multiplicative factor of order one.

To write down the high-temperature asymptotics of
(\ref{eq:n=4SU2SchurAsy1}) more explicitly, we need the following
small-$\beta$ estimate \cite{Ardehali:2015b}:
\begin{equation}
\begin{split}
\ln \Gamma(q^r;q,q)\sim&-\frac{\pi^2}{3\beta}(r-1)+\left((r-1)\ln(1-e^{-2\pi i r})-\frac{1}{2\pi i}Li_2(e^{-2\pi i r})+\frac{i\pi (r-1)^2}{2}-\frac{i\pi}{12}\right)\\
&+\beta\left(\frac{r^3}{6}-\frac{r^2}{2}+\frac{5r}{12}-\frac{1}{12}\right).\label{eq:GammaFVest}
\end{split}
\end{equation}

Combining (\ref{eq:n=4SU2SchurAsy1}) and (\ref{eq:GammaFVest}), and
using $Li_2(-1)=-\pi^2/12$, we find
\begin{equation}
\begin{split}
\ln Z^{Schur}_{SU(N)\ \mathcal{N}=4}(\beta)=
(N-1)\ln(\frac{2\pi}{\beta}) -(N-1)\ln 2-\ln N! + o(1)\quad\quad
(\text{as $\beta\to 0$}).\label{eq:SUNschurIndexAsy}
\end{split}
\end{equation}
This asymptotic relation is confirmed in appendix~\ref{app:drukker}
using a very different approach.\\

Interestingly, a comparison of the Schur partition function of the
SU($2$) $\mathcal{N}=4$ theory in (\ref{eq:N=4indexSchurSU2}) and
the SUSY partition function (with $p=q$) of SO($3$) SQCD with two
flavors in (\ref{eq:SONindex}), reveals that the two precisely
coincide. In fact, the $\mathcal{N}=2$ partition function of the
SU($2$) $\mathcal{N}=4$ theory with $v=(pq)^{1/6}$, coincides with
the SUSY partition function of SO($3$) SQCD with two flavors even
when $p\neq q$. It would be nice to have a deeper understanding of
this coincidence.\\

\subsection{The example of the non-Lagrangian $E_6$
SCFT}\label{sec:e6}

For non-Lagrangian theories the $\mathcal{N}=2$ partition function
can not be defined via path-integration. Nonetheless, the
$\mathcal{N}=2$ index is well-defined from the Hamiltonian
perspective of (\ref{eq:IndexN=2Def}). When the `t~Hooft anomalies
of the theory are known, one can then compute the $\mathcal{N}=2$
Bobev-Bullimore-Kim polynomial $E_{\mathrm{susy}}(b,m_v)$, and
\emph{define} the $\mathcal{N}=2$ partition function via
(\ref{eq:LagEquivN=2Z}). This procedure can be done, for instance,
for the $E_6$ SCFT \cite{Minahan:1996}, whose $\mathcal{N}=2$ index
and $E_{\mathrm{susy}}(b,m_v)$ are both known.

It turns out that our methods do not apply directly to the
$\mathcal{N}=2$ partition function of the $E_6$ SCFT. We instead
consider an equivariant deformation of the $\mathcal{N}=2$ partition
function, which is computed by path-integration in presence of a
\emph{real} background U($1$)$_w$ gauge field $m_w$ along
$S^1_\beta$, that couples a conserved U($1$) flavor charge in the
theory. We denote the resulting partition function by
$Z^{\mathcal{N}=2}_{E_6}(b,\beta,m_v;m_w)$. This equivariant
partition function is related to an equivariant $\mathcal{N}=2$
index $\mathcal{I}^{E_6}(b,\beta,m_v;m_w)$, in which $w:=e^{i\beta
m_w}$ plays the role of an additional fugacity for the U($1$)$_w$
charge. An equation similar to (\ref{eq:LagEquivN=2Z}), but with an
equivariant Bobev-Bullimore-Kim polynomial
$E_{\mathrm{susy}}(b,m_v;m_w)$, mediates
$Z^{\mathcal{N}=2}_{E_6}(b,\beta,m_v;m_w)$ and
$\mathcal{I}^{E_6}(b,\beta,m_v;m_w)$. Explicitly \cite{Bobev:2015}
\begin{equation}
\begin{split}
E_{\mathrm{susy}}^{E_6}(b,m_v;m_w)=&\frac{i}{6}(\frac{98}{27})(\omega+\frac{3}{2}m_v)^3+i\left(\frac{b^2+b^{-2}}{24}\right)(-\frac{22}{3})(\omega+\frac{3}{2}m_v)\\
&+\frac{i}{2}(4)m_w^2(\omega+\frac{3}{2}m_v).\label{eq:EcE6}
\end{split}
\end{equation}
Note that the effect of nonzero $m_v$ is accounted for precisely by
the shift (\ref{eq:omegaShift}). Setting $m_w=0$ and comparing with
(\ref{eq:EcBequiv}) reveals, for example, that
$\mathrm{Tr}R^3=98/27$. The effect of nonzero equivariant parameters
such as $m_w$ is easily obtained in general by shifting $R\omega$ in
(\ref{eq:EcBequiv}) to $R\omega+Q_w m_w$. The $m_w$-dependent terms
in the equivariant Bobev-Bullimore-Kim polynomial then encode
various `t~Hooft anomalies associated to the U($1$)$_w$ current. The
second line of (\ref{eq:EcE6}), for instance, indicates that
$\mathrm{Tr}RQ_w^2=4$.

The $\mathcal{N}=2$ index of the $E_6$ SCFT (also known as the $T_3$
theory) is computed in \cite{Gadde:2010e6}, and is given by
\begin{equation}
\begin{split}
\mathcal{I}^{E_6}(b,\beta,m_v;m_w)=&\frac{(p;p)(q;q)}{2\Gamma((pq)^{1/3}vw^{\pm
2})\Gamma((pq)^{-2/3}v)}\int_{-1/2}^{1/2}\mathrm{d}x_s
\frac{\Gamma((pq)^{-1/3}v^{1/2}w^{\pm1}s^{\pm 1})}{\Gamma(s^{\pm
2})}\hat{\mathcal{I}}(s)\\
&+\frac{1}{2}\frac{\Gamma(w^{-2})}{\Gamma((pq)^{1/3}vw^{-2})}
\left(\hat{\mathcal{I}}(s=(pq)^{-1/3}v^{1/2}w)+\hat{\mathcal{I}}(s=(pq)^{1/3}v^{-1/2}w^{-1})\right)\\
&+\frac{1}{2}\frac{\Gamma(w^{2})}{\Gamma((pq)^{1/3}vw^{2})}
\left(\hat{\mathcal{I}}(s=(pq)^{-1/3}v^{1/2}w^{-1})+\hat{\mathcal{I}}(s=(pq)^{1/3}v^{-1/2}w)\right),\label{eq:e6index}
\end{split}
\end{equation}
where
\begin{equation}
\begin{split}
\hat{\mathcal{I}}(s)=\frac{(p;p)^2(q;q)^2}{3!}\Gamma((pq)^{1/3}v)^2
\int \mathrm{d}^2 x\  &\left(\prod_{1\le i< j\le
3}\frac{\Gamma((pq)^{1/3}v (z_i/z_j)^{\pm 1})}{\Gamma((z_i/z_j)^{\pm
1})}\right)\\
&\ \ \ \ \left(\prod_{i=1}^3
\Gamma((pq)^{1/3}v^{-1/2}(s^{-1/3}z_i)^{\pm1})\right)^2,
\end{split}
\end{equation}
with the integral over the square $-1/2\le x_1,x_2\le 1/2$. The
parameters are related via $s=e^{2\pi i x_s},z_i=e^{2\pi i x_i}$,
and are constrained to satisfy $\sum_{i=1}^{3}x_i\in\mathbb{Z}$.

Luckily, the expression (\ref{eq:e6index}) involves the Pochhammer
symbols and elliptic gamma functions that we are already familiar
with. The method of Rains hence applies immediately.

As $\beta\to0$, the integrals in each of the three lines of
(\ref{eq:e6index}) take the form (\ref{eq:LagN=2IndexSimp1}), with
some effective potentials and associated Rains functions that can be
easily obtained.  For example, the Rains function associated to the
integral on the first line is\footnote{Note that to write the first
term in the following Rains function, we are applying
(\ref{eq:GammaOffCenter2}) with $r=-2/3$. This extrapolation of
(\ref{eq:GammaOffCenter2}) can be justified (when
$b,b^{-1}\neq\sqrt{2}$) by an argument similar to the one in
Appendix~\ref{app:rains}. In fact the derivation of
(\ref{eq:GammaOffCenter2}) in Appendix~\ref{app:rains} indicates
that constraining $r$ to the range $]0,2[$ is too conservative.
Similarly, (\ref{eq:GammaToHypGamma}) has to be extrapolated to
obtain (\ref{eq:e6indexAsy}).}
\begin{equation}
\begin{split}
L_h^{1^{st}\
\mathrm{line}}(x_1,x_2,x_s)=&2(1+\frac{2}{3})\vartheta(x_s)-\vartheta(2x_s)\\
&+(1-2/3-1)\sum_{i<j}\vartheta(x_i-x_j)+
6(1-\frac{2}{3})\sum_{i}\vartheta(x_i-x_s/3),
\end{split}
\end{equation}
with the second line of the above function coming from the integrand
of $\hat{\mathcal{I}}(s)$. Rains's generalized triangle inequality
(\ref{eq:RainsGTI}) then implies that $L_h^{1^{st}\ \mathrm{line}}$
is minimized at $x_1=x_2=x_s=0$. We can thus prune the integral
tails, use our central estimate (\ref{eq:GammaToHypGamma}), and then
employ the inequality (\ref{eq:RainsGTIav}) to ensure that tails
completion is safe. The evaluation of the asymptotics thus proceeds
similarly to the cases in subsection \ref{sec:Z3finite}. Note that
the nonzero real parameter $m_w$ leads to a nonzero phase $\Theta$
in (\ref{eq:LagIndexSimp1}), even though the analog of $Q_h$ for the
integrand of (\ref{eq:e6index}) vanishes. But as in
subsection~\ref{sec:Z3finite} the nonzero phase does not present an
obstacle to our analysis, because the Rains function is minimized
only at the origin.

All in all, we find the small-$\beta$ asymptotics
\begin{equation}
\begin{split}
\ln\mathcal{I}^{E_6}(b,\beta,m_v;m_w)\sim
i\frac{\pi^2}{3\beta}(\mathrm{Tr}R)(\omega+\frac{3}{2}m_v)+\ln
Z_{3d}^{E_6}(b,m_v;m_w)+\beta
E_{\mathrm{susy}}^{E_6}(b,m_v;m_w),\label{eq:e6indexAsy}
\end{split}
\end{equation}
with $\mathrm{Tr}R=-22/3$, and with some $Z_{3d}^{E_6}(b,m_v;m_w)$
whose derivation we omit. From (\ref{eq:LagEquivN=2Z}) we then
conclude
\begin{equation}
\begin{split}
\ln Z^{\mathcal{N}=2}_{E_6}(b,\beta,m_v;m_w)\sim
i\frac{\pi^2}{3\beta}(\mathrm{Tr}R)(\omega+\frac{3}{2}m_v)+\ln
Z_{3d}^{E_6}(b,m_v;m_w),\quad\quad (\text{as $\beta\to
0$})\label{eq:e6N=2ZAsy}
\end{split}
\end{equation}
just as if the $E_6$ SCFT was a Lagrangian $\mathcal{N}=2$ theory
with finite $Z_{S^3}$.\\

\section{Discussion}\label{sec:Dis}

In this work we have studied the SUSY partition function of 4d
supersymmetric gauge theories with a U($1$)$_R$ symmetry, and with a
(compact) semi-simple gauge group. More precisely, we have also
assumed the R-charges of the chiral multiplets to be inside the
interval\footnote{Otherwise, it seems like the SUSY partition
function would be ill-defined. On $S^3\times S^1$, the scalars
inside a chiral multiplet have a curvature coupling, which gives
their Kaluza-Klein zero-modes a mass. This mass would become
non-positive (yielding a non-compact Higgs branch, or a tachyonic
direction) if the R-charge of the multiplet does not belong to
$]0,2[$. Nonetheless, it may be possible to use meromorphic
continuation (of the path-integral, or of the Romelsberger
prescription \cite{Romelsberger:2007pre}) to consistently assign
SUSY partition functions to theories containing chiral multiplets
with $r\notin]0,2[$.\label{footnote:mero}} $]0,2[$, and we have
taken the cancelation of the following anomalies for granted: $i)$
the gauge$^3$ anomaly; $ii)$ the U(1)$_R$-gauge-gauge anomaly;
$iii)$ the gauge-gravitational-gravitational anomaly; and $iv)$ the
gauge-U(1)$_R$-U(1)$_R$ anomaly.

A major role in our analysis is played by the Rains function
$L_h(x_1,\dots,x_{r_G})$ of the SUSY gauge theory, defined in
Eq.~(\ref{eq:LhDef}) (see (\ref{eq:varthetaDef}) for the definition
of the function $\vartheta$ appearing in $L_h$). According to
Eq.~(\ref{eq:VeffEquiv1}), $L_h$ is proportional to
$V^{\mathrm{eff}}$.

Another important role was played above by the function
$Q_h(x_1,\dots,x_{r_G})$ of the SUSY gauge theory, defined in
Eq.~(\ref{eq:QhDef}) (see (\ref{eq:kappaDef}) for the definition of
the function $\kappa$ appearing in $Q_h$). Only theories with chiral
matter content may have nonzero $Q_h$. Such nonzero $Q_h$ can make
the high-temperature analysis of the SUSY partition function
difficult.

Let us now recapitulate some of our main findings, and then move on
to exploring the unresolved problems and open directions related to
the subject of this work.

\begin{itemize}
\item It is sometimes said in the literature that ``as $\beta\to 0$, the SUSY partition
function of a 4d theory reduces (after its divergent Cardy-like
piece is stripped off) to the squashed-three-sphere partition
function of the 3d theory obtained by reducing the 4d theory on
$S^1_\beta$''. As already emphasized in
\cite{Aharony:2013a,DiPietro:2014}, this statement is not generally
true. In section~\ref{sec:N=1Asy}, we have obtained the condition
under which the above statement \emph{is true} in a SUSY gauge
theory with a semi-simple gauge group: the Rains function of the 4d
theory must have a unique minimum at the origin of
$\mathfrak{h}_{cl}$ (corresponding to $x_1=\dots=x_{r_G}=0$). In
particular, this condition is satisfied in all the SU($N$) $ADE$
SQCD theories discussed in \cite{Kutasov:2014,Intriligator:2003ADE},
and also the Sp($2N$) SQCD theories discussed in \cite{Dolan:2008}.

\item In \cite{Ardehali:2015a,Ardehali:2015b} prescriptions were put
forward for extracting the central charges of a finite-$N$ 4d SCFT
from its superconformal index. The example of the SO($3$) SQCD with
two flavors, that we studied in section~\ref{sec:N=1Asy}, shows that
the finite-$N$ prescriptions of \cite{Ardehali:2015a,Ardehali:2015b}
are not valid in general. On the other hand, the said prescriptions
can be applied (for extracting $c$ and $a$ as in
(\ref{eq:cFromTrRs}) and (\ref{eq:aFromTrRs})) successfully to SUSY
gauge theories with a semi-simple gauge group, with non-chiral
matter content (hence $Q_h=0$), and with
$\mathrm{dim}\mathfrak{h}_{qu}=0$. (In fact all the `t~Hooft
anomalies of such theories can be extracted from the
high-temperature asymptotics of their equivariant Romelsberger index
$\mathcal{I}(b,\beta;m_a)$; see the comments below
(\ref{eq:ISSindexAsy3}).) Moreover, even nonzero $Q_h$ (arising from
chiral matter content) does not present an obstruction to the said
prescriptions if the Rains function of the theory has a unique
minimum at the origin of $\mathfrak{h}_{cl}$ (see the comments below
(\ref{eq:LagIndexSimp3Romelsberger})).

\item We have shown that the leading high-temperature asymptotics of $\ln Z^{\mathrm{SUSY}}(\beta)$
is \emph{not universal} for SUSY gauge theories with a semi-simple
gauge group, in the following sense. If the Rains function of the
theory is not minimized at the origin of $\mathfrak{h}_{cl}$, the
distance between $\mathfrak{h}_{qu}$ and the origin can serve as an
order parameter for labeling the infinite-temperature phase of the
theory on $S_b^3\times S^1_\beta$. [Note that at any finite
temperature, a finite-$N$ gauge theory on $S_b^3\times S^1_\beta$
can not be assigned a phase, because the spatial manifold of the
theory is compact. In the infinite-temperature limit, however, a
phase emerges. The possibility of emergence of a thermodynamic
ensemble in the high-temperature limit of a relativistic
finite-volume system can be most easily understood in free QFTs; the
Fock space of a free QFT becomes populated without a bound as
$\beta\to 0$.] If this order parameter is nonzero, the leading
high-temperature asymptotics of $\ln Z^{\mathrm{SUSY}}(\beta)$ may
differ---and would certainly differ if the theory is
non-chiral---from the generic Cardy-like asymptotics in
(\ref{eq:dk}).
\end{itemize}

The remarks in the last bullet point above suggest the following
interpretation for the asymptotic relations we found in the ISS and
the BCI$_5$ models. Let's begin with the BCI$_5$ theory.
Figure~\ref{fig:BCI} indicates that this theory has an
infinite-temperature phase which partially breaks the gauge group
SO($5$). Indeed the expression for $Y^{BCI_5}_{S^3}(b)$ in
(\ref{eq:Ybci}) suggests that in this \emph{Higgsed phase}, the 3d
theory effectively consists of an SO($3$) vector multiplet with a
chiral matter multiplet in the five-dimensional representation, and
an SQED theory. For the ISS model, the expression for
$Y^{ISS}_{S^3}(b)$ in (\ref{eq:Yiss}) suggests again a Higgsed phase
at infinite temperature, this time with only an SQED effective 3d
theory. (It might be possible to interpret the exponential function
in the integrand of (\ref{eq:Yiss}) as an induced FI parameter.)\\

\subsection{Open problems}
We have not treated chiral theories (with $Q_h\neq 0$) in full
generality. The following problem is thus the most important loose
end of the present work.
\begin{description}
  \item[Problem 1)] Restricting still to SUSY gauge theories with a
  semi-simple gauge group, find a general expression similar to
  (\ref{eq:LagIndexSimp6noTheta}), that is valid for theories with $Q_h\neq0$.
\end{description}
A related puzzle is the following.
\begin{description}
  \item[Problem 1.1)] Find a SUSY gauge theory with a semi-simple
  gauge group, in which $Q_h$ is nonzero on the minimum set of
  $L_h$. (Or prove that such a theory does not exist.)\\
\end{description}

Even focusing on non-chiral theories (hence $Q_h=0$), we have not
been able to clarify some of the intriguing phenomena we observed in
our explicit examples. For instance, our case by case investigation
suggests that $\mathrm{Tr}R>0$ when $L_{h\ \mathrm{min}}<0$ (the ISS
and the BCI$_{\ge5}$ models), and that $\mathrm{Tr}R=0$ when $L_h$
vanishes identically (the SO($3$) SQCD and the $\mathcal{N}=4$ SYM).
It is highly desirable to know if these correlations are general or
not. We can phrase this as follows.
\begin{description}
  \item[Problem 2)] Is there a general correlation between the sign
  of $\mathrm{Tr}R$ in a SUSY gauge theory with a semi-simple gauge
  group, and the sign of the theory's $L_{h\ \mathrm{min}}$?
\end{description}

A possibly related problem is the connection between the finiteness
of $Z_{S^3}(b)$ and the validity of the Di~Pietro-Komargodski
formula. As discussed in the introduction, we suspect (but have not
been able to show) that all theories with finite $Z_{S^3}(b)$
satisfy the Di~Pietro-Komargodski formula. The following problem
phrases the question in terms of the functions $L_{h}$ and
$\tilde{L}_{S^3}$ (see Eq.~(\ref{eq:hypURains}) for the definition
of $\tilde{L}_{S^3}$).
\begin{description}
  \item[Problem 3)] Prove (or disprove) that in a SUSY gauge theory with a semi-simple
  gauge group, if the function $\tilde{L}_{S^3}$ (and thus $L_{h}$) is strictly
  positive in some punctured neighborhood of the origin, then $L_{h}$
  is positive semi-definite.\\
\end{description}

Another important direction for extending the present work is the
following.
\begin{description}
  \item[Problem 4)] Extend the results of the present paper to SUSY
  gauge theories with a compact gauge group.
\end{description}
The added difficulty would of course be in analyzing the extra
U($1$) factors in the gauge group.\\

\subsection{Two new simple tests of supersymmetric
dualities}\label{sec:tests}

Dual QFTs must have equal partition functions. As a trivial
corollary, the high-temperature asymptotics of the SUSY partition
functions of dual 4d SUSY QFTs must match.

Assume now that both sides of the duality are 4d SUSY gauge theories
(with a U($1$)$_R$ symmetry, and free of various harmful anomalies)
with a semi-simple gauge group, and with $Q_h=0$. The relation
(\ref{eq:LagIndexSimp6noTheta}) then yields two quantities to be
matched between the theories: $L_{h\ \mathrm{min}}$ and
$\mathrm{dim}\mathfrak{h}_{qu}$. Comparison of $L_{h\ \mathrm{min}}$
can rule out for instance the confinement scenario for the SU($2$)
ISS model: on the gauge theory (UV) side, as discussed in
subsection~\ref{sec:c<a}, we have $L_{h\ \mathrm{min}}=-2/15$, while
on the mesonic (IR) side\footnote{Following \cite{Vartanov:2010}, we
are assuming that a SUSY partition function can be consistently
assigned to the proposed IR theory, even though the IR chiral
multiplet would have R-charge $12/5\notin]0,2[$. This assignment
requires an analytic continuation of the kind mentioned in
footnote~\ref{footnote:mero}. The duality test in
\cite{Vartanov:2010} can then be thought of as comparing the
low-temperature asymptotics of the supposedly dual SUSY partition
functions; the low-temperature test goes beyond `t~Hooft anomaly
matching already at the leading order (in the large-$\beta$
expansion) there, because on the IR side the relation
(\ref{eq:equivZlowT}) and the comments below it do not apply. See
\cite{Gerchkovitz:2013} for an alternative take on this problem. I
thank L.~Di~Pietro and Z.~Komargodski for correspondence on this
point, and for explaining to me related subtleties that were
overlooked in an early draft of the present paper.} we have no gauge
group and thus $L_{h}=0$.

As another example, consider the recent $E_7$ SQCD duality of
\cite{Kutasov:2014,Kutasov:2014ed}. In that case a direct
examination reveals that $L_{h\
\mathrm{min}}=\mathrm{dim}\mathfrak{h}_{qu}=0$, both on the electric
and the magnetic side. Their proposal hence passes both our tests.

Our numerical investigation indicates that the magnetic Pouliot
theory with $N_f=7$ and its electric dual \cite{Pouliot:1995} also
both have $L_{h\ \mathrm{min}}=\mathrm{dim}\mathfrak{h}_{qu}=0$, and
thus their duality passes our tests. Note that on the magnetic side,
since the Rains function is minimized only at the origin of
$\mathfrak{h}_{cl}$, the discussion of subsection~\ref{sec:Z3finite}
applies, and therefore the nonzero $Q_h$ does not present an
obstruction to performing the tests in this case.\\

The case of the interacting $\mathcal{N}=1$ SCFTs with $c<a$ (namely
the IR fixed points of the ISS model, and the BCI$_{2N+1}$ model
with $1<N<5$) is particularly interesting. A dual description for
these theories is currently lacking. Our results for $L_{h\
\mathrm{min}}$ and $\mathrm{dim}\mathfrak{h}_{qu}$ on the electric
side might help to test future proposals for magnetic duals of these
theories.\\

\subsection{Holography and the asymptotics of 4d superconformal indices}

Studying the high-temperature asymptotics of \emph{the large-$N$
limit} of the superconformal indices of 4d SCFTs has already proven
fruitful for holography. It has led to a rather general solution to
the problem of Holographic Weyl Anomaly in the traditional
AdS$_5$/CFT$_4$ scenarios \cite{Ardehali:2015a,Ardehali:2015b}. More
precisely, at the leading order ($O(N^2)$), the holographic Weyl
anomaly in AdS$_5$/CFT$_4$ was addressed by Henningson-Skenderis
\cite{Henningson:1998gx} and Gubser \cite{Gubser:1998} back in 1998.
But in the traditional scenarios, with AdS$_5$ times a toric
Sasaki-Einstein 5-manifold (SE$_5$) on the gravity side, the anomaly
has a subleading $O(N^0)$ piece whose AdS/CFT matching was open in
the general case until the works
\cite{Beccaria:2014,Ardehali:2015a,Ardehali:2015b}; in
\cite{Ardehali:2015a} the matching of the subleading piece was
established (crucially relying on results of \cite{Beccaria:2014}
and \cite{Eager:2012hx}) for the cases where the toric SE$_5$ is
smooth and the SCFT does not have matter in the adjoint
representation of the gauge group(s); in \cite{Ardehali:2015b} the
matching was shown for the general case, assuming $i)$ that the
boundary single-trace index and the bulk single-particle index
match, and $ii)$ that an (essentially combinatorial) conjecture
proposed and supported in
\cite{Agarwal:2013} is valid.\\

In the present paper we analyzed the high-temperature asymptotics of
the Romelsberger indices of various gauge theories at \emph{finite
$N$}. The finite-$N$ indices of holographic SCFTs are expected to
encode information about micro-states of the supersymmetric
Giant~Gravitons of the dual string theories \cite{Bourdier:2015}.
Take for instance the SU($N$) $\mathcal{N}=4$ SYM. One of the novel
results of the present paper is the following high-temperature
asymptotics for the superconformal index of this theory (see
Eqs.~(\ref{eq:N=4indexAsy1}) and (\ref{eq:N=4indexAsy})):
\begin{equation}
\mathcal{I}(b=1,\beta)=\sum_{\mathrm{operators}}(-1)^F
e^{-\beta(\Delta-\fft12r)}\approx (\frac{1}{\beta})^{N-1}.
\end{equation}
The above \emph{canonical} relation can be transformed to the
\emph{micro-canonical} ensemble to yield the asymptotic
(fermion-number weighted) degeneracy of the protected high-energy
operators in the $\mathcal{N}=4$ theory:
\begin{equation}
N(E)\approx E^{N-2},
\end{equation}
with $E=\Delta-r/2$. This result should presumably be reproduced by
geometric quantization of the $1/16$ BPS Giant~Gravitons of IIB
theory on AdS$_5\times S^5$, along the lines of \cite{Biswas:2006}.
It would be interesting to see if this expectation pans out.\\

\subsection{Crossed channel: quantum Coulomb branch dynamics on $R^3\times S^1$}

Take a 4d $\mathcal{N}=1$ SUSY gauge theory with a U($1$)
R-symmetry, and with a semi-simple gauge group. Its SUSY partition
function $Z^{\mathrm{SUSY}}(b,\beta)$ was so far defined by a
path-integral on $S_b^3\times S_\beta^1$, with $S_b^3$ the
unit-radius squashed three-sphere. We now replace the $S_b^3$ with
the round three-sphere $S_{r_3}^3$ of arbitrary radius $r_3>0$. The
path-integral on the new space gives
$Z^{\mathrm{SUSY}}(\beta;r_3)=Z^{\mathrm{SUSY}}(b=1,\beta/r_3)$;
i.e. the resulting partition function only depends on the ratio
$\beta/r_3$ \cite{Festuccia:2011}. Thus, as far as
$Z^{\mathrm{SUSY}}(\beta;r_3)$ is concerned, shrinking the $S^1$ is
equivalent to decompactifying the $S^3$. We hence fix $\beta$, and
send $r_3$ to infinity. In this limit we expect the unlifted
zero-modes on $S_{r_3}^3\times S_\beta^1$ to roughly correspond to
the quantum zero-modes on $R^3\times S^1$. Therefore at high
temperatures the unlifted holonomies of the theory on
$S_{r_3}^3\times S_\beta^1$ should be in correspondence with (a real
section of) the quantum Coulomb branch of the 3d $\mathcal{N}=2$
theory obtained from reducing the 4d theory on the circle of
$R^3\times S^1$. In particular, we expect
$\mathrm{dim}\mathfrak{h}_{qu}$ to be equal to the (complex-)
dimension of the quantum Coulomb branch of the 3d theory. (Recall
that the Coulomb branch of the 3d theory consists not just of the
holonomies around the $S^1$, but also of the dual 3d photons; hence
our references above to ``a real section'' and
``complex-dimension''.)

We do not expect to recover the $R^3\times S^1$ Higgs branch from
the zero-modes on $S_{r_3}^3\times S_\beta^1$, because for any
(arbitrarily small) curvature on the $S^3$, curvature couplings
presumably lift the Higgs-type zero-modes on $S_{r_3}^3\times
S_\beta^1$.

From the point of view of $R^3\times S^1$, picking one of the $R^3$
directions as time\footnote{The following discussion is in the
spirit of the arguments in \cite{Shaghoulian:2015}, though our
treatment is not as precise. We are approaching $R^3$ from $S^3$,
rather than from $T^3$ (as in \cite{Shaghoulian:2015}). While on
$T^3$ each of the circles can be picked as the time direction,
picking a time direction along the $S^3$ makes the spatial sections
time-dependent, rendering our arguments in the paragraph of this
footnote somewhat hand-wavy. I thank E.~Shaghoulian for several
helpful conversations related to the subject of the present
subsection.}, we can relate $\mathcal{E}^{DK}_0$ to the Casimir
energy associated to the spatial manifold $R^2\times S^1$: we
reintroduce $r_3$ in $\mathcal{E}^{DK}_0$ (by replacing its $\beta$
with $\beta/r_3$), set in it $b=1$, interpret $\tilde{\beta}:=2\pi
r_3$ as the circumference of the crossed channel thermal circle, and
write
\begin{equation}
\mathcal{E}_0^{DK}(\beta;r_3)=\tilde{\beta}E_0^{R^2\times
S^1}(\beta),\quad\text{with}\quad E_0^{R^2\times
S^1}(\beta)=\frac{\pi}{6\beta}\mathrm{Tr}R.\label{eq:dkEcross}
\end{equation}
Now $E_0^{R^2\times S^1}(\beta)$ admits an interpretation as the
(regularized) Casimir energy associated to the spatial $R^2\times
S_\beta^1$. Similarly, resurrecting the $r_3$ in $V^{\mathrm{eff}}$,
and setting in it $b=1$, we obtain what can be loosely regarded as
$\tilde{\beta}$ times the quantum effective potential on (a real
section of) the crossed channel Coulomb branch. From this
perspective, the two tests we advocated in
subsection~\ref{sec:tests} would not really be new, but would
correspond to the comparison of low-energy properties on $R^3\times
S^1$.

The discussion in the previous three paragraphs is rather intuitive,
and should be considered suggestive at best. It is desirable to have
it made more precise. Nevertheless, in the examples of the SU($N$),
Sp($2N$), and SO($2N+1$) SQCD theories, and the SU($N$)
$\mathcal{N}=4$ SYM, we see that (upon quotienting by the Weyl
group) $\mathfrak{h}_{qu}$ \emph{does indeed resemble} (a real
section of) the $R^3\times S^1$ quantum Coulomb branch; see
\cite{Aharony:2013a,Aharony:2013b} and \cite{Seiberg:1997}. We
therefore conjecture that the relation between $\mathfrak{h}_{qu}$
and the unlifted Coulomb branch on $R^3\times S^1$ continues to
remain valid, at least for all the theories with a positive
semi-definite Rains function. In particular, we predict that, when
placed on $R^3\times S^1$, all the SU($N$) $ADE$ SQCD and the
Pouliot theories (in the appropriate range of their parameters such
that all their $r_\chi$ are in $]0,2[$) have no quantum Coulomb
branch, and the $\mathbb{Z}_2$ and $\mathbb{Z}_3$ orbifolds of the
SU($N$) $\mathcal{N}=4$ theory have an $(N-1)$-dimensional unlifted
Coulomb branch\footnote{For the $\mathbb{Z}_2$ orbifold theory, the
statement can not apply to the $\mathcal{N}=2$ phase, which has its
classical Coulomb branch protected by supersymmetry. We conjecture
that in such $\mathcal{N}=2$ theories the Rains function encodes
information about the softly-broken (to $\mathcal{N}=1$) phase.}.\\

For theories whose Rains function is not positive semi-definite, on
the other hand, it seems like this connection with $R^3\times S^1$
fails: the Rains function of the SU($2$) ISS model does not have a
flat direction, and appears to suggest a Higgs vacuum for the theory
on $R^3\times S^1$; however, the study of Poppitz and Unsal
\cite{Poppitz:2009} indicates that this theory possesses an unlifted
Coulomb branch on $R^3\times S^1$, and in particular does not
necessarily break the gauge group at low energies. It would be nice
to understand if this conflict is only a manifestation of the
sloppiness of our intuitive arguments above, or it has a more
interesting origin.\\

\acknowledgments

The project reported on here was a direct outcome of the author's
collaborations and discussions with Finn~Larsen, Jim~Liu, and
Phil~Szepietowski, whose ideas, help, and feedback have contributed
to this work at various levels and various stages. I am grateful to
them, as well as to C.~Beem, J.~Bourdier, N.~Drukker, H.~Elvang,
J.~Felix, A.~Gadde, D.~Mayerson, J.~McGreevy, D.~Poland, S.~Razamat,
C.~Uhlemann, and especially F.~Bouya, S.~Chapman, K.~Intriligator,
G.~Knodel, U.~Kol, Y.~Nakayama, K.~Ohmori, L.~Pando~Zayas,
E.~Shaghoulian, and Jaewon~Song for helpful conversations and
correspondences related to the subject matter of this paper. I am
particularly indebted to Peter~Miller and Eric~Rains for valuable
discussions on the asymptotic analysis used in this work. I also
thank L.~Di~Pietro and Z.~Komargodski for their insightful comments
and helpful feedback on a draft of this paper. The plots in this
paper are all produced by Mathematica. This project was supported
partly by the physics department at University of Michigan, and
partly by the US Department of Energy under grant DE-SC0007859.

Finally, my profound thanks go to A.~Peters for all the support,
encouragement, and inspiration I have received from her while this
work was in progress. This paper is dedicated to her.

\appendix

\section{Derivation of the elliptic gamma function estimates}\label{app:rains}

Define the non-compact quantum dilogarithm $\psi_b$ (c.f. the
function $e_b(x)$ in \cite{Faddeev:2001}; $\psi_b(x)=e_b(-i x)$) via
\begin{equation}
\psi_b(x):=e^{-i\pi
x^2/2+i\pi(b^2+b^{-2})/24}\Gamma_h(ix+\omega;\omega_1,\omega_2),\label{eq:hyperbolicGammaPsi}
\end{equation}
where
\begin{equation}
\omega_1:=i b,\quad\omega_2:=i b^{-1},\quad\text{and}\quad
\omega:=(\omega_1+\omega_2)/2.
\end{equation}
For generic choice of $b$, the zeros of $\psi_b(x)^{\pm 1}$ are of
first order, and lie at $\pm((b+b^{-1})/2+b\mathbb{Z}^{\ge
0}+b^{-1}\mathbb{Z}^{\ge 0})$. Upon setting $b=1$ we get the
function $\psi(x)$ of \cite{Felder:1999}; i.e.
$\psi_{b=1}(x)=\psi(x)$.

From the asymptotics of the hyperbolic gamma function (see e.g.
\cite{Rains:2009}), it follows that for fixed $\mathrm{Re}(x)$ and
fixed $b>0$
\begin{equation}
\ln\psi_b(x)\sim 0,\quad\quad\quad (\text{as } \beta\to 0, \text{
for } \mathrm{Im}(x)= -1/\beta)\label{eq:psiAsy}
\end{equation}
with a transcendentally small error, of the type $e^{-1/\beta}$.\\

An identity due to Narukawa \cite{Narukawa:2004} implies (see also
Appendix~A of \cite{Ardehali:2015b})
\begin{equation}
\begin{split}
\Gamma(x;\sigma,\tau):=e^{2i\pi
Q_{+}(x;\sigma,\tau)}\psi_b(\text{\footnotesize{$-\frac{2\pi i
x}{\beta}-\frac{b+b^{-1}}{2}$}})\prod_{n=1}^{\infty}\frac{\psi_b(-\frac{2\pi
in}{\beta}-\frac{2\pi i
x}{\beta}-\frac{b+b^{-1}}{2})}{\psi_b(-\frac{2\pi
in}{\beta}+\frac{2\pi i
x}{\beta}+\frac{b+b^{-1}}{2})},\label{eq:GammaFVSq}
\end{split}
\end{equation}
where
\begin{equation}
\begin{split}
Q_{+}(x;\sigma,\tau)=&-\frac{x^3}{6\tau\sigma}+\frac{\tau+\sigma+1}{4\tau\sigma}x^2-\frac{\tau^2+\sigma^2+3\tau\sigma+3\tau+3\sigma+1}{12\tau\sigma}x\\
&+\frac{1}{24}(\tau+\sigma+1)(1+\tau^{-1}+\sigma^{-1}).\label{eq:QpDef}
\end{split}
\end{equation}\\

The two relations (\ref{eq:psiAsy}) and (\ref{eq:GammaFVSq})
immediately imply (\ref{eq:GammaAsyPosExt}). Moreover, the three
relations (\ref{eq:hyperbolicGammaPsi}), (\ref{eq:psiAsy}), and
(\ref{eq:GammaFVSq}) imply (\ref{eq:GammaToHypGamma}) and
(\ref{eq:GammaToHypGammaVec}).

To derive (\ref{eq:GammaOffCenter2}) we need the following fact: for
fixed $r\in]0,2[$ and fixed $b>0$, as $\beta\to 0$ the function
$\ln\psi_b(-\frac{2\pi i \{x\}}{\beta}+(r-1)\frac{b+b^{-1}}{2})$ is
uniformly bounded over ($x\in)$ $\mathbb{R}$. It suffices of course
to establish this fact in the ``fundamental domain'' $x\in [0,1[$.
To obtain the uniform bound, divide this interval into
$[0,N_0\beta]$ and $[N_0\beta,1[$, with $N_0$ chosen as follows.
Since $\psi_b(-2\pi i N+(r-1)\frac{b+b^{-1}}{2})\to 1$ as
$N\to\infty$, there is a large enough $N_0$, so that for all $N>N_0$
we have $\psi_b(-2\pi i N+(r-1)\frac{b+b^{-1}}{2})\approx 1$, with
an error of say $.1$. With this choice of $N_0$ it is clear that
$\ln\psi_b(-\frac{2\pi i x}{\beta}+(r-1)\frac{b+b^{-1}}{2})$ is
uniformly bounded over $[N_0\beta,1[$ (for all $\beta$ smaller than
$1/N_0$). On the other hand, since $\ln\psi_b(-2\pi i
x+(r-1)\frac{b+b^{-1}}{2})$ is continuous, it is guaranteed to be
uniformly bounded on the compact domain $[0,N_0]$; re-scaling
$x\mapsto\frac{x}{\beta}$ this implies the uniform bound on
$\ln\psi_b(-\frac{2\pi i x}{\beta}+(r-1)\frac{b+b^{-1}}{2})$ over
$[0,N_0\beta]$, and we are done. Note that for
$\ln\psi_b(-\frac{2\pi i \{x\}}{\beta}+(r-1)\frac{b+b^{-1}}{2})$ to
not diverge at $x\in\mathbb{Z}$, we need
$r(\frac{b+b^{-1}}{2})\notin
b\mathbb{Z}^{\le0}+b^{-1}\mathbb{Z}^{\le0}$ and
$(r-2)(\frac{b+b^{-1}}{2})\notin
b\mathbb{Z}^{\ge0}+b^{-1}\mathbb{Z}^{\ge0}$; our constraint
$r\in]0,2[$ takes care of these.

To obtain (\ref{eq:GammaOffCenter2Vec}), we can apply the argument
of the previous paragraph, except that we do not get the uniform
bound on $[0,N_0\beta]$: our ``continuous function with a compact
support'' argument fails when $r=0$, because $\psi_b(-\frac{2\pi i
\{x\}}{\beta}-\frac{b+b^{-1}}{2})$ diverges at $x\in\mathbb{Z}$.
This is why (\ref{eq:GammaOffCenter2Vec}) applies uniformly only on
$x\in\mathbb{R}\backslash\mathbb{Z}^{(\beta)}$, with
$\mathbb{Z}^{(\beta)}$ an $O(\beta)$ neighborhood of $\mathbb{Z}$.

\section{Asymptotics of the Schur partition function of the $\mathcal{N}=4$ theory}\label{app:drukker}

The Schur partition function of the $\mathcal{N}=4$ theory is
exceptionally well under control, because of its connection with the
partition function of a free-fermion system on a circle. Employing
this connection, expressions for the Schur index of the
$\mathcal{N}=4$ theory were obtained in \cite{Bourdier:2015}, that
we asymptotically analyze in this appendix. We write down the
\emph{all-orders} small-$\beta$ expansion of the log of this
partition function.

Recall that the Rains function of the $\mathcal{N}=4$ theory
vanishes, and hence for this theory
$\mathrm{dim}\mathfrak{h}_{qu}=\mathrm{dim}\mathfrak{h}_{cl}$, which
when the gauge group is SU($N$) equals $N-1$. In the body of the
paper we were not able to obtain all-orders asymptotics for the
partition functions of theories with
$\mathrm{dim}\mathfrak{h}_{qu}>0$. The partition function studied in
this appendix is the only example with
$\mathrm{dim}\mathfrak{h}_{qu}>0$ for which we can write down
all-orders asymptotics.\\

Before spelling out the said partition function, we introduce the
mathematical technique required for its asymptotic analysis. This
technique is explained by Zagier in \cite{Zagier:2006}, but its
proof was omitted there. We now present the method, outline its
proof (mirroring a similar one in \cite{Zagier:2006}), and along the
way fix some typos in \cite{Zagier:2006}.

Consider a real function $G(\beta)$ defined in terms of another real
function $f(\beta)$ as
\begin{equation}
G(\beta)=\sum_{m=0}^{\infty}f((m+a)\beta),\label{eq:ZagierFormShift}
\end{equation}
with some $a\in[0,1[$. Assume that $f(\beta)$ has the small-$\beta$
asymptotic development
\begin{equation}
f(\beta)\sim \sum_{n=0}^{\infty}f_{n} \beta^{n},\label{eq:fAsy0}
\end{equation}
and assume that the integral $\int_0^\infty f(\beta)\mathrm{d}\beta$
exists, and that all the derivatives of $f(\beta)$ vanish faster
than $1/\beta^{1+\varepsilon}$ (with some $\varepsilon>0$) as
$\beta\to\infty$. Then, according to Zagier \cite{Zagier:2006}, the
small-$\beta$ asymptotics of $G(\beta)$ is given by
\begin{equation}
G(\beta)\sim\frac{I_f}{\beta}+\sum_{n=0}^{\infty}f_n \zeta(-n,a)\
\beta^{n},\label{eq:ZagierResultShift}
\end{equation}
with $I_{f}:=\int_{0}^{\infty}f(x)\mathrm{d}x$.

The proof goes as follows. Start with the Euler-MacLaurin formula
(see Chapter~8 of \cite{Olver:1974})
\begin{equation}
\begin{split}
\sum_{m=0}^{M-1}f(m+a)=&\int_0^M f(t)\mathrm{d}t
+\sum_{n=0}^{N-1}\frac{B_{n+1}(a)}{(n+1)!}(f^{(n)}(M)-f^{(n)}(0))\\
&+(-1)^{N+1}\int_0^M \frac{B_N(\{t-a\})}{N!}f^{(N)}(t)
\mathrm{d}t.\label{eq:EulerMacLaurin}
\end{split}
\end{equation}

In the above equation, we have assumed $0\le a<1$, we have used the
Bernoulli polynomials $B_i(x)$, and employed the fractional-part
function $\{z\}=z-\lfloor z \rfloor$.

Scaling the argument of $f(\ast)$ as $f(\ast\cdot x)$, taking the
limit $M\to\infty$, and recalling that for all $n\ge 0$ we have
$f^{(n)}(M)\to 0$ as $M\to\infty$, we arrive at
\begin{equation}
\begin{split}
\sum_{m=0}^{\infty}f((m+a)x)=&\frac{1}{x}\int_0^\infty
f(t)\mathrm{d}t+\sum_{n=0}^{N-1}f_n\zeta(-n,a) x^n\\
&+\left[(-1)^{N+1}\int_0^\infty \frac{B_N(\{t/x-a\})}{N!}f^{(N)}(t)
\mathrm{d}t\right] x^{N-1}.\label{eq:EulerMacLaurinScaled}
\end{split}
\end{equation}
We have used the Hurwitz zeta $\zeta(-n,a)=B_{n+1}(a)/(n+1)$ instead
of the Bernoulli polynomials. Recall also that the coefficients of
the asymptotic expansion of $f(x)$ around zero are given by
$f_n=f^{(n)}(0)/n!$.

Since $N$ can be taken to be arbitrarily large, Eq.
(\ref{eq:EulerMacLaurinScaled}) establishes
(\ref{eq:ZagierResultShift}).\\

Armed with the above technique, we now analyze the Schur partition
function of the SU($N$) $\mathcal{N}=4$ SYM. The Schur index of this
theory is observed in \cite{Bourdier:2015} to be proportional to the
partition function of a free-fermion system on a circle. This
free-fermion partition function $Z(N)$ is determined via
\begin{equation}
Z(N)=\sum_{\sum_{\ell}\ell
m_{\ell}=N}\prod_{\ell}(-1)^{(\ell-1)m_\ell}\frac{Z_\ell^{m_\ell}}{m_\ell
! \ell^{m_\ell}},\label{eq:ZNandZl}
\end{equation}
in terms of the so-called spectral traces $Z_\ell$, given by
\begin{equation}
Z_\ell=\sum_{p\in\mathbb{Z}}\left(\frac{1}{q^{\frac{p}{2}-\frac{1}{4}}+q^{-\frac{p}{2}+\frac{1}{4}}}\right)^\ell.
\end{equation}

The claim in \cite{Bourdier:2015} (see also \cite{Bourdier:2015b})
is that
\begin{equation}
\mathcal{I}^{Schur}_{SU(N)\
\mathcal{N}=4}(\beta)=\frac{q^{-(N^2-1)/8}}{\Delta_N}\frac{\eta^2(\tau/2)}{\eta^4(\tau)}Z(N),\label{eq:drukkerFiniteN}
\end{equation}
where $\Delta_N$ is given for odd $N$ by
\begin{equation}
\Delta(N)=\frac{\vartheta_2}{\vartheta_3}=\frac{2\eta^4(2\tau)\eta^2(\tau/2)}{\eta^6(\tau)},
\end{equation}
while for even $N$ we have $\Delta_{N}=1$. (See \cite{Bourdier:2015}
for the definition of the functions $\vartheta_2,\vartheta_3$.)

Combining (\ref{eq:drukkerFiniteN}) and (\ref{eq:lagSchurZ}) we
obtain
\begin{equation}
Z^{Schur}_{SU(N)\
\mathcal{N}=4}(\beta)=\frac{1}{\Delta_N}\frac{\eta^2(\tau/2)}{\eta^4(\tau)}Z(N).\label{eq:ZdrukkerFiniteN}
\end{equation}

To analyze the high-temperature asymptotics of $Z^{Schur}_{SU(N)\
\mathcal{N}=4}(\beta)$ we first note that
\begin{equation}
\ln\Delta(N)\sim 0\quad \quad (\beta\to 0),\label{eq:DeltaAsy}
\end{equation}
irrespective of whether $N$ is even or odd. This follows from the
asymptotics of $\eta(\tau)$, and we leave its verification to the
interested reader.

Next, we rewrite the spectral traces, in a way that makes the
application of Zagier's method straightforward
\begin{equation}
\begin{split}
Z_\ell&=\sum_{p\ge1}\left(\frac{1}{q^{\frac{p}{2}-\frac{1}{4}}+q^{-\frac{p}{2}+\frac{1}{4}}}\right)^\ell
+\sum_{p\ge0}\left(\frac{1}{q^{-\frac{p}{2}-\frac{1}{4}}+q^{\frac{p}{2}+\frac{1}{4}}}\right)^\ell\\
&=2\sum_{p\ge0}\left(\frac{1}{q^{-\frac{p}{2}-\frac{1}{4}}+q^{\frac{p}{2}+\frac{1}{4}}}\right)^\ell\\
&=\sum_{p\ge0}f_\ell((p+1/2)\beta),
\end{split}
\end{equation}
with
\begin{equation}
f_\ell(\beta)=2\left(\frac{1}{e^{\beta/2}+e^{-\beta/2}}\right)^\ell.
\end{equation}

Applying (\ref{eq:ZagierResultShift}), and using
\begin{equation}
\begin{split}
\zeta(s,1/2)=(2^s-1)\zeta(s),\label{eq:zetaHalfHur}
\end{split}
\end{equation}
we find that
\begin{equation}
\begin{split}
Z_\ell\sim\frac{I_{f_\ell}}{\beta},\label{eq:ZlAsy}
\end{split}
\end{equation}
to all orders in $\beta$, and with
\begin{equation}
I_{f_\ell}=2\int_{0}^{\infty}\mathrm{d}x\
\left(\frac{1}{e^{x/2}+e^{-x/2}}\right)^\ell=\frac{1}{2^{\ell-2}}\frac{(\ell-2)!!}{(\ell-1)!!}\times
\begin{cases}
\frac{\pi}{2}\quad &(\ell \text{ odd}),\\
1\quad &(\ell \text{ even}).
\end{cases}
\end{equation}
Note that $I_{f_1}=\pi$.

Since $Z_{\ell}$ is (asymptotically) inversely proportional to
$\beta$, the leading behavior of $Z(N)$ is found from
(\ref{eq:ZNandZl}) to be
\begin{equation}
Z(N)\approx \frac{(\pi/\beta)^N}{N!},\label{eq:ZNasy}
\end{equation}
with an error that is down by a factor of $\beta$.

Combining (\ref{eq:ZNasy}), (\ref{eq:DeltaAsy}), and
(\ref{eq:ZdrukkerFiniteN}), we arrive at
\begin{equation}
\ln Z^{Schur}_{SU(N)\ \mathcal{N}=4}(\beta)=
(N-1)\ln(\frac{2\pi}{\beta})-(N-1)\ln 2-\ln N!+o(1),\label{eq:INasy}
\end{equation}
in perfect accord with (\ref{eq:SUNschurIndexAsy}).\\

Note that since (\ref{eq:ZlAsy}) is all-orders exact, we can combine
it with (\ref{eq:ZNandZl}) and (\ref{eq:ZdrukkerFiniteN}) to write
the all-orders asymptotic relation
\begin{equation}
\ln Z^{Schur}_{SU(N)\ \mathcal{N}=4}(\beta)\sim
\ln(P_{N-1}(\beta)/\pi),\label{eq:INasyFull}
\end{equation}
with $P_{N-1}(\beta)$ the degree $N-1$ polynomial in $1/\beta$
defined by
\begin{equation}
P_{N-1}(\beta)=\sum_{\sum_{\ell}\ell
m_{\ell}=N}\left(\frac{1}{\beta}\right)^{\sum
m_\ell-1}\prod_{\ell}(-1)^{(\ell-1)m_\ell}\frac{I_{f_\ell}^{m_\ell}}{m_\ell
! \ell^{m_\ell}}.\label{eq:PN}
\end{equation}

The SU($2$) case is easy to analyze explicitly. Since $I_{f_1}=\pi$
and $I_{f_2}=1$, we have
\begin{equation}
\ln Z^{Schur}_{SU(2)\ \mathcal{N}=4}(\beta)\sim
\ln(\frac{\pi}{2\beta}-\frac{1}{2\pi}).\label{eq:ZSchur2asyFull}
\end{equation}

\end{document}